\documentclass[twocolumn,tighten,twocolappendix]{aastex7}
\usepackage{array}
\usepackage{booktabs}
\usepackage{tabularx}
\usepackage{ragged2e}
\usepackage{scalerel}
\usepackage{appendix}
\usepackage{blindtext}
\usepackage{booktabs}
\usepackage{longtable}
\usepackage[referable]{threeparttablex}
\usepackage{wrapfig}
\usepackage{longtable,tabu}
\usepackage{booktabs}
\usepackage{amsmath}
\usepackage{longtable}
\usepackage{tabularx}
\usepackage{amsmath}
\usepackage{array}
\usepackage{booktabs}
\usepackage{ragged2e}
\usepackage{graphicx}

\usepackage{natbib}
\usepackage{float}
\usepackage{tablefootnote}
\usepackage{apjfonts}
\usepackage{lipsum}
\usepackage{afterpage}
\usepackage{url}
\usepackage{gensymb}

\usepackage{hyperref}
\hypersetup{colorlinks=true, citecolor=blue, linkcolor=blue, urlcolor=blue}

\usepackage[T1]{fontenc}
\usepackage{beramono}
\usepackage{listings}

\DeclareUnicodeCharacter{2212}{-}
\DeclareUnicodeCharacter{0177}{\^y}
\shorttitle{On the TF Relation for AGN Hosts}
\shortauthors{Robinson et al.}

\begin{document}
\title{On the Tully-Fisher Relation for Active Galaxies -- I: Evidence of Larger Scatter}

\author[0000-0002-4262-4845]{Justin H. Robinson}
\affiliation{\normalfont \centering Department of Physics and Astronomy, Georgia State University, Atlanta, GA 30303, USA} 
\email{jrobinson138@gsu.edu}

\author[0000-0002-4917-7873]{Mitchell Revalski}
\affiliation{\normalfont \centering Space Telescope Science Institute, 3700 San Martin Drive, Baltimore, MD 21218, USA}
\email{mrevalski@stsci.edu}

\author[0000-0002-2816-5398]{Misty C. Bentz}
\affiliation{\normalfont \centering Department of Physics and Astronomy, Georgia State University, Atlanta, GA 30303, USA}
\email{bentz@gsu.edu}

\author[0000-0002-6465-3639]{D. Michael Crenshaw}
\affiliation{\normalfont \centering Department of Physics and Astronomy, Georgia State University, Atlanta, GA 30303, USA}
\email{dcrenshaw@gsu.edu}

\author[0000-0003-0509-1776]{H\'el\`ene M. Courtois}
\affiliation{\normalfont \centering Universit\'e Claude Bernard Lyon 1, IUF, IP2I Lyon, 4 rue Enrico Fermi, 69622 Villeurbanne, France}
\email{h.courtois@ip2i.in2p3.fr}

\author[0009-0000-3960-167X]{Veronica Lahue}
\affiliation{\normalfont \centering Department of Chemistry and Physics, Troy University, Troy, AL 36081, USA}
\email{vlahue@troy.edu}

\author[0000-0001-7238-7062]{Julian Falcone}
\affiliation{\normalfont \centering Department of Physics and Astronomy, University of South Carolina, Columbia, SC 29208, USA}
\email{jfalcone2@sc.edu}

\author[0000-0003-0792-863X]{Rachael L.\ Merritt}
\affiliation{\normalfont \centering Department of Physics, University of Colorado Boulder, Boulder, CO 80309, USA}
\affiliation{\normalfont \centering JILA, National Institute of Standards and Technology and the University of Colorado, Boulder, CO 80309, USA }
\email{Rachael.Merritt@colorado.edu}

\author{Ishita Chintala}
\affiliation{\normalfont \centering School of Physics, Georgia Institute of Technology, 837 State Street, Atlanta, GA 30332-0430,
USA}
\email{ichintala3@gatech.edu}

\author{Thomas Kay}
\affiliation{\normalfont \centering Department of Chemistry and Physics, Troy University, Troy, AL 36081, USA}
\email{tkay@troy.edu}

\author{Mira Menon}
\affiliation{\normalfont \centering LAMP High School, Montgomery, AL, 36111, USA}
\email{astro.grav@icloud.com}

\author{Naveen Ali}
\affiliation{\normalfont \centering School of Physics, Georgia Institute of Technology, 837 State Street, Atlanta, GA 30332-0430,
USA}
\email{nali68@gatech.edu}

\author{Bradley Clemons}
\affiliation{\normalfont \centering Department of Physics and Astronomy, Georgia State University, Atlanta, GA 30303, USA}
\email{jrobinson138@gsu.edu}

\author{Atul Gautam}
\affiliation{\normalfont \centering School of Physics, Georgia Institute of Technology, 837 State Street, Atlanta, GA 30332-0430,
USA}
\email{agautam70@gatech.edu}

\author{Thomas Gregoire}
\affiliation{\normalfont \centering Department of Physics and Astronomy, Georgia State University, Atlanta, GA 30303, USA}
\email{jrobinson138@gsu.edu}

\begin{abstract}
We present an investigation of the Tully-Fisher (TF) relation solely for galaxies hosting an active galactic nucleus (AGN). Using 22 galaxies with primary, $z-$independent distances, we find that active galaxies exhibit significantly larger scatter about all TF relations compared to each respective calibration for (largely) inactive galaxies. The larger scatter persists despite removal of the AGN contamination from the photometry of the Type 1 AGNs via 1) careful surface brightness decompositions or 2) employing SEDs to constrain the light contribution of the AGN. These results suggest that the influence of an AGN on its host galaxy's surface brightness may extend beyond the nucleus. We also calculate the percentage difference between TF and primary distances, and find that TF-based distances are biased towards overestimation of the primary distances to active galaxies by anywhere from $\sim$$5 - 10$\% for the optical/near-infrared and $\sim$15\% for distances predicted from inverting the Baryonic TF (BTF) relation. As TF-based distances (especially the $I−$band) are relied on heavily for analysis and modeling of the local peculiar velocity ($V_{\mathrm{PEC}}$) field, we suggest that active galaxies be removed from future $V_{\mathrm{PEC}}$ modeling samples.

\end{abstract}

\received{}
\accepted{June 19, 2026}

\section{Introduction} \label{sec:intro}

The Tully-Fisher (TF) relation \citep{tf1977} is an empirical correlation that places galaxy rotational velocity on a mass scale, with higher-mass late-type galaxies generally achieving faster rotational velocities. As stellar light traces mass, the observed empirical relationship is that of maximum rotational velocity (measured either from resolved rotation curves or the deprojected line width of the unresolved \ion{H}{1} 21\,cm emission line) and intrinsic brightness. Measurement of rotational velocity then predicts a corresponding absolute magnitude in a given bandpass, and comparison with the apparent magnitude constrains a distance. The Cosmicflows program (\citealt{cosmicflows1,cosmicflows2,cosmicflows3,cosmicflows4}; CF) is home to the largest catalog of $\sim$10,000 TF-based distances. Each calibration establishes a line-width-luminosity relation in each photometric bandpass by employing galaxy clusters \citep{tc2012,sorce2013,neill2014}, 
allowing a universal TF slope to be determined from multiple, independent correlations of rotation rate and apparent luminosities while also decoupling it from the contribution of peculiar velocity ($V_{\mathrm{PEC}}$). The calibrating sample includes $\sim$600 galaxies from 20 different clusters \citep{Kourkchi2020}, with zero-points in absolute magnitude provided by primary, $z-$independent distances as measured from Leavitt’s Law (Cepheid period–luminosity relationship; \citealt{lp1912}), tip of the red giant branch stars (TRGB; \citealt{iben1983,salaris1997}) surface brightness fluctuations (SBF; \citealt{tonry2001,Blakeslee2010}), type Ia supernovae (SN Ia; \citealt{phillips1993,Hamuy1996,Riess1996,Perlmutter1997,Jha2007}) type II supernovae (SN II; \citealt{Barbon1979,Doggett1985,woosley1987,Schlegel1990}), or geometric distances. 

The TF relation is also routinely employed for observational constraints on the Hubble constant $H_{0}$ (e.g., \citealt{Sakai2000,tp2000,Freedman2001,Masters2006}, CF, \citealt{Kourkchi2020}). Recent efforts have used the relation in combination with the CF program's model of the local $V_{\mathrm{PEC}}$ field \citep{shaya2017,graziani2019} for improved determinations of $H_{0}$ \citep{Boubel2024,Scolnic2024}. Although the TF relation remains a critical rung on the Cosmic Distance Ladder, the inherent scatter \citep{cosmicflows1,tc2012} limits its utility, necessitating its role as a secondary distance measurement technique rather than a primary one. However, the sheer quantity of TF-based distances compensates for the limited quality of individual distances, and the development of such large samples has allowed robust studies of the physical drivers of the observed scatter about, and sources of deviation from, the TF relation. 

It has been shown that the TF slope, scatter, or both can be affected by galaxy disc sizes \citep{reyes2011}, the ratio of dark to luminous matter \citep{Pizagno2007}, star formation rate \citep{Barton2001,Buchalter2001}, and/or cluster and group environments (\citealt{Ouellette2017} and references therein). Star formation rate and history specifically have long been shown to directly affect the scatter \citep{bd2001,Kannappan2002}, where bursts of star formation can elevate galaxy surface brightnesses above that inferred by their rotational velocities. Deviations in blue and near-infrared wavelengths have also been attributed to the presence of an active galactic nucleus (AGN), specifically nuclei classified with Seyfert activity \citep{Torres-Flores2013}. The analysis of \cite{Ristea2024} found that the scatter of the TF relation may be due to internal feedback processes such as star formation or AGN activity, processes that have been shown to be connected in the literature \citep{page2001,alexander2005,Netzer2007,Silverman2009,Santini2012,Azadi2015,Mountrichas2023_2}. 

In addition to the possible impact on host galaxy surface brightnesses and thus the absolute magnitude of a galaxy on the TF relation, AGN feedback can also affect galaxy-scale disk kinematics, which consequently may impact maximum rotational velocity on the TF relation. The study by \cite{Frosst2022} of 100 zoom-in simulations of isolated galaxies found evidence of significant disruption in disk kinematics compared to inactive galaxies when supermassive black hole (SMBH) feedback was included as a parameter. Observationally, recent efforts to quantify AGN-driven ionized mass outflows in the narrow-line region (NLR) have measured outflows to span kiloparsec scales \citep{of2006,revalski2018,Revalski2018_2,Revalski2021,Revalski2022,Revalski2025}. Disruption of disk-scale kinematics has even been observed in low-mass galaxies \citep{Manzano-King2019,Manzano-King2020} and in hosts of low-ionization nuclear emission-line regions (LINERS; Sankey et al. 2026, in prep). As the presence of an AGN has been shown to significantly affect both the surface brightness and the disk kinematics of their host galaxies, there is then a need for investigation of the adherence of active galaxies on the canonical TF relation and thus the accuracy of TF-based distances to active galaxies.

A direct investigation of active galaxies on the TF relation has been precluded by lack of quality distances. Only in the last $\sim$10 years have AGN hosts been routinely targeted for $z-$independent distance techniques other than the TF method \citep{sabbi2018,misty2019,yuan2020,yuan4051,anand2021,gravity2021,riess2022,markham2026}. With nearly 30 $z-$independent distances for AGN hosts now available, we present here an investigation of active galaxies on the TF relation. In \S\ref{section:data}, we present the current sample of AGN host galaxies with 1) $z-$independent, non-TF-based distances, 2) \ion{H}{1} 21\,cm spectral observations, 3) broad-band optical, near-infrared, and/or infrared photometry, including a brief description of the two-dimensional surface brightness decompositions of the nuclear contamination from the optical and near-infrared images of the Seyfert 1 hosts. In \S\ref{section:construction_tf}, we describe our derivations of deprojected rotational velocity and corrected absolute magnitude following the TF prescriptions in the literature to construct TF relations for active galaxies, which we present in \S\ref{section:results}. Finally, in \S\ref{section:discussion}, we place the TF and Baryonic TF (BTF) relations for active galaxies in context with known sources of deviation from and/or scatter about the canonical TF relations.



\section{Data}\label{section:data}
We construct here the current sample of active galaxies that 1) have optical, near-infrared, and/or infrared imaging, 2) have been targeted for \ion{H}{1} 21\,cm spectroscopic observations, and 3) have primary, $z-$independent, non-TF-based distance measurements. The 28 galaxies that comprise the sample are listed in Table~\ref{basics}, in addition to their nuclear activity classification and distance to each AGN, and we describe here all photometric and spectroscopic data employed in this investigation.

We note here that we do not expect interacting systems to lie on the canonical TF relation, as galaxy disks are likely to be dynamically disrupted from typical rotation by interactions, collisions, and/or mergers. The interacting galaxies of this sample are tabulated in the bottom panel of each table and are as follows: NGC 3227, NGC 3786, NGC 4438, NGC 5194, NGC 7469, and NGC 7674, and are not included in any quoted statistics in this work. Therefore, the sample size used to investigate AGN hosts on the TF relation is 22.

\begin{deluxetable*}{lccccccc}
\tabletypesize{\footnotesize}
\tablecaption{AGN Host Galaxy Sample Characteristics}
\tablehead{
\colhead{Target} & \colhead{Redshift} & \colhead{Hubble} & \colhead{Distance} & \colhead{Distance} & \colhead{Ref} & \colhead{AGN} & \colhead{Ref}\\
\colhead{Name} & \colhead{($z$)} & \colhead{Type} &  \colhead{(Mpc)} & \colhead{Method} & & \colhead{Type} & \\
\colhead{(1)} & \colhead{(2)} &\colhead{(3)} & \colhead{(4)} & \colhead{(5)} & \colhead{(6)} & \colhead{(7)} & \colhead{(8)}
}
\startdata
NGC 1068 & 0.00379 & (R)SA(rs)b & 10.72 $^{+ 0.52 }_{- 0.52 }$ & PL & \cite{markham2026} & 2 & 1\\
NGC 1566 & 0.00499 & SAB(s)bc & 18.00 $^{+ 2.00 }_{- 2.00 }$ & TRGB & \cite{sabbi2018} & 1 & 2 \\
NGC 3147 & 0.00941 & SA(rs)bc & 39.30 $^{+ 1.63 }_{- 1.63 }$ & SNIa & CF2 & 2 & 3 \\
NGC 3783* & 0.00973 & (R')SB(r)a & 39.90 $^{+ 14.50 }_{- 11.90 }$  & Geo.\ & \cite{misty2021} & 1  & 4\\
& & & & & \cite{gravity2021} \\
NGC 3982 & 0.00370 & SAB(r)b & 22.12 $^{+ 0.73 }_{- 0.73 }$ & PL & \cite{riess2022} & 1.9 & 5 \\
NGC 4051* & 0.00235 & SAB(rs)bc & 16.60 $^{+ 0.30 }_{- 0.30 }$ & PL & \cite{yuan4051} & 1 & 4  \\
NGC 4138 & 0.00308 & SA0(r) & 13.80 $^{+ 1.65 }_{- 1.65 }$ & SBF & \cite{tonry2001} & 2 & 2  \\
& & & & & \cite{Blakeslee2010} \\
NGC 4151* & 0.00333 & (R')SAB(rs)ab & 15.80 $^{+ 0.40 }_{- 0.40 }$ & PL & \cite{yuan2020} & 1 & 4 \\
NGC 4258 & 0.00149 & SAB(s)bc & 7.54 $^{+ 0.20 }_{- 0.20 }$ & Geo.\ & \cite{Humphreys2013} & L & 6 \\
& & & & & \cite{riess2016} \\
NGC 4303 & 0.00523 & SAB(rs)bc & 16.47 $^{+ 0.27 }_{- 0.27 }$ & PL & \cite{markham2026} & 2 & 5  \\
NGC 4395 & 0.00107 & SA(s)m & 4.10 $^{+ 0.40 }_{- 0.40 }$ & PL & \cite{thim2004} & 1 & 5 \\
NGC 4565 & 0.00409 & SA(s)b & 11.90 $^{+ 0.30 }_{- 0.20 }$ & TRGB & \cite{Radburn-Smith2011} & 2 & 3  \\
NGC 4639 & 0.00337 & SAB(rs)bc & 20.25 $^{+ 0.66 }_{- 0.66 }$ & PL & \cite{riess2016} & 1 & 7  \\
NGC 4939 & 0.01038 & SA(s)bc & 43.75 $^{+ 1.21 }_{- 1.21 }$ & SNII & \cite{jaeger2017,jaeger2017_2} & 2 & 2  \\
NGC 4945 & 0.00187 & SB(s)cd & 3.47 $^{+ 0.05 }_{- 0.05 }$ & TRGB & \cite{anand2021} & 2 & 2 \\
NGC 5055 & 0.00168 & SA(rs)bc & 8.87 $^{+ 0.31 }_{- 0.31 }$ & TRGB & \cite{McQuinn2017} & L & 8 \\
NGC 5643 & 0.00400 & SAB(rs)c & 12.42 $^{+ 0.63 }_{- 0.63 }$ & TRGB & \cite{anand2021} & 2 & 2 \\
NGC 5728 & 0.00930 & SAB(r)a & 42.27 $^{+ 1.75 }_{- 1.75 }$ & SNIa & \cite{Burns2018} & 1.9 & 2  \\
NGC 6814* & 0.00521 & SAB(rs)bc & 21.60 $^{+ 0.40 }_{- 0.40 }$ & PL & \cite{misty2019} & 1 & 4  \\
NGC 6951 & 0.00476 & SAB(rs)bc & 18.79 $^{+ 0.87 }_{- 0.87 }$ & SNIa & CF2 & L & 9 \\
M 58 & 0.00507 & SAB(rs)b & 21.00 $^{+ 2.00 }_{- 2.00 }$ & SNIa & \cite{Ruiz-Lapuente1996} & L & 3  \\
M 104 & 0.00364 & SA(s)a & 9.55 $^{+ 0.13 }_{- 0.31 }$ & TRGB & \cite{McQuinn2016} & L & 8 \\
\tableline
\multicolumn{8}{l}{\text{Interacting Galaxies}} \\
\tableline
NGC 3227* & 0.00382 & SAB(s)a pec & 23.70 $^{+ 2.60 }_{- 2.60 }$ & SBF & \cite{tonry2001} & 1 & 4 \\
& & & & & \cite{Blakeslee2010} \\
NGC 3786 & 0.00908 & SAB(rs)a pec & 49.68 $^{+ 4.92 }_{- 4.92 }$ & SNIa &\cite{koshida2017} &  1.9 & 2\\
NGC 4438 & 0.00023 & SA0/a(s) pec & 16.50 $^{+ 1.30 }_{- 1.30 }$ & SBF & \cite{cantiello2018} & L & 5  \\
NGC 5194 & 0.00154 & SA(s)bc pec & 7.40 $^{+ 0.60 }_{- 0.60 }$ & TRGB & \cite{sabbi2018} & 2 & 8, 10  \\
NGC 7469* & 0.01644 & (R')SAB(rs)a & 61.90 $^{+ 3.30 }_{- 3.30 }$ & SNIa & \cite{koshida2017} & 1 & 1 \\
& & & & & \cite{Ganeshalingam2013} \\
NGC 7674 & 0.02928 & SA(r)bc pec & 117.00 $^{+ 3.77 }_{- 3.77 }$ & SNIa & \cite{Weyant2014} & 2 & 1  \\
\enddata
\tablecomments{The sample of active galaxies that have 1) optical, near-infrared, and/or infrared imaging, 2) \ion{H}{1} 21\,cm spectra, and 3) $z-$independent, non-TF-based distances. Galaxy names marked with asterisks indicate galaxies consistent with the sample of Paper I (see \S\ref{section:data}). Column 2 lists the spectroscopic redshift of the \ion{H}{1} 21\,cm line (adopted from Paper I for all consistent galaxies and from RC3 for all remaining targets). Column 3 lists the Hubble morphological classification (adopted from Paper I for all consistent galaxies and from NED for all remaining targets). Distances to each AGN host are listed in Column 4 with their respective measurement methods in Column 5; TRGB stands for Tip of the Red Giant Branch stars, SNIa for Supernova Type Ia, SBF for Surface Brightness Fluctuations, Geo.\ for geometric-based distance, PL for Cepheid Period-Luminosity relation, and SNII for Supernova Type II). Column 6 lists the references for each distance measurement. Column 7 lists the nuclear activity classification, the references for each classification listed in Column 8 and are as follows: 1) \cite{Osterbrock1993}, 2) \cite{koss2022}, 3) \cite{panessa2002}, 4) \cite{kw1974}, 5) \cite{vv2006}, 6) \cite{cecil2000}, 7) \cite{ho1997}, 8) \cite{ho1997}, 9) \cite{perez2000}, 10) \cite{ford1985}. The bottom panel denotes interacting systems, which are not included in any quoted statistics in this work.}
\label{basics}
\end{deluxetable*}
\subsection{Photometry}\label{section:photometry}
\subsubsection{Optical and Near-Infrared Photometry}
For the majority of our sample, we adopt apparent $B$ and $V$ magnitudes from the Third Reference Catalog of Bright Galaxies (RC3; \citealt{dv1991}) as was done for previous TF calibrations (e.g., \citealt{tf1985,tp2000,cosmicflows1}). Apparent magnitudes in the $R$ and $I$ are adopted from the CF program's Homogenized Photometry, tabulated in the Extragalactic Distance Database (EDD; \citealt{edd}). These measurements were generally adopted from photometric campaigns carried out by members of the CF program using the University of Hawaii 2.2\,m telescope, the Canada–
France–Hawaii Telescope, the United Kingdom Infrared
Telescope, the 1\,m and 3.9\,m telescopes
at Siding Spring Observatory, the Kitt Peak National Observatory and CTIO 0.9\,m telescopes, and the MDM Observatory 1.3\,m
McGraw-Hill telescope. Finally, all infrared magnitudes in the $W_{1}$ and $W_{2}$ filters are adopted from the Wide-field Infrared Survey Explorer (WISE; \citealt{wise}).

\begin{deluxetable*}{lcccccccccc}
\tabletypesize{\footnotesize}
\tablecaption{Observed Photometric and Spectroscopic Galaxy Properties}
\tablehead{
\colhead{Target} & \colhead{$B$} & \colhead{$V$} & \colhead{$R$} & \colhead{$I$} & \colhead{$W_{1}$} & \colhead{$W_{2}$} & \colhead{$q_{d}$}  & \colhead{$W_{20}$} & \colhead{$W_{\mathrm{m}50}$} & \colhead{Flux}
\\
& (mag) & (mag) & (mag) & (mag) & (mag) & (mag) & & (km s$^{-1}$) & (km s$^{-1}$) & (Jy km s$^{-1}$) \\
\colhead{(1)} & \colhead{(2)} &\colhead{(3)} & \colhead{(4)} & \colhead{(5)} & \colhead{(6)} & \colhead{(7)} & \colhead{(8)} & \colhead{(9)} & \colhead{(10)} & \colhead{(11)}
}
\startdata
NGC 1068 & 9.61 $\pm$ 0.10 & 8.87 $\pm$ 0.10 & 8.46 $\pm$ 0.10 & 7.80 $\pm$ 0.10 & 4.17 $\pm$ 0.20 & 3.04 $\pm$ 0.30 & 0.85 $\pm$ 0.04 & 299.0 $\pm$ 8.0 & 300.0 $\pm$ 8.0 & 31.2 \\
NGC 1566 & 10.33 $\pm$ 0.03 & 9.73 $\pm$ 0.03 & 9.18 $\pm$ 0.10 & 8.64 $\pm$ 0.11 & 8.91 $\pm$ 0.10 & 8.62 $\pm$ 0.10 & 0.79 $\pm$ 0.04 & 229.0 $\pm$ 6.0 & 213.0 $\pm$ 13.0 & 141.6 \\
NGC 3147 & 11.43 $\pm$ 0.16 & 10.61 $\pm$ 0.16 & \nodata & \nodata & 8.99 $\pm$ 0.10 & 8.88 $\pm$ 0.10 & 0.89 $\pm$ 0.06 & 418.0 $\pm$ 9.0 & 393.0 $\pm$ 15.0 & 24.1 \\
NGC 3783* & 12.89 $\pm$ 0.20 & 12.09 $\pm$ 0.20 & 11.50 $\pm$ 0.20 & \nodata & 8.38 $\pm$ 0.10 & 7.38 $\pm$ 0.10 & 0.96 $\pm$ 0.02 & 152.5 $\pm$ 1.5 & 147.0 $\pm$ 8.0 & 10.2 \\
NGC 3982 & 11.78 $\pm$ 0.16 & 11.59 $\pm$ 0.04 & 11.20 $\pm$ 0.10 & 10.77 $\pm$ 0.10 & 10.34 $\pm$ 0.10 & 10.07 $\pm$ 0.10 & 0.87 $\pm$ 0.06 & 232.0 $\pm$ 7.0 & 226.0 $\pm$ 18.0 & 15.3 \\
NGC 4051* & 10.84 $\pm$ 0.20 & 10.11 $\pm$ 0.20 & 9.93 $\pm$ 0.20 & 9.37 $\pm$ 0.10 & 8.87 $\pm$ 0.10 & 7.97 $\pm$ 0.10 & 0.58 $\pm$ 0.05 & 264.5 $\pm$ 0.1 & 245.0 $\pm$ 8.0 & 30.7 \\
NGC 4138 & 12.16 $\pm$ 0.15 & 11.32 $\pm$ 0.16 & 10.72 $\pm$ 0.10 & 10.09 $\pm$ 0.10 & 9.40 $\pm$ 0.10 & 9.27 $\pm$ 0.10 & 0.66 $\pm$ 0.06 & 329.0 $\pm$ 10.0 & 322.0 $\pm$ 18.0 & 14.2 \\
NGC 4151* & 11.29 $\pm$ 0.20 & 10.80 $\pm$ 0.20 & 10.18 $\pm$ 0.20 & 9.73 $\pm$ 0.20 & 6.74 $\pm$ 0.10 & 5.93 $\pm$ 0.10 & 0.94 $\pm$ 0.05 & 139.6 $\pm$ 0.4 & 131.0 $\pm$ 8.0 & 36.3 \\
NGC 4258 & 9.10 $\pm$ 0.07 & 8.41 $\pm$ 0.04 & 7.88 $\pm$ 0.10 & 7.32 $\pm$ 0.09 & 8.53 $\pm$ 0.10 & 8.19 $\pm$ 0.10 & 0.39 $\pm$ 0.02 & 442.0 $\pm$ 5.0 & 385.0 $\pm$ 20.0 & 162.0 \\
NGC 4303 & 10.18 $\pm$ 0.09 & 9.65 $\pm$ 0.09 & 9.17 $\pm$ 0.10 & 8.81 $\pm$ 0.10 & 9.02 $\pm$ 0.10 & 8.99 $\pm$ 0.10 & 0.89 $\pm$ 0.05 & 178.0 $\pm$ 5.0 & 163.0 $\pm$ 8.0 & 84.9 \\
NGC 4395 & 10.55 $\pm$ 0.20 & 10.06 $\pm$ 0.20 & 9.72 $\pm$ 0.10 & 9.17 $\pm$ 0.10 & 12.62 $\pm$ 0.10 & 11.84 $\pm$ 0.10 & 0.83 $\pm$ 0.04 & 134.0 $\pm$ 5.0 & 120.0 $\pm$ 8.0 & 284.3 \\
NGC 4565 & 10.42 $\pm$ 0.07 & 9.58 $\pm$ 0.09 & 9.02 $\pm$ 0.10 & 8.25 $\pm$ 0.10 & 7.96 $\pm$ 0.10 & 8.20 $\pm$ 0.10 & 0.20 $\pm$ 0.00 & 528.0 $\pm$ 5.0 & 514.0 $\pm$ 8.0 & 161.0 \\
NGC 4639 & 12.24 $\pm$ 0.10 & 11.54 $\pm$ 0.10 & 11.09 $\pm$ 0.10 & 10.37 $\pm$ 0.17 & 10.08 $\pm$ 0.10 & 10.00 $\pm$ 0.10 & 0.68 $\pm$ 0.05 & 303.0 $\pm$ 7.0 & 299.0 $\pm$ 8.0 & 16.4 \\
NGC 4939 & 11.90 $\pm$ 0.20 & 11.26 $\pm$ 0.21 & 10.85 $\pm$ 0.10 & 10.34 $\pm$ 0.10 & 10.05 $\pm$ 0.10 & 9.69 $\pm$ 0.10 & 0.51 $\pm$ 0.04 & 463.0 $\pm$ 5.0 & 459.0 $\pm$ 18.0 & 45.1 \\
NGC 4945 & 9.30 $\pm$ 0.20 & \nodata & 7.27 $\pm$ 0.10 & 7.32 $\pm$ 0.70 & 7.77 $\pm$ 0.10 & 6.50 $\pm$ 0.10 & 0.20 $\pm$ 0.01 & 384.0 $\pm$ 9.0 & 374.0 $\pm$ 13.0 & 322.6 \\
NGC 5055 & 9.31 $\pm$ 0.10 & 8.59 $\pm$ 0.10 & 8.02 $\pm$ 0.10 & 7.52 $\pm$ 0.14 & 8.37 $\pm$ 0.10 & 8.33 $\pm$ 0.10 & 0.58 $\pm$ 0.03 & 400.0 $\pm$ 5.0 & 390.0 $\pm$ 1.0 & 483.8 \\
NGC 5643 & 10.74 $\pm$ 0.14 & 10.00 $\pm$ 0.14 & 9.75 $\pm$ 0.10 & 9.12 $\pm$ 0.10 & 9.41 $\pm$ 0.10 & 8.76 $\pm$ 0.10 & 0.87 $\pm$ 0.04 & 209.0 $\pm$ 8.0 & 202.0 $\pm$ 13.0 & 57.8 \\
NGC 5728 & 12.27 $\pm$ 0.14 & 11.35 $\pm$ 0.14 & \nodata & 10.20 $\pm$ 0.10 & 9.36 $\pm$ 0.10 & 9.03 $\pm$ 0.10 & 0.58 $\pm$ 0.04 & 413.0 $\pm$ 8.0 & 410.0 $\pm$ 19.0 & 10.2 \\
NGC 6814* & 12.16 $\pm$ 0.30 & 11.18 $\pm$ 0.20 & 10.62 $\pm$ 0.20 & 9.91 $\pm$ 0.20 & 9.35 $\pm$ 0.10 & 8.81 $\pm$ 0.10 & 0.98 $\pm$ 0.02 & 96.2 $\pm$ 0.1 & 84.0 $\pm$ 8.0 & 27.2 \\
NGC 6951 & 11.64 $\pm$ 0.15 & 10.65 $\pm$ 0.16 & \nodata & \nodata & 8.98 $\pm$ 0.10 & 8.78 $\pm$ 0.10 & 0.83 $\pm$ 0.06 & 344.0 $\pm$ 10.0 & 328.0 $\pm$ 12.0 & 32.3 \\
M 58 & 10.48 $\pm$ 0.08 & 9.66 $\pm$ 0.08 & 9.04 $\pm$ 0.10 & 8.41 $\pm$ 0.10 & 8.49 $\pm$ 0.10 & 8.28 $\pm$ 0.10 & 0.79 $\pm$ 0.04 & 373.0 $\pm$ 7.0 & 370.0 $\pm$ 11.0 & 8.3 \\
M 104 & 8.98 $\pm$ 0.06 & 8.00 $\pm$ 0.06 & 7.43 $\pm$ 0.10 & 6.61 $\pm$ 0.10 & 6.59 $\pm$ 0.10 & 7.23 $\pm$ 0.10 & 0.41 $\pm$ 0.02 & 778.0 $\pm$ 8.0 & 770.0 $\pm$ 17.0 & 8.6 \\
\tableline
\multicolumn{11}{l}{\text{Interacting Galaxies}} \\
\tableline
NGC 3227* & 11.85 $\pm$ 0.20 & 11.00 $\pm$ 0.20 & 10.48 $\pm$ 0.20 & 9.75 $\pm$ 0.10 & 8.66 $\pm$ 0.10 & 7.95 $\pm$ 0.10 & 0.42 $\pm$ 0.02 & 436.3 $\pm$ 3.8 & 428.0 $\pm$ 9.0 & 14.6 \\
NGC 3786 & 13.24 $\pm$ 0.18 & 12.94 $\pm$ 0.02 & \nodata & \nodata & 10.28 $\pm$ 0.10 & 9.88 $\pm$ 0.10 & 0.59 $\pm$ 0.04 & 460.0 $\pm$ 29.0 & 466.0 $\pm$ 30.0 & 10.4 \\
NGC 4438 & 11.02 $\pm$ 0.07 & 10.17 $\pm$ 0.07 & \nodata & \nodata & 8.46 $\pm$ 0.10 & 8.36 $\pm$ 0.10 & 0.37 $\pm$ 0.02 & 261.0 $\pm$ 15.0 & 244.0 $\pm$ 100.0 & 7.2 \\
NGC 5194 & 9.08 $\pm$ 0.15 & 8.48 $\pm$ 0.15 & 7.86 $\pm$ 0.10 & 7.29 $\pm$ 0.10 & 8.84 $\pm$ 0.10 & 8.66 $\pm$ 0.10 & 0.62 $\pm$ 0.03 & 195.0 $\pm$ 6.0 & \nodata & 250.0 \\
NGC 7469* & 13.05 $\pm$ 0.20 & 12.51 $\pm$ 0.20 & 12.10 $\pm$ 0.20 & 11.12 $\pm$ 0.20 & 8.43 $\pm$ 0.10 & 7.64 $\pm$ 0.10 & 0.81 $\pm$ 0.02 & 208.8 $\pm$ 16.9 & 215.0 $\pm$ 16.0 & 1.0 \\
NGC 7674 & 13.92 $\pm$ 0.10 & 13.23 $\pm$ 0.10 & \nodata & \nodata & 9.27 $\pm$ 0.10 & 8.11 $\pm$ 0.10 & 0.91 $\pm$ 0.08 & 344.0 $\pm$ 127.0 & 135.0 $\pm$ 50.0 & 2.2 \\
\enddata 
\tablecomments{Integrated galaxy apparent magnitudes, disk axis ratios, and 21\,cm spectral characteristics. Galaxy names marked with asterisks indicate galaxies consistent with the sample of Paper I (see \S\ref{section:data}), where the AGN contamination has been modeled and removed (See \S\ref{section:photometry}). Uncertainties on the optical and near-infrared magnitudes are either adopted from 1) Paper I, 2) RC3, or 3) EDD. Column 8 lists the observed minor ($b$) to major ($a$) disk axis ratio, where $q_{d}=b/a$. Columns 9 and 10 list the two definitions of the projected line width of the unresolved \ion{H}{1} 21\,cm emission line used in this work (See \S\ref{section:wmax}). Column 11 lists the integrated flux of \ion{H}{1} 21\,cm emission of each galaxy. The bottom panel denotes interacting systems, which are not included in any quoted statistics in this work.}
\label{appmags_21cm}
\end{deluxetable*}

The brightness contribution from Type 1 AGNs has been measured to be upwards of 30\% of the total integrated galaxy light \citep{misty2013}, and as such can play a significant role in distance determinations that rely on galaxy apparent brightness. We assume the galaxies in our sample that are hosts to Type 2 (or Type 1.9) AGN contribute a negligible amount of light to the integrated galaxy magnitudes, where the dusty torus is expected to obscure the light from the accretion disk and broad line region clouds according to the unified model \citep{antonucci1993,up1995,netzer2015,padovani2017}. For the majority of the Type 1 hosts, we look to our previous TF distance measurements for a sample of Seyfert 1 galaxies (\citealt{me2}; hereafter Paper I), where we constrained and removed the light contamination from the unobscured nuclei via careful surface brightness decompositions. We adopt the photometric and spectroscopic data presented in Paper I for the following galaxies and label them with asterisks in all tables: NGC 3227, NGC 3783, NGC 4051, NGC 4151, NGC 6814, and NGC 7469.

The imaging for the six galaxies listed above was collected from multiple observatories between $2003-2016$. Each have medium $V$-band images from HST \citep{misty2009a,misty2013,misty2018}, in addition to subsamples observed with the 3.5\,m WIYN telescope\footnote{The WIYN Observatory is a joint facility of the University of Wisconsin–
Madison, Indiana University, the National Optical Astronomy Observatory and
the University of Missouri.} \citep{misty2018}, and the MDM
Observatory 1.3 m McGraw-Hill Telescope \citep{misty2009b}. New imaging presented in Paper I was obtained with the 3.5\,m Apache Point
Observatory (APO) Astrophysical Research Consortium
(ARC) telescope, the Cerro Tololo Inter-American Observatory/Small and
Moderate Aperture Research Telescope System (CTIO/
SMARTS) 0.9\,m telescope, and the 0.5\,m ARC Small Aperture Telescope
(ARCSAT). All of the ground-based images were reduced in IRAF\footnote{IRAF is distributed by the National Optical Astronomy Observatory, which
is operated by the Association of Universities for Research in Astronomy under
a cooperative agreement with the National Science Foundation.}.

For the six Type 1 AGNs above, we removed the nuclear contamination from the observed magnitudes via two-dimensional surface brightness decompositions using the $\textsc{Galfit}$ \citep{peng2002,peng2010} software. A detailed description of the decompositions are available in \cite{misty2009a}, \cite{misty2013}, \cite{misty2018}, and Paper I, but in brief, the $\textsc{Galfit}$ software employs combinations of analytical surface brightness components to model a galaxy image. These components can correspond to physical morphological features such as bars, rings, bulges, etc. The light profile of an AGN, which is an unresolved point source in each image, is fit with the point spread function (PSF) model of each image. Example decompositions and residuals for HST images can be found in \cite{misty2013} and \cite{misty2018}, and Figure~1 of Paper I for ground-based images. 

We adopt all AGN-free galaxy apparent magnitudes from Paper I in the Johnson-Cousins $B$, $V$, $R$, and $I$ bandpasses for the six Seyfert 1 galaxies above. Observed photometric properties for our sample are tabulated in Table~\ref{appmags_21cm}. The Type 1 nuclei that remain are NGC 1566 and NGC 4639. While the nuclear activity of NGC 1566 has varied widely, its bolometric luminosity has remained low \citep{Woo2002}. Additionally, the AGN in NGC 1566 was in a low-luminosity state in the late 70s through the early 80s \citep{Baribaud1992}, which is most likely when the $B$ and $V-$band observations recorded in RC3 and adopted here were conducted. Thus, we do not expect the nuclear light of NGC 1566 to significantly contaminate its integrated galaxy luminosity. The nucleus in NGC 4639 was classified as a low-luminosity AGN (LLAGN; \citealt{ho1997}), with a nuclear $B-$band apparent magnitude of 20.27 measured by \cite{Ho2001} via HST images. As such, we do not expect the nuclear contribution of the AGN in NGC 4639 to significantly contaminate integrated apparent magnitudes.

For all galaxies with the AGN contamination modeled and removed, we adopt a typical uncertainty of 0.2\,mag, consistent with Paper I and \cite{misty2018} based on our experience with $\textsc{Galfit}$. For NGC 6814, poor seeing conditions and bright sky background in the resulted in a larger adopted uncertainty of 0.3\,mag on its $B-$band magnitude to account for the additional uncertainty in separating the disk light from the sky. For all other apparent magnitudes, we adopt the same uncertainties reported in each respective photometric database listed above. 

\subsubsection{Infrared Photometry}\label{section:ir_photometry}
For WISE apparent magnitudes, the angular resolution of its detector ($\sim$$6\arcsec$) is too low to separate the AGN from the bulge via surface brightness decompositions, and thus any nuclear contamination remains in the integrated $W_{1}$ and $W_{2}$ apparent magnitudes (see  \S\ref{discussion:IR}). To attempt to constrain and remove the light contribution of the nuclei, we look to the spectral energy distributions (SEDs) of a sample of Seyfert 1 AGNs constructed by \cite{rachael}. These SEDs are the most accurate multi-wavelength SEDs constructed for these AGNs to-date, as they 1) consist of simultaneous optical/UV/X-ray observations supplemented by ground-based near-IR data and 2) have undergone surface brightness decompositions on high-resolution HST images to isolate and remove the host-galaxy light contribution from the SEDs. The near-infrared data employed by \cite{rachael} was observed with the WIYN Observatory and the Vista Hemisphere Survey \citep{mcmahon2013}.

With SEDs of a subsample of Seyfert 1 AGNs in-hand, we extrapolate the SED to the wavelengths observed by WISE and use the IRAF task $\textsc{synphot}$ to calculate the flux of the AGN SED through the WISE $W_{1}$ and $W_{2}$ throughputs. Finally, we subtract the light contribution of the AGN via its SED from the respective integrated WISE luminosity in each bandpass. We indicate the subtraction as downward-facing blue arrows in Figure~\ref{mainplot}, where a longer arrow represents a larger amount of light contributed by the AGN in that filter, and provide further discussion in \S\ref{discussion:IR}.

\begin{deluxetable*}{lccccccc}
\tablecaption{Galaxy Disk Inclinations and Corrected Apparent Magnitudes}
\tablehead{
\colhead{Target} & \colhead{Inclination} & \colhead{$m_{B}^{b,i,k}$} & \colhead{$m_{V}^{b,i,k}$} & \colhead{$m_{R}^{b,i,k}$} & \colhead{$m_{I}^{b,i,k}$} & \colhead{$m_{W1}^{b,k,a}$} & \colhead{$m_{W2}^{b,k,a}$}  \\
& (deg) & (mag) & (mag) & (mag) & (mag) & (mag) & (mag)  \\
\colhead{(1)} & \colhead{(2)} &\colhead{(3)} & \colhead{(4)} & \colhead{(5)} & \colhead{(6)} & \colhead{(7)} & \colhead{(8)}
}
\startdata
NGC 1068 & 32.40 $\pm$ 4.40 & 9.33 $\pm$ 0.11 & 8.66 $\pm$ 0.11 & 8.28 $\pm$ 0.10 & 7.66 $\pm$ 0.10 & 4.12 $\pm$ 0.20 & 2.99 $\pm$ 0.30 \\
NGC 1566 & 38.32 $\pm$ 3.56 & 10.13 $\pm$ 0.04 & 9.60 $\pm$ 0.04 & 9.04 $\pm$ 0.10 & 8.53 $\pm$ 0.11 & 8.87 $\pm$ 0.10 & 8.56 $\pm$ 0.10 \\
NGC 3147 & 27.57 $\pm$ 7.98 & 11.19 $\pm$ 0.18 & 10.42 $\pm$ 0.18 & \nodata & \nodata & 8.93 $\pm$ 0.10 & 8.81 $\pm$ 0.10 \\
NGC 3783$^{\star}$ & 16.61 $\pm$ 4.18 & 12.39 $\pm$ 0.20 & 11.71 $\pm$ 0.20 & 10.75 $\pm$ 0.20 & \nodata & 8.30 $\pm$ 0.10 & 7.29 $\pm$ 0.10 \\
NGC 3982 & 30.10 $\pm$ 7.21 & 11.61 $\pm$ 0.17 & 11.47 $\pm$ 0.06 & 11.09 $\pm$ 0.11 & 10.68 $\pm$ 0.11 & 10.29 $\pm$ 0.10 & 10.01 $\pm$ 0.10 \\
NGC 4051$^{\star}$ & 56.24 $\pm$ 3.75 & 10.46 $\pm$ 0.21 & 9.88 $\pm$ 0.20 & 9.66 $\pm$ 0.20 & 9.16 $\pm$ 0.10 & 8.83 $\pm$ 0.10 & 7.92 $\pm$ 0.10 \\
NGC 4138 & 50.01 $\pm$ 4.87 & 11.77 $\pm$ 0.17 & 11.05 $\pm$ 0.16 & 10.45 $\pm$ 0.11 & 9.88 $\pm$ 0.11 & 9.36 $\pm$ 0.10 & 9.22 $\pm$ 0.10 \\
NGC 4151$^{\star}$ & 20.38 $\pm$ 8.59 & 11.14 $\pm$ 0.20 & 10.69 $\pm$ 0.20 & 10.09 $\pm$ 0.20 & 9.66 $\pm$ 0.20 & 6.70 $\pm$ 0.10 & 5.87 $\pm$ 0.10 \\
NGC 4258 & 70.09 $\pm$ 1.30 & 8.24 $\pm$ 0.08 & 7.79 $\pm$ 0.05 & 7.27 $\pm$ 0.10 & 6.83 $\pm$ 0.09 & 8.49 $\pm$ 0.10 & 8.14 $\pm$ 0.10 \\
NGC 4303 & 27.57 $\pm$ 6.70 & 10.01 $\pm$ 0.10 & 9.53 $\pm$ 0.10 & 9.06 $\pm$ 0.10 & 8.72 $\pm$ 0.10 & 8.97 $\pm$ 0.10 & 8.93 $\pm$ 0.10 \\
NGC 4395 & 34.51 $\pm$ 4.07 & 10.41 $\pm$ 0.20 & 9.99 $\pm$ 0.20 & 9.62 $\pm$ 0.10 & 9.10 $\pm$ 0.10 & 12.58 $\pm$ 0.10 & 11.79 $\pm$ 0.10 \\
NGC 4565 & 89.88 $\pm$ 18.16 & 8.89 $\pm$ 0.07 & 8.43 $\pm$ 0.09 & 7.92 $\pm$ 0.10 & 7.36 $\pm$ 0.10 & 7.91 $\pm$ 0.10 & 8.15 $\pm$ 0.10 \\
NGC 4639 & 48.77 $\pm$ 3.80 & 11.85 $\pm$ 0.11 & 11.27 $\pm$ 0.11 & 10.82 $\pm$ 0.11 & 10.16 $\pm$ 0.17 & 10.03 $\pm$ 0.10 & 9.95 $\pm$ 0.10 \\
NGC 4939 & 61.18 $\pm$ 2.57 & 11.13 $\pm$ 0.21 & 10.68 $\pm$ 0.21 & 10.32 $\pm$ 0.11 & 9.90 $\pm$ 0.11 & 9.98 $\pm$ 0.10 & 9.62 $\pm$ 0.10 \\
NGC 4945 & 89.88 $\pm$ 51.32 & 7.48 $\pm$ 0.20 & \nodata & 6.03 $\pm$ 0.10 & 6.37 $\pm$ 0.70 & 7.70 $\pm$ 0.10 & 6.43 $\pm$ 0.10 \\
NGC 5055 & 56.59 $\pm$ 1.98 & 8.78 $\pm$ 0.11 & 8.20 $\pm$ 0.11 & 7.64 $\pm$ 0.10 & 7.22 $\pm$ 0.14 & 8.33 $\pm$ 0.10 & 8.28 $\pm$ 0.10 \\
NGC 5643 & 30.10 $\pm$ 4.81 & 10.02 $\pm$ 0.14 & 9.47 $\pm$ 0.15 & 9.30 $\pm$ 0.10 & 8.80 $\pm$ 0.10 & 9.33 $\pm$ 0.10 & 8.68 $\pm$ 0.10 \\
NGC 5728 & 56.59 $\pm$ 2.97 & 11.39 $\pm$ 0.15 & 10.70 $\pm$ 0.15 & \nodata & 9.75 $\pm$ 0.11 & 9.28 $\pm$ 0.10 & 8.95 $\pm$ 0.10 \\
NGC 6814$^{\star}$ & 11.72 $\pm$ 5.88 & 11.46 $\pm$ 0.30 & 10.65 $\pm$ 0.20 & 10.20 $\pm$ 0.20 & 9.62 $\pm$ 0.20 & 9.27 $\pm$ 0.10 & 8.73 $\pm$ 0.10 \\
NGC 6951 & 34.51 $\pm$ 6.11 & 10.10 $\pm$ 0.16 & 9.49 $\pm$ 0.17 &\nodata & \nodata & 8.87 $\pm$ 0.10 & 8.68 $\pm$ 0.10 \\
M 58 & 38.32 $\pm$ 3.56 & 10.10 $\pm$ 0.09 & 9.37 $\pm$ 0.09 & 8.79 $\pm$ 0.11 & 8.21 $\pm$ 0.10 & 8.44 $\pm$ 0.10 & 8.22 $\pm$ 0.10 \\
M 104 & 68.76 $\pm$ 1.35 & 7.74 $\pm$ 0.08 & 7.00 $\pm$ 0.08 & 6.57 $\pm$ 0.11 & 5.92 $\pm$ 0.10 & 6.53 $\pm$ 0.10 & 7.17 $\pm$ 0.10 \\
\tableline
\multicolumn{8}{l}{\text{Interacting Galaxies}} \\
\tableline
NGC 3227$^{\star}$ & 67.86 $\pm$ 1.44 & 11.03 $\pm$ 0.20 & 10.40 $\pm$ 0.20 & 9.90 $\pm$ 0.20 & 9.28 $\pm$ 0.10 & 8.61 $\pm$ 0.10 & 7.90 $\pm$ 0.10 \\
NGC 3786 & 55.58 $\pm$ 3.07 & 12.63 $\pm$ 0.19 & 12.48 $\pm$ 0.06 & \nodata & \nodata & 10.22 $\pm$ 0.10 & 9.82 $\pm$ 0.10 \\
NGC 4438 & 71.36 $\pm$ 1.25 & 10.40 $\pm$ 0.08 & 9.82 $\pm$ 0.08 & \nodata & \nodata & 8.42 $\pm$ 0.10 & 8.31 $\pm$ 0.10 \\
NGC 5194 & 53.47 $\pm$ 2.18 & 8.74 $\pm$ 0.15 & 8.29 $\pm$ 0.15 & 7.62 $\pm$ 0.10 & 7.11 $\pm$ 0.10 & 8.80 $\pm$ 0.10 & 8.61 $\pm$ 0.10 \\
NGC 7469$^{\star}$ & 36.76 $\pm$ 2.02 & 12.61 $\pm$ 0.20 & 12.21 $\pm$ 0.20 & 11.80 $\pm$ 0.20 & 10.88 $\pm$ 0.20 & 8.34 $\pm$ 0.10 & 7.55 $\pm$ 0.10 \\
NGC 7674 & 24.75 $\pm$ 12.03 & 13.60 $\pm$ 0.11 & 13.01 $\pm$ 0.10 & \nodata & \nodata & 9.16 $\pm$ 0.10 & 8.00 $\pm$ 0.10 \\
\enddata
\tablecomments{Galaxy names marked with asterisks indicate galaxies consistent with the sample of Paper I (see \S\ref{section:data}). Column 2 lists the inclination angles of each galaxy disk derived from the observed disk axis ratios as described in \S\ref{section:wmax}. Columns 3-8 list the apparent magnitudes in each bandpass corrected for Galactic extinction, inclination-dependent extinction, redshift, and/or aperture (See \S\ref{section:corrected_mags}). The bottom panel denotes interacting systems, which are not included in any quoted statistics in this work.}
\label{corrected_mags}
\end{deluxetable*}

\begin{deluxetable*}{lcccccccc}
\tablecaption{Maximum Rotational Velocities and Derived Absolute Magnitudes}
\tablehead{
\colhead{Target} & \colhead{$W_{R}^{i}$} & \colhead{$W_{mx}^{i}$} & \colhead{$M_{B}^{b,i,k}$} & \colhead{$M_{V}^{b,i,k}$} & \colhead{$M_{R}^{b,i,k}$} & \colhead{$M_{I}^{b,i,k}$} & \colhead{$M_{W1}^{b,k,a}$} & \colhead{$M_{W2}^{b,k,a}$}\\
& (km\,s$^{-1}$) & (km\,s$^{-1}$) & (mag) & (mag) & (mag) & (mag) & (mag) & (mag)  \\
\colhead{(1)} & \colhead{(2)} &\colhead{(3)} & \colhead{(4)} & \colhead{(5)} & \colhead{(6)} & \colhead{(7)} & \colhead{(8)} & \colhead{(9)}
}
\startdata
NGC 1068 & 487.30 $\pm$ 60.84 & 536.18 $\pm$ 60.86 & -20.82 $\pm$ 0.15 & -21.49 $\pm$ 0.15 & -21.87 $\pm$ 0.15 & -22.50 $\pm$ 0.15 & -26.03 $\pm$ 0.22 & -27.17 $\pm$ 0.32 \\
NGC 1566 & 309.66 $\pm$ 26.12 & 316.81 $\pm$ 32.08 & -21.14 $\pm$ 0.25 & -21.68 $\pm$ 0.24 & -22.24 $\pm$ 0.26 & -22.75 $\pm$ 0.27 & -22.41 $\pm$ 0.26 & -22.71 $\pm$ 0.26 \\
NGC 3147 & 820.98 $\pm$ 219.89 & 809.61 $\pm$ 221.41 & -21.79 $\pm$ 0.20 & -22.55 $\pm$ 0.20 & \nodata & \nodata & -24.04 $\pm$ 0.13 & -24.16 $\pm$ 0.13 \\
NGC 3783* & 426.29 $\pm$ 104.49 & 480.52 $\pm$ 107.86 & -20.62 $\pm$ 0.81 & -21.29 $\pm$ 0.81 & -22.26 $\pm$ 0.81 & \nodata & -24.70 $\pm$ 0.80 & -25.71 $\pm$ 0.80 \\
NGC 3982 & 388.66 $\pm$ 85.44 & 412.88 $\pm$ 91.62 & -20.11 $\pm$ 0.18 & -20.26 $\pm$ 0.09 & -20.64 $\pm$ 0.13 & -21.05 $\pm$ 0.13 & -21.44 $\pm$ 0.12 & -21.72 $\pm$ 0.12 \\
NGC 4051* & 272.78 $\pm$ 11.92 & 282.83 $\pm$ 15.31 & -20.64 $\pm$ 0.21 & -21.22 $\pm$ 0.21 & -21.45 $\pm$ 0.21 & -21.95 $\pm$ 0.11 & -22.27 $\pm$ 0.11 & -23.18 $\pm$ 0.11 \\
NGC 4138 & 379.86 $\pm$ 30.08 & 395.38 $\pm$ 35.87 & -18.92 $\pm$ 0.31 & -19.65 $\pm$ 0.31 & -20.25 $\pm$ 0.28 & -20.83 $\pm$ 0.28 & -21.34 $\pm$ 0.28 & -21.48 $\pm$ 0.28 \\
NGC 4151* & 318.73 $\pm$ 128.70 & 352.96 $\pm$ 130.56 & -19.85 $\pm$ 0.21 & -20.30 $\pm$ 0.21 & -20.91 $\pm$ 0.21 & -21.33 $\pm$ 0.21 & -24.29 $\pm$ 0.11 & -25.12 $\pm$ 0.11 \\
NGC 4258 & 429.69 $\pm$ 6.38 & 398.45 $\pm$ 21.56 & -21.14 $\pm$ 0.10 & -21.60 $\pm$ 0.07 & -22.12 $\pm$ 0.12 & -22.54 $\pm$ 0.11 & -20.91 $\pm$ 0.11 & -21.26 $\pm$ 0.11 \\
NGC 4303 & 311.43 $\pm$ 70.40 & 326.78 $\pm$ 71.75 & -21.07 $\pm$ 0.11 & -21.55 $\pm$ 0.10 & -22.02 $\pm$ 0.11 & -22.36 $\pm$ 0.11 & -22.11 $\pm$ 0.11 & -22.15 $\pm$ 0.11 \\
NGC 4395 & 187.73 $\pm$ 20.72 & 197.67 $\pm$ 23.59 & -17.65 $\pm$ 0.29 & -18.08 $\pm$ 0.29 & -18.44 $\pm$ 0.23 & -18.97 $\pm$ 0.23 & -15.48 $\pm$ 0.23 & -16.27 $\pm$ 0.23 \\
NGC 4565 & 490.00 $\pm$ 5.01 & 500.31 $\pm$ 8.01 & -21.49 $\pm$ 0.09 & -21.94 $\pm$ 0.11 & -22.45 $\pm$ 0.11 & -23.03 $\pm$ 0.11 & -22.46 $\pm$ 0.11 & -22.23 $\pm$ 0.11 \\
NGC 4639 & 352.47 $\pm$ 22.50 & 383.23 $\pm$ 23.09 & -19.68 $\pm$ 0.13 & -20.26 $\pm$ 0.13 & -20.72 $\pm$ 0.13 & -21.38 $\pm$ 0.19 & -21.50 $\pm$ 0.12 & -21.58 $\pm$ 0.12 \\
NGC 4939 & 485.06 $\pm$ 13.25 & 505.82 $\pm$ 23.77 & -22.08 $\pm$ 0.22 & -22.52 $\pm$ 0.22 & -22.89 $\pm$ 0.13 & -23.31 $\pm$ 0.12 & -23.22 $\pm$ 0.13 & -23.58 $\pm$ 0.13 \\
NGC 4945 & 346.00 $\pm$ 9.02 & 357.65 $\pm$ 13.02 & -20.22 $\pm$ 0.21 & \nodata & -21.67 $\pm$ 0.11 & -21.35 $\pm$ 0.70 & -20.00 $\pm$ 0.10 & -21.27 $\pm$ 0.10 \\
NGC 5055 & 433.68 $\pm$ 11.56 & 449.07 $\pm$ 9.96 & -20.96 $\pm$ 0.13 & -21.54 $\pm$ 0.13 & -22.10 $\pm$ 0.13 & -22.53 $\pm$ 0.16 & -21.41 $\pm$ 0.10 & -21.46 $\pm$ 0.10 \\
NGC 5643 & 344.62 $\pm$ 52.11 & 370.40 $\pm$ 56.11 & -20.45 $\pm$ 0.18 & -21.00 $\pm$ 0.18 & -21.17 $\pm$ 0.15 & -21.67 $\pm$ 0.15 & -21.14 $\pm$ 0.15 & -21.79 $\pm$ 0.15 \\
NGC 5728 & 449.26 $\pm$ 18.11 & 473.37 $\pm$ 27.46 & -21.74 $\pm$ 0.18 & -22.43 $\pm$ 0.17 & \nodata & -23.39 $\pm$ 0.14 & -23.85 $\pm$ 0.13 & -24.18 $\pm$ 0.13 \\
NGC 6814* & 372.16 $\pm$ 184.20 & 388.09 $\pm$ 187.82 & -20.21 $\pm$ 0.30 & -21.02 $\pm$ 0.20 & -21.48 $\pm$ 0.20 & -22.06 $\pm$ 0.20 & -22.40 $\pm$ 0.11 & -22.95 $\pm$ 0.11 \\
NGC 6951 & 540.10 $\pm$ 85.60 & 558.44 $\pm$ 86.39 & -21.27 $\pm$ 0.19 & -21.88 $\pm$ 0.19 & \nodata & \nodata & -22.95 $\pm$ 0.35 & -23.14 $\pm$ 0.35 \\
M 58 & 540.31 $\pm$ 44.01 & 575.04 $\pm$ 46.09 & -21.51 $\pm$ 0.23 & -22.24 $\pm$ 0.22 & -22.83 $\pm$ 0.23 & -23.40 $\pm$ 0.23 & -23.18 $\pm$ 0.22 & -23.39 $\pm$ 0.22 \\
M 104 & 793.91 $\pm$ 11.25 & 812.59 $\pm$ 19.64 & -22.16 $\pm$ 0.09 & -22.90 $\pm$ 0.08 & -23.33 $\pm$ 0.11 & -23.99 $\pm$ 0.11 & -23.37 $\pm$ 0.10 & -22.73 $\pm$ 0.10 \\
\tableline
\multicolumn{8}{l}{\text{Interacting Galaxies}} \\
\tableline
NGC 3227* & 430.02 $\pm$ 6.01 & 448.88 $\pm$ 10.66 & -20.85 $\pm$ 0.31 & -21.47 $\pm$ 0.31 & -21.98 $\pm$ 0.31 & -22.60 $\pm$ 0.26 & -23.26 $\pm$ 0.26 & -23.97 $\pm$ 0.26 \\
NGC 3786 & 511.57 $\pm$ 39.85 & 543.76 $\pm$ 40.92 & -20.85 $\pm$ 0.29 & -21.00 $\pm$ 0.22 & \nodata & \nodata & -23.26 $\pm$ 0.24 & -23.66 $\pm$ 0.24 \\
NGC 4438 & 235.69 $\pm$ 15.73 & 245.23 $\pm$ 105.41 & -20.69 $\pm$ 0.19 & -21.27 $\pm$ 0.19 & \nodata & \nodata & -22.67 $\pm$ 0.20 & -22.78 $\pm$ 0.20 \\
NGC 5194 & 198.74 $\pm$ 8.92 & \nodata & -20.61 $\pm$ 0.23 & -21.05 $\pm$ 0.23 & -21.73 $\pm$ 0.20 & -22.24 $\pm$ 0.20 & -20.55 $\pm$ 0.20 & -20.74 $\pm$ 0.20 \\
NGC 7469* & 288.43 $\pm$ 30.00 & 202.16 $\pm$ 17.93 & -21.34 $\pm$ 0.23 & -21.75 $\pm$ 0.23 & -22.16 $\pm$ 0.23 & -23.05 $\pm$ 0.23 & -25.61 $\pm$ 0.15 & -26.41 $\pm$ 0.15 \\
NGC 7674 & 730.99 $\pm$ 450.27 & 290.60 $\pm$ 351.93 & -21.80 $\pm$ 0.16 & -22.40 $\pm$ 0.15 & \nodata & \nodata & -26.18 $\pm$ 0.12 & -27.35 $\pm$ 0.12 \\
\enddata
\tablecomments{Galaxy names marked with asterisks indicate galaxies consistent with the sample of Paper I (see \S\ref{section:data}). Columns 2 and 3 list both definitions of the deprojected \ion{H}{1} 21\,cm line width, both of which are statistically equal to twice the maximum rotation rate (see \S\ref{section:wmax}). Columns 4-9 list corrected absolute magnitudes of each AGN host in each bandpass derived from their corrected apparent magnitudes (Table \ref{corrected_mags}) and primary distances (Table \ref{basics}). The bottom panel denotes interacting systems, which are not included in any quoted statistics in this work.}
\label{table:absmags}
\end{deluxetable*}

\subsection{AGN Type}
We adopt consistent nuclear classifications for the six galaxies from Paper I, and label them with asterisks in each table. We adopt all other nuclear activity classifications from the second data release of the Swift BAT AGN Spectroscopic Survey (BASS; \citealt{koss2022}), which consists of hard X-ray-selected AGNs from the Swift BAT 70-month sample \citep{Baumgartner2013}, and list them in Column 7 of Table~\ref{basics}. We denote AGNs with a LINER classification as type L.


\subsection{Spectroscopy}\label{section:spectroscopy}
For the six galaxies shared between the samples of this work and Paper I, we adopt all \ion{H}{1} 21\,cm spectroscopic measurements presented in Paper I from observations with the 100\,m Robert C.\ Byrd Green Bank Telescope\footnote{The Green Bank Observatory is a facility of the National Science Foundation
operated under cooperative agreement by Associated Universities,
Inc.} (GBT). For the remaining galaxies, we adopt the emission line measurements tabulated in EDD. Adopted emission line widths and integrated flux measurements are listed in Table~\ref{appmags_21cm}.

\subsection{Distances}\label{section:distances}
To maintain consistency with previous TF relation calibrations, we only adopt distance measurements via Cepheids, TRGB stars, SBF, SN Ia, SN II, or geometric distances. For galaxies with more than one distance measurement technique available, we defaulted to the primary distance (i.e., Cepheids or TRGB stars). For galaxies where multiple SN Ia distance measurements are available, we employ the five-source SN Ia distance from CF2, which draws from the main SN compilation of UNION2 \citep{Amanullah2010} in combination with four additional compilations \citep{Prieto2006,Jha2007,Hicken2009,Folatelli2010}, calibrated to the CF2 zero-point (\citealt{cosmicflows2}; see Figures 1 and 5 of \citealt{Courtois2012}).

\section{Construction of TF Relations}\label{section:construction_tf}
With 21\,cm emission line widths, apparent magnitudes, and primary $z-$independent distances in hand, we construct here TF relations solely populated by AGN host galaxies that span optical to infrared passbands. 

\subsection{\ion{H}{1} 21cm Line Widths}\label{section:wmax}
Measurement of the width of the unresolved \ion{H}{1} 21\,cm emission line profile can be directly translated to a galaxy's maximum rotation rate \citep{epstein1964,roberts1969}. The TF relation has employed two definitions of the line width in the literature; the original definition measured the width of the profile at 20\% of the peak flux, denoted $W_{20}$ \citep{tf1985}. In later work, the definition was updated such that the width of the profile is measured at 50\% of the mean flux, denoted $W_{m50}$ \citep{courtois2009}. The $W_{20}$ definition was used for calibrations of the $B$, $V$, and $R$ bands, while the $W_{m50}$ definition was used for the calibrations of the $I$ and WISE passbands. We detail each width definition and translation to deprojected, maximum rotational velocity ($W_{\lambda}^{i}$) below.

For both definitions of the line width, the observed width must be corrected to statistically agree with twice the maximum rotational velocity, accounting for the inclusion of both redshifted and blueshifted gas motions in the unresolved emission line profile. For the $W_{20}$ definition, the prescription is described by \cite{tf1985} as

\begin{equation}
W_{R}^{2} = W_{20}^{2} + W_{t}^{2} - 2W_{20}W_{t}[1 - e^{-(W_{20}/W_{c})^{2}}]
- 2W_{t}^{2}e^{-(W_{20}/W_{c})^{2}}
\end{equation}

\noindent where $W_{\mathrm{t}} = 38$\,km\,s$^{-1}$ describes random gas motions and $W_{\mathrm{c}} = 120$\,km\,s$^{-1}$ characterizes the transition from dual-horned \ion{H}{1} profiles to Gaussian-shaped profiles. The updated definition of the width measurement given by \cite{courtois2009} first includes corrections for instrumental and redshift broadening, defined as

\begin{equation}
    W_{m50}^{c} = \frac{W_{m50}}{(1+z)} - 2\Delta v\lambda
\end{equation} 

\noindent where $z$ is the redshift of the \ion{H}{1} line (see Column 2 of Table~\ref{basics}), $\Delta v$ is the smoothed resolution of the spectrum, and $\lambda$ is an empirically determined constant equal to 0.25. The translation to maximum rotational velocity was also updated as

\begin{equation}
\begin{split}
W_{mx}^{2} = (W_{m50}^{c})^{2} + (W_{t,m50})^{2}[1 - 2e^{-(W_{m50}^{c}/W_{c,m50})^{2}}]\\
& \hspace{-4cm} - 2W_{m50}^{c}W_{t,m50}[1 - e^{-(W_{m50}^{c}/W_{c,m50})^{2}}]
\end{split}
\end{equation}

\noindent where $W_{c,m50}$ = 100 km s$^{-1}$ and $W_{t,m50}$ = 9 km s$^{-1}$. The translated width (either $W_{R}$ or $W_{mx}$) is then deprojected to edge-on orientation by
\begin{equation}
    W^{i}_{\lambda} = W_{\lambda} / \sin(i)
\end{equation}
where $i$ is the inclination of the galaxy disk. 

Galaxy disk inclinations are generally derived from the disk axis ratios. We adopt the axis ratios constrained from surface brightness decompositions for all Seyfert 1 host galaxies in the sample of Paper I and tabulate them in Table~\ref{appmags_21cm}. In the case of NGC 4151, the spatially resolved \ion{H}{1} 21\,cm distribution observed by \cite{mundell1999} reveals a much more face-on disk inclined at 21$^\circ$ with respect to typical measurements of the disk axis ratio in the optical of $\sim$0.6 \citep{dv1976,dv1991,misty2018}. We adopt 21$^\circ$ as the inclination of NGC 4151, consistent with Paper I. For all other galaxies, we adopt the axis ratios reported by RC3, and tabulate all axis ratios in Table~\ref{appmags_21cm}. To derive disk inclinations from axis ratios, we follow the same prescription outlined in Paper I and the main TF works in the literature \citep{tf1977,tp2000,cosmicflows1,cosmicflows2}:

\begin{equation}
    \cos(i) = [(q_{d}^{2} - q_{0,d}^{2})/(1 - q_{0,d}^{2})]^{1/2}
\end{equation}

\noindent where $q_{d}=b/a$ is the axis ratio of the galaxy disk and $q_{0,d}$ is the intrinsic axis ratio of an edge-on galaxy, given as a single, global flattening value of $q_{0,d} = 0.2$ by \cite{tp2000}.

\subsection{TF Calibrations}\label{section:calibrations}
Canonical TF relations have been published for the Johnson-Cousins $B$, $V$, $R$, and $I$ optical and near-infrared filters (\citealt{tp2000,cosmicflows1}; Paper I; \citealt{tc2012}), and the WISE $W_{1}$ and $W_{2}$ infrared passbands \citep{Kourkchi2020}. The universal fits for each bandpass are: 

\begin{equation}
    M_{B}^{b,i,k} = -19.99 - 7.27(\log W_{R}^{i} - 2.5)
\end{equation}
\begin{equation}
    M_{V}^{b,i,k} = −20.39 - 7.62(\log W_{R}^{i} - 2.5)
\end{equation}
\begin{equation}
    M_{R}^{b,i,k} = -21.00 - 7.65(\log W_{R}^{i} - 2.5)
\end{equation}
\begin{equation}
    M_{I}^{b,i,k} = -21.39 - 8.81(\log W_{mx}^{i} - 2.5)
\end{equation}
\begin{equation}
    M_{W1}^{b,k,a} = −20.36 − 9.47(\log  W_{mx}^{i} - 2.4)
\end{equation}
\begin{equation}
    M_{W2}^{b,k,a} = −19.76 − 9.66(\log  W_{mx}^{i} - 2.4)
\end{equation}

\begin{figure*}
\centering
\includegraphics[trim={2.2cm 2cm 0cm 2cm},clip,scale=0.55]{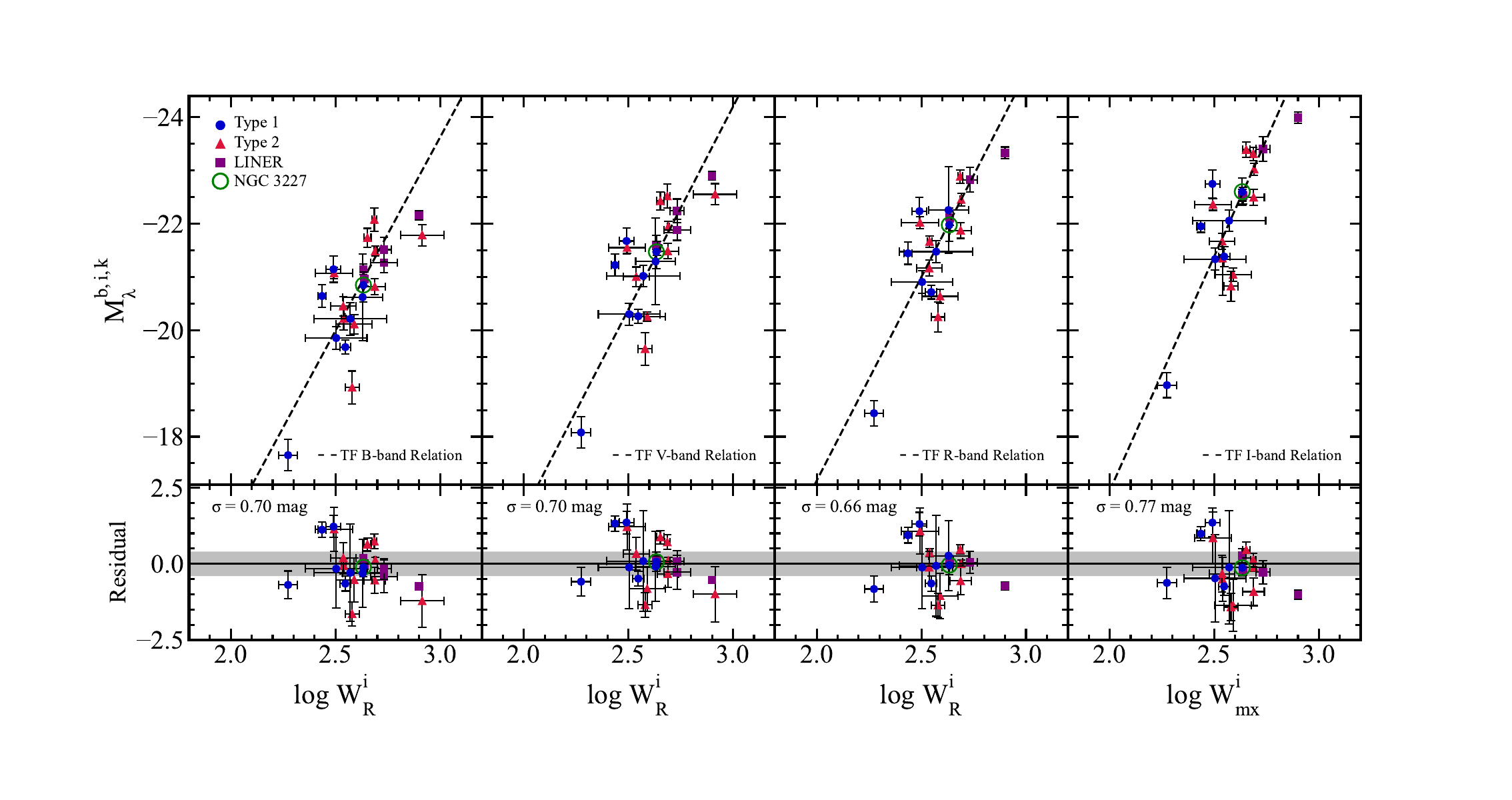}
\\
\includegraphics[trim={0cm 1.5cm 0cm 2cm},clip,scale=0.55]{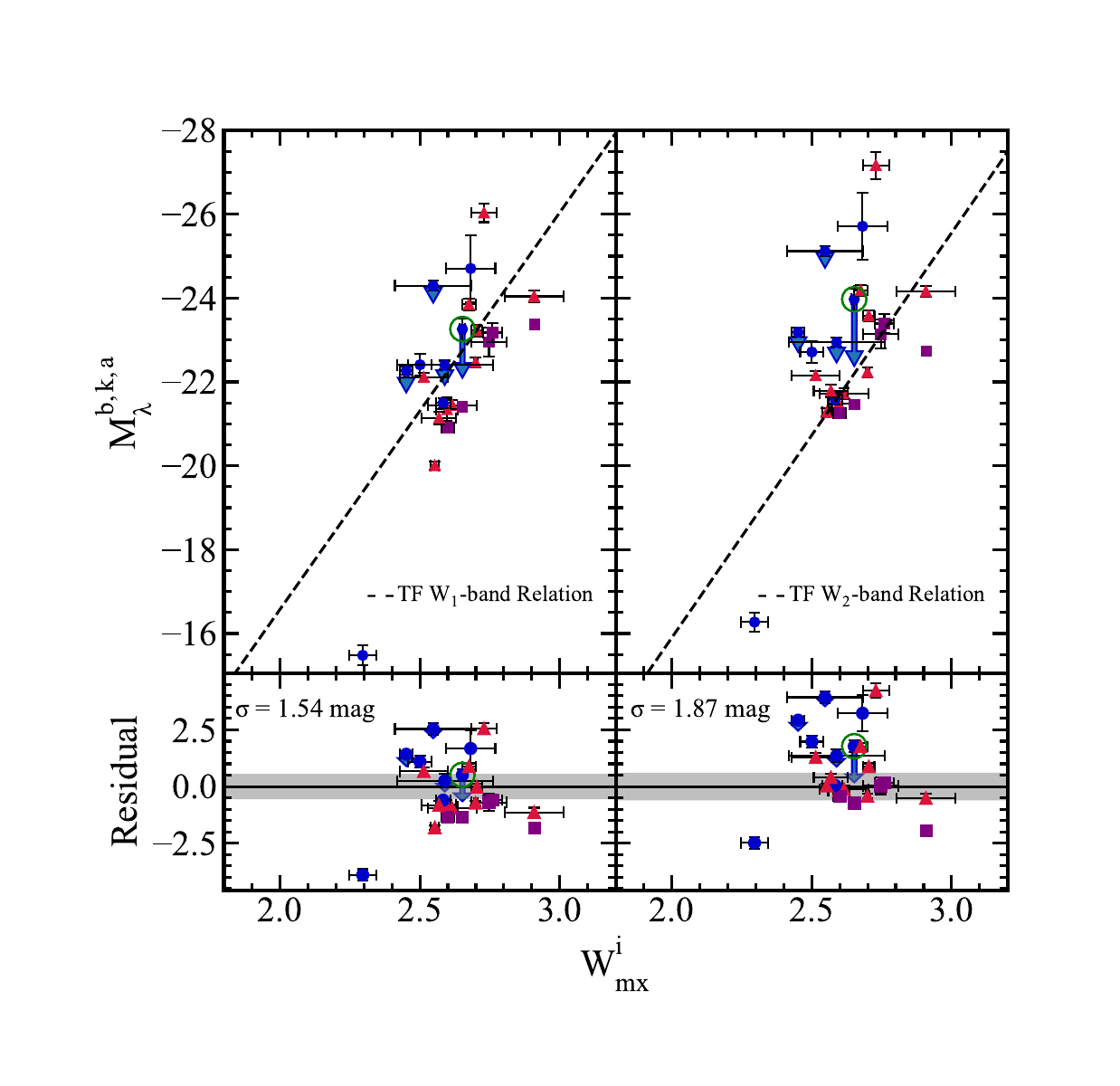}
\caption{Top row: TF relations in the $B$, $V$, $R$, and $I$ passbands for the sample of active galaxies (see Table~\ref{basics}). The superscripts on the absolute magnitude denote that Galactic extinction ($b$), inclination-dependent extinction ($i$), and redshift corrections ($k$) have been applied (see \S\ref{section:corrected_mags}). Blue circles indicate galaxies hosting a Seyfert 1 AGN, red triangles a Seyfert 1.9 or 2 AGN, purple squares a LINER, and the open green circle indicates the interacting AGN host NGC 3227. No interacting systems were included in any quoted statistics in this work. The dashed lines in each panel are the TF calibrations for inactive galaxies, not lines of best fit. The bottom panels display the residuals (TF-predicted absolute magnitude minus the corrected absolute magnitude) for each filter, where the black horizontal line indicates perfect agreement with the TF prediction, and the grey bar indicates the typical scatter of $\pm$0.4\,mag from inactive galaxies. The scatter ($\sigma$) of each residual distribution is indicated in each of the bottom panels. Bottom row; Same as top panel, but for the $W_{1}$ and $W_{2}$ passbands. The typical scatter, indicated by the gray bars in the bottom panels, for the $W_{1}$ and $W_{2}$ filters is $\pm$0.56\,mag and $\pm$0.61\,mag, respectively. The arrows quantify the amount of light removed from the integrated galaxy brightnesses by extrapolating the brightness contribution of the SED \citep{rachael} of each AGN through the WISE bandpasses via synthetic photometry with the IRAF task $\textsc{synphot}$ (See \S\ref{discussion:IR}).
}\label{mainplot}
\end{figure*}

\noindent where $b$ is the correction for Galactic extinction along the line of sight, $i$ is the attenuation of a galaxy's light due to a spiral disk's inclination relative to the observer's line of sight, and $k$ accounts for the $k-$correction, or the effect of the spectroscopic redshift on each galaxy's observed magnitude in each passband. We estimate the Galactic extinction along the line of sight in each bandpass using the \cite{schlafly2011} recalibration of the Milky Way dust map of \cite{schlegel1998}.

The attenuation due to an individual galaxy's inclination, or the inclination-dependent extinction, is given by ${A_{i}^{\lambda} = \gamma_{\lambda}\log(a/b)}$ (\citealt{tully1998}; \citealt{tp2000}; \citealt{cosmicflows1}; \citealt{tc2012}; Paper I), where $\lambda$ is the passband, $a/b$ is the ratio of major to minor axes of the galaxy disk, and $\gamma_{\lambda}$ is defined as
\begin{equation}\label{b_inc}
    \gamma_{B} = 1.57 + 2.75(\log W_{R}^{i} - 2.5)
\end{equation}
\begin{equation}\label{v_inc}
    \gamma_{V} = 1.01 + 2.94(\log W_{\textsc{R}}^{i} - 2.5)
\end{equation}
\begin{equation}
    \gamma_{R} = 1.15 + 1.88(\log W_{R}^{i} - 2.5)
\end{equation}
\begin{equation}
    \gamma_{I} =0.92 + 1.63(\log W_{mx}^{i} - 2.5)
\end{equation}

\noindent A new methodology was constructed to constrain the inclination-dependent extinction with the advent of the Sloan Digital Sky Survey (SDSS) and WISE TF calibrations \citep{Kourkchi2020}, where $A_{i} = \gamma_{\lambda}\mathcal{F}_{\lambda}(i)$. In this formulation, $\mathcal{F}_{\lambda}(i)$ is a function of galaxy inclination, and $\gamma_{\lambda}$ is a new third-degree polynomial term constructed as a linear combination of observables. However, as stated by \cite{Kourkchi2020_2}, the inclination-dependent extinction is prominent at optical wavelengths, but minimal ($\sim$0.05\,mag) at wavelengths observed at $W_{1}$ (3.4\,$\mu$m) and negligible ($\sim$0.02\,mag) at $W_{2}$ (4.6\,$\mu$m). Thus, we assume the inclination corrections on the WISE magnitudes investigated here are negligible.

The redshift corrections, $k$, for each passband (\citealt{tp2000}, \citealt{Chilingarian2010}; Paper I) are as follows:

\begin{equation}\label{b_k}
    A_{k}^{B} = (3.6 - 0.36T)z
\end{equation}
\begin{equation}\label{v_k}
    A_{k}^{V} = (2.23 - 0.22T)z
\end{equation}
\begin{equation}
    A_{k}^{R} = [4.24(R - I) - 1.10]z
\end{equation}
\begin{equation}
    A_{k}^{I} = 0.302z + 8.768z^{2} - 68.680z^{3} + 181.904z^{4}
\end{equation}
where $T$ is the galaxy morphological type (1, 3, 5, and 7 corresponding to Sa, Sb, Sc, and Sd) and $z$ is the redshift. The $k-$corrections in this work utilize the morphological classifications listed in Table~\ref{basics}, which we adopt from the NASA Extragalactic Database (NED). For the WISE passbands, the $k-$corrections are effectively the same for the two passbands \citep{kourkchi2019} and are calculated as $A_{k}^{W1,2} = -2.27z$ \citep{os1968,Huang2007}.

An aperture correction, $a$, is applied to WISE magnitudes where point sources were employed for photometric calibration within apertures of a much smaller ($\sim$8\farcs25) fixed radius compared to the galaxies. We follow the fixed corrections on $W_{1}$ and $W_{2}$ apparent magnitudes given by the WISE Explanatory Supplement\footnote{\url{http://wise2.ipac.caltech.edu/docs/release/allsky/expsup/sec4_4c.html}} and adopted by the TF calibrations of \cite{kourkchi2019}, \cite{Kourkchi2020}, and \cite{Kourkchi2020_2}, given as $A_{a}^{W1} = -0.034$\,mag and $A_{a}^{W2} = -0.041$\,mag.

\subsection{Corrected Galaxy Magnitudes}\label{section:corrected_mags}
With magnitude corrections defined, each apparent magnitude is adjusted by the following formalism:

\begin{equation}
    m_{\lambda}^{*} = m_{\lambda}^{\mathrm{(total)}} - A_{b}^{\lambda} - A_{i}^{\lambda} - A_{k}^{\lambda} - A_{a}^{\lambda} 
\end{equation}

\noindent where $m_{\lambda}^{\mathrm{(total)}}$ is the observed, uncorrected apparent magnitude, and $A_{b}^{\lambda}$, $A_{i}^{\lambda}$, $A_{k}^{\lambda}$, and $A_{a}^{\lambda}$ are the Galactic extinction, inclination, redshift, and aperture corrections respectively. The aperture corrections are only applied to the WISE magnitudes. All corrected apparent magnitudes are listed in Table~\ref{corrected_mags}. We then derive the corrected absolute magnitudes in each passband, $M_{\lambda}^{\mathrm{*}}$, using $m_{\lambda}^{\mathrm{*}}$ and the primary, $z-$independent distances detailed in \S\ref{section:distances} (see Table~\ref{basics}), and tabulate all absolute magnitudes in Table \ref{table:absmags}.

\section{Results}\label{section:results}
Employing the derived maximum rotational velocities and corrected absolute magnitudes for our sample, we construct here TF relations for AGN host galaxies in the $B$, $V$, $R$, $I$, $W_{1}$, and $W_{2}$ passbands, and compare each distribution to the predictions from each respective canonical TF relation. With integrated \ion{H}{1} 21\,cm fluxes and observed galaxy colors in-hand, we also examine active galaxies on the BTF relation.

Excluding interacting systems, we calculate the scatter ($\sigma$) about each calibration along the magnitude axis, consistent with the main CF works in the literature, as follows:

\begin{equation}\label{equation:vertscatter}
\sigma^{2} = \left(\frac{1}{N} \sum_{y} \sigma_{y}^{2}\right) - \sigma_{u}^{2}
\end{equation}

\noindent where $\sigma_{y}$ is the deviation of the derived absolute magnitude of each galaxy using its primary distance from the TF-predicted absolute magnitude, $\sigma_{u}$ is the mean of the uncertainties of $\sigma_{y}$, and $N$ is the total number of data points. 

\subsection{TF Relations}
We plot the absolute magnitudes listed in Table \ref{table:absmags}, derived from the corrected apparent magnitudes listed in Table~\ref{corrected_mags} and primary distances in Table~\ref{basics}, as a function of deprojected \ion{H}{1} 21\,cm line width in the $B$, $V$, $R$, $I$, $W_{1}$, and $W_{2}$ filters in Figure~\ref{mainplot}. The dashed lines in each panel are the canonical TF relations for each respective passband, they are not lines of best fit. The nuclear activity of each AGN, also tabulated in Table~\ref{basics}, is denoted in the legend, where blue circles indicate Type 1 AGNs, red triangles indicate Type 1.9 and Type 2 AGNs, purple squares indicate LINERs, and the interacting active galaxy NGC 3227 is labeled with an open green circle. The arrows on select Type 1 AGNs in the bottom row denote our attempts to constrain and remove the contribution of the nucleus from the integrated galaxy magnitudes (see \S\ref{discussion:IR}). 

We plot the residual (TF-predicted $M_{\lambda}^{b,i,k}$ minus the derived $M_{\lambda}^{b,i,k}$) under each panel, where the gray bars represent the typical scatter of 0.4\,mag for the canonical optical and near-infrared relations \citep{tp2000,cosmicflows1,tc2012}, 0.56\,mag at $W_{1}$, and 0.61\,mag at $W_{2}$ \citep{Kourkchi2020}. We calculate the scatter $\sigma$ about the TF relations exhibited by the sample of active galaxies as 0.70\,mag for the $B-$band (22 galaxies), 0.70\,mag for the $V-$band (21 galaxies), 0.66\,mag for the $R-$band (19 galaxies), 0.77\,mag for the $I-$band (18 galaxies), 1.54\,mag for the $W_{1}-$band (22 galaxies) and 1.87\,mag for the $W_{2}-$band (22 galaxies).

\begin{figure*}
\centering
\includegraphics[trim={0.1cm 10cm 0cm 10cm},clip,scale=0.26]{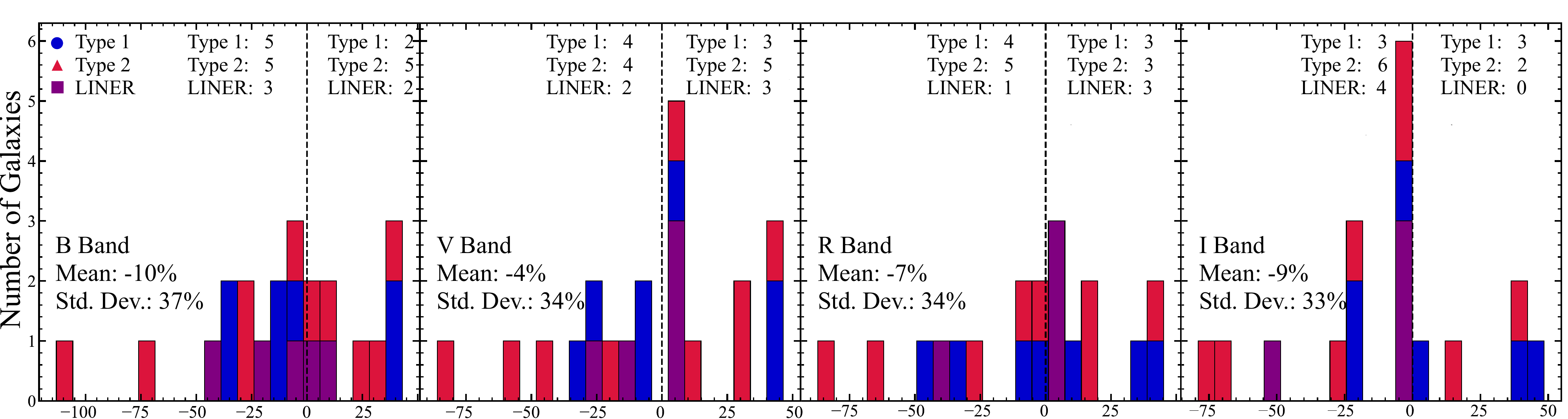}
\\
\includegraphics[trim={0cm 8.5cm 1cm 10.5cm},clip,scale=0.26]{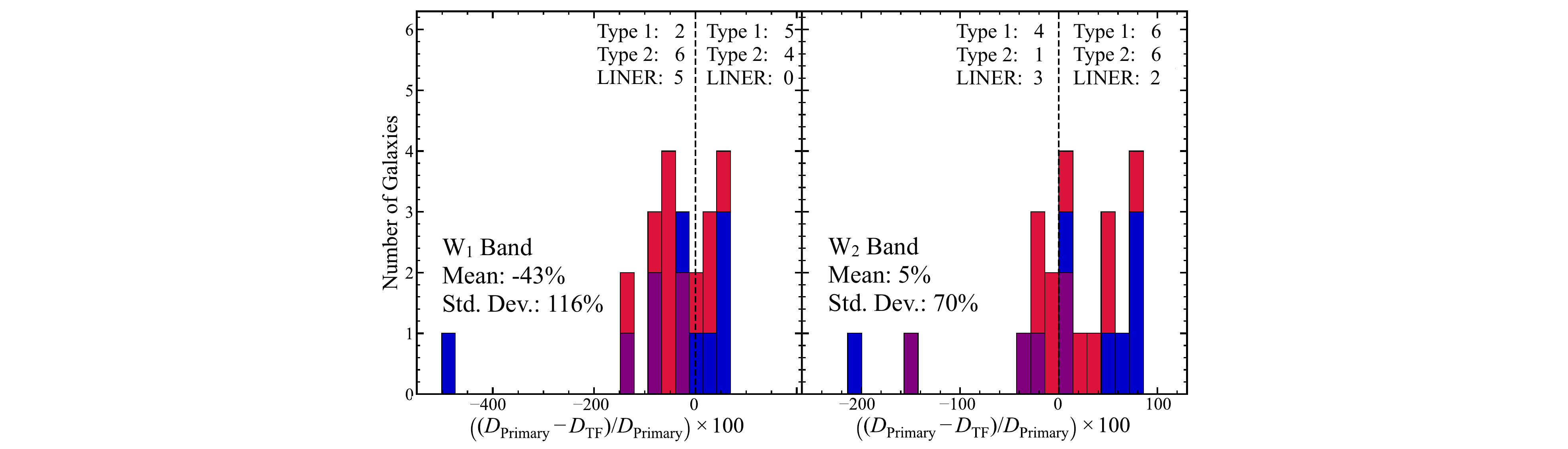}
\caption{Top column: Histograms of the percentage differences of active galaxies in the $B$, $V$, $R$, and $I$ passbands. Vertical dashed lines in each panel denote perfect agreement with predictions from TF in that bandpass. Galaxies that populate the area left of each vertical dashed line have TF distances larger than a primary distance for the same galaxy, and those that populate the right of the line have distances smaller than a primary distance. AGN activity type is labeled as blue for Type 1, red for Type 2, and purple for LINER. The mean and standard deviation of each distribution is labeled in each panel. Bottom row: same as above but for the $W_{1}$ and $W_{2}$ passbands.}\label{hist1}
\end{figure*}

To explore the behavior of each type separately, we display histograms for our sample of active galaxies as a function of the percentage difference between primary and TF-predicted distance in each bandpass in Figure \ref{hist1}. The vertical dashed line indicates perfect agreement between a galaxy's primary distance and its TF-determined distance, the left side of the dashed line indicates TF distances that are overestimated relative to primary distances, and the right side underestimated. We also calculate the mean and standard deviations of each distribution in Figure \ref{hist1}, and display them in each panel.

\begin{deluxetable*}{lcccccc}
\tablecaption{Deviations from TF Predictions}
\tablehead{
\colhead{Target} & \colhead{Dev.\ $M_{B}^{b,i,k}$} & \colhead{Dev.\ $M_{V}^{b,i,k}$} & \colhead{Dev.\ $M_{R}^{b,i,k}$} & \colhead{Dev.\ $M_{I}^{b,i,k}$} & \colhead{Dev.\ $M_{W1}^{b,k,a}$} & \colhead{Dev.\ $M_{W2}^{b,k,a}$}\\
 & (mag) & (mag) & (mag) & (mag) & (mag) & (mag)  \\
\colhead{(1)} & \colhead{(2)} &\colhead{(3)} & \colhead{(4)} & \colhead{(5)} & \colhead{(6)} & \colhead{(7)} 
}
\startdata
NGC 1068 & -0.54 $\pm$ 0.42 & -0.33 $\pm$ 0.44 & -0.56 $\pm$ 0.44 & -0.91 $\pm$ 0.46 & 2.55 $\pm$ 0.23 & 4.22 $\pm$ 0.33 \\
NGC 1566 & 1.22 $\pm$ 0.36 & 1.36 $\pm$ 0.37 & 1.31 $\pm$ 0.38 & 1.35 $\pm$ 0.47 & 1.09 $\pm$ 0.27 & 1.98 $\pm$ 0.27 \\
NGC 3147 & -1.22 $\pm$ 0.87 & -0.99 $\pm$ 0.91 & \nodata & \nodata & -1.13 $\pm$ 0.20 & -0.51 $\pm$ 0.20 \\
NGC 3783* & -0.31 $\pm$ 1.12 & -0.09 $\pm$ 1.15 & 0.26 $\pm$ 1.15 & \nodata & 1.68 $\pm$ 0.81 & 3.23 $\pm$ 0.81 \\
NGC 3982 & -0.53 $\pm$ 0.72 & -0.82 $\pm$ 0.73 & -1.05 $\pm$ 0.74 & -1.37 $\pm$ 0.86 & -0.96 $\pm$ 0.19 & -0.12 $\pm$ 0.18 \\
NGC 4051* & 1.12 $\pm$ 0.25 & 1.32 $\pm$ 0.25 & 0.94 $\pm$ 0.25 & 0.99 $\pm$ 0.24 & 1.42 $\pm$ 0.11 & 2.93 $\pm$ 0.11 \\
NGC 4138 & -1.64 $\pm$ 0.40 & -1.35 $\pm$ 0.40 & -1.36 $\pm$ 0.39 & -1.42 $\pm$ 0.45 & -0.88 $\pm$ 0.29 & -0.18 $\pm$ 0.28 \\
NGC 4151* & -0.16 $\pm$ 1.29 & -0.12 $\pm$ 1.35 & -0.12 $\pm$ 1.36 & -0.48 $\pm$ 1.43 & 2.54 $\pm$ 0.25 & 3.93 $\pm$ 0.25 \\
NGC 4258 & 0.18 $\pm$ 0.11 & 0.19 $\pm$ 0.09 & 0.10 $\pm$ 0.13 & 0.26 $\pm$ 0.23 & -1.34 $\pm$ 0.12 & -0.44 $\pm$ 0.12 \\
NGC 4303 & 1.13 $\pm$ 0.72 & 1.21 $\pm$ 0.75 & 1.07 $\pm$ 0.76 & 0.85 $\pm$ 0.85 & 0.67 $\pm$ 0.18 & 1.29 $\pm$ 0.18 \\
NGC 4395 & -0.69 $\pm$ 0.45 & -0.59 $\pm$ 0.47 & -0.83 $\pm$ 0.44 & -0.62 $\pm$ 0.51 & -3.90 $\pm$ 0.25 & -2.48 $\pm$ 0.25 \\
NGC 4565 & 0.11 $\pm$ 0.10 & 0.10 $\pm$ 0.11 & -0.00 $\pm$ 0.12 & -0.12 $\pm$ 0.13 & -0.73 $\pm$ 0.11 & -0.42 $\pm$ 0.11 \\
NGC 4639 & -0.65 $\pm$ 0.24 & -0.49 $\pm$ 0.25 & -0.65 $\pm$ 0.25 & -0.74 $\pm$ 0.30 & -0.60 $\pm$ 0.13 & 0.05 $\pm$ 0.13 \\
NGC 4939 & 0.74 $\pm$ 0.23 & 0.72 $\pm$ 0.24 & 0.47 $\pm$ 0.15 & 0.13 $\pm$ 0.22 & -0.02 $\pm$ 0.14 & 0.88 $\pm$ 0.14 \\
NGC 4945 & -0.06 $\pm$ 0.22 & \nodata & 0.37 $\pm$ 0.14 & -0.51 $\pm$ 0.71 & -1.81 $\pm$ 0.11 & 0.03 $\pm$ 0.10 \\
NGC 5055 & -0.02 $\pm$ 0.16 & 0.10 $\pm$ 0.16 & 0.05 $\pm$ 0.16 & -0.20 $\pm$ 0.18 & -1.34 $\pm$ 0.10 & -0.74 $\pm$ 0.10 \\
NGC 5643 & 0.19 $\pm$ 0.51 & 0.33 $\pm$ 0.53 & -0.12 $\pm$ 0.52 & -0.32 $\pm$ 0.60 & -0.82 $\pm$ 0.18 & 0.40 $\pm$ 0.18 \\
NGC 5728 & 0.64 $\pm$ 0.22 & 0.88 $\pm$ 0.22 & \nodata & 0.46 $\pm$ 0.26 & 0.88 $\pm$ 0.14 & 1.76 $\pm$ 0.14 \\
NGC 6814* & -0.29 $\pm$ 1.59 & 0.09 $\pm$ 1.65 & -0.07 $\pm$ 1.66 & -0.12 $\pm$ 1.86 & 0.25 $\pm$ 0.30 & 1.36 $\pm$ 0.29 \\
NGC 6951 & -0.41 $\pm$ 0.54 & -0.28 $\pm$ 0.56 & \nodata & \nodata & -0.70 $\pm$ 0.36 & 0.03 $\pm$ 0.36 \\
M 58 & -0.17 $\pm$ 0.34 & 0.08 $\pm$ 0.35 & 0.05 $\pm$ 0.36 & -0.28 $\pm$ 0.38 & -0.59 $\pm$ 0.23 & 0.16 $\pm$ 0.23 \\
M 104 & -0.74 $\pm$ 0.10 & -0.54 $\pm$ 0.09 & -0.73 $\pm$ 0.12 & -1.01 $\pm$ 0.14 & -1.82 $\pm$ 0.11 & -1.96 $\pm$ 0.11 \\
\tableline
\multicolumn{7}{l}{\text{Interacting Galaxies}} \\
\tableline
NGC 3227* & -0.11 $\pm$ 0.32 & 0.06 $\pm$ 0.32 & -0.04 $\pm$ 0.32 & -0.13 $\pm$ 0.28 & 0.51 $\pm$ 0.26 & 1.78 $\pm$ 0.26 \\
NGC 3786 & -0.66 $\pm$ 0.38 & -0.98 $\pm$ 0.34 & \nodata & \nodata & -0.27 $\pm$ 0.24 & 0.66 $\pm$ 0.24 \\
NGC 4438 & 1.63 $\pm$ 0.28 & 1.85 $\pm$ 0.29 & \nodata & \nodata & 2.41 $\pm$ 0.34 & 3.12 $\pm$ 0.33 \\
NGC 5194 & 2.09 $\pm$ 0.27 & 2.20 $\pm$ 0.28 & 2.27 $\pm$ 0.25 & \nodata & \nodata & \nodata \\
NGC 7469* & 1.64 $\pm$ 0.40 & 1.67 $\pm$ 0.41 & 1.46 $\pm$ 0.42 & 3.37 $\pm$ 0.41 & 6.15 $\pm$ 0.17 & 7.56 $\pm$ 0.17 \\
NGC 7674 & -0.83 $\pm$ 1.95 & -0.77 $\pm$ 2.04 & \nodata & \nodata & 5.22 $\pm$ 0.60 & 6.97 $\pm$ 0.59 \\
\enddata
\tablecomments{Columns 4-9 list the deviation (TF-predicted absolute magnitude minus the corrected absolute magnitude) of each AGN host in each bandpass. The bottom panel denotes interacting systems, which are not included in any quoted statistics in this work.}

\label{table:deviations}
\end{deluxetable*}


\subsection{The Baryonic TF Relation}\label{section:btfr}
The TF relation is an empirical correlation of rotational velocity and intrinsic brightness, which is essentially a relationship between a late-type galaxy's angular momentum and mass. As more massive objects generally spin faster, the more fundamental form of the TF relation is arguably the baryonic TF relation, a tightly-correlated relationship between maximum rotational velocity and baryonic matter (the sum of stellar and gas contributions to luminous mass; \citealt{mcgaugh2000,mcgaugh2005,lelli2016,iorio2017}). The BTF relation, originally calibrated using the peak of the observed rotation curve, has also been calibrated for numerous measurement methods using the unresolved \ion{H}{1} 21\,cm emission line \citep{lelli2019}. 

Each calibration of \cite{lelli2019} follows the form $\log(y) = s\log(x) + I$, where $y$ is the baryonic mass ($M_{\mathrm{BARY}}$) in units of $M_{\odot}$, $x$ is a rotational velocity measurement in km\,s$^{-1}$, and $s$ and $I$ are the slopes and intercepts respectively. We select the $W_{20}$ line width calibration (as all galaxies in our sample have $W_{20}$ measurements; see Table~\ref{appmags_21cm}) from \cite{lelli2019}, with $s=3.75 \pm 0.08$ and $I=1.99 \pm 0.18$. 

$M_{\mathrm{BARY}}$ is calculated as $1.33M_{\mathrm{HI}} + M_{\star}$, where $M_{\mathrm{HI}}$ and $M_{\star}$ are \ion{H}{1} and stellar masses of each galaxy, respectively. SMBH mass is not included as a parameter of $M_{\mathrm{BARY}}$ estimates in this work, as the typical $M_{\mathrm{BH}}/M_{\mathrm{BARY}}$ fraction for this sample is 0.0014. The factor of 1.33 accounts for the contribution of helium to a galaxy's total gas mass (see \citealt{mcgaugh2012} and references therein). We convert the \ion{H}{1} fluxes tabulated in Table~\ref{appmags_21cm} to \ion{H}{1} masses by:

\begin{equation}
    M_{\mathrm{H\,I}} = 2.36 \times 10^5 \cdot D^2 \int S_{\nu} \, dv
\end{equation}

\noindent \citep{roberts1962}, where $D$ is the primary distance in Mpc and $S_{v}\,dv$ is the integrated \ion{H}{1} flux in units of Jy\,km\,s$^{-1}$. Gas mass ($M_{\mathrm{GAS}}$) is calculated as 1.33$M_{\mathrm{HI}}$, and the $B - V$ colors of our sample were then employed along with the prescriptions of stellar mass-to-light (M/L) ratios as a function of galaxy color of \cite{schombert2018} to constrain $M_{\star}$.

We plot all active galaxies in our sample on the BTF relation in the top panel of Figure~\ref{fig:bary}. The dashed line is the BTF calibration for the $W_{20}$ definition of the \ion{H}{1} line width \citep{lelli2019}, and the shapes of the data points indicated in the legend are as follows: filled circles denote galaxies hosting a Type 1 nucleus, filled triangles indicate a Type 2 or 1.9 nucleus, filled squares indicate host of a LINER, and the open circle denotes the interacting galaxy NGC 3227. The fraction of gas mass to total baryonic mass for each galaxy is indicated by the color bar, where gas-dominated systems are represented by bluer colors and stellar-dominated systems by redder colors. 

The scatter quoted by BTF calibrations of \cite{lelli2019} is the intrinsic scatter calculated along the orthogonal direction with respect to the linear fit ($\sigma_{\perp,\mathrm{i}}$). The total orthogonal scatter ($\sigma_{\perp,\mathrm{t}}$) includes both the intrinsic and observed scatter ($\sigma_{\perp,\mathrm{o}}$), thus the intrinsic scatter is calculated as:

\begin{equation}
\sigma_{\perp,\mathrm{i}}^{2} = \left(\frac{1}{N} \sum_{\perp} \sigma_{\perp,\mathrm{t}}^{2}\right) - \sigma_{\perp,\mathrm{o}}^{2}
\end{equation}

\noindent where $\sigma_{\perp,\mathrm{t}}$ is the deviation of the derived baryonic masses of each galaxy perpendicular to linear fit, $\sigma_{\perp,\mathrm{o}}$ is the mean of the observational uncertainties projected onto the orthogonal direction, and $N$ is the total number of data points.

We plot $\sigma_{\perp,\mathrm{t}}$ in the bottom panel of  Figure~\ref{fig:bary}, with the gray bar indicating the intrinsic scatter of the $W_{20}$ BTF calibration of 0.035\,dex. We also provide calculations of $\sigma$ (scatter along the y-axis according to Equation~\ref{equation:vertscatter}), the total orthogonal scatter ($\sigma_{\perp,\mathrm{t}}$) and the intrinsic orthogonal scatter ($\sigma_{\perp,\mathrm{i}}$). For active galaxies about the BTF relation, we find a vertical scatter of 0.35\,dex, total orthogonal scatter of 0.10\,dex, and intrinsic orthogonal scatter of 0.06\,dex (21 galaxies).

\section{Discussion}\label{section:discussion}
Among the drivers of outlying behavior about the TF relation, which include the ratio of dark matter to baryonic matter \citep{Pizagno2007}, galaxy disc sizes \citep{reyes2011}, and environmental effects such as local cluster environments (\citealt{Ouellette2017} and references therein), the correlation of scatter about the TF relation and star formation has been of focus \citep{Barton2001,Buchalter2001,bd2001,Kannappan2002,Torres-Flores2013,Ristea2024}. However, the effect of the presence of an AGN on the TF relation is relatively unexplored.

While the samples used to calibrate the canonical TF relations did not explicitly flag AGNs for removal \citep{tp2000,cosmicflows1,tc2012,Kourkchi2020}, the selection criteria generally exclude active galaxies. Primarily, all works above select spirals with inclinations greater than $45\degree$. As Type 1 AGNs have been observed to be preferentially hosted by face-on ($<45\degree$) galaxies \citep{Keel1980,mr1995,McLeod1995,Simcoe1997,Gkini2021}, this criterion naturally excludes a significant amount of Seyfert 1 hosts. The nuclear flux from unobscured Type 1 AGNs represents the primary expected source of photometric scatter in TF relations, whereas the high levels of nuclear obscuration inherent in Type 2 systems are expected to largely mitigate such contamination.

While certain studies, such as \cite{Tiley2016}, employ explicit Baldwin-Phillips-Terlevich (BPT; \citealt{Baldwin1981}) diagnostics to exclude likely Seyfert candidates, the primary optical and near-infrared TF calibrations do not. Studies that utilize SDSS photometry, such as \cite{Mocz2012} and, most importantly, the TF calibrations of \cite{Kourkchi2020}, exclude galaxies with photometric or spectroscopic flags. This methodology naturally removes systems characterized by either point-source-dominated cores or significantly disturbed emission-line profiles, effectively implicitly removing high-luminosity AGNs from the calibration samples in addition to a large fraction of low- to moderate-luminosity AGNs. Given the above, analyses of active galaxies on the TF relation relative to each calibration serves as a good comparison between active galaxies and an (approximately) inactive galaxy sample.

\subsection{Active Galaxies on the TF Relation}
In all filters, we find evidence that this sample of AGN hosts exhibits larger $\sigma$ than is exhibited by a largely inactive galaxy sample, where the scatter exhibited by active galaxies is $\sim$75$\%$ larger in the optical and near-infrared, and $\sim$180$\%$ larger in the infrared. Once more, interacting systems are not included in any quoted statistics or scatter. The larger scatter may suggest that while the TF relation can predict accurate distances to a large distribution of active galaxies, the predicted distance to any one active galaxy is subject to higher uncertainty.

We note here that this sample does not represent a statistically significant sample of active galaxies, and as such we are limited in our analysis of their universal behavior on the TF relation. The sample is also biased towards high-mass, stellar-dominated, high-surface brightness galaxies that span only $\sim$2 decades in absolute magnitude, whereas TF calibrations will typically span $\sim$5-6 decades in absolute magnitude. A statistically significant sample would require relatively equal subsamples of nuclear activity types and a variety of galaxy cluster memberships to mitigate contributions of $V_{\mathrm{PEC}}$, which would reasonably dictate a sample of $\sim$50$-$100 active galaxies. As such, it remains to be seen how lower-mass, gas-rich AGN host galaxies align with the canonical TF relations. 

\subsubsection{Optical and Near-Infrared Passbands}
We compare here the scatter exhibited by active galaxies on the TF relations for the $B$, $V$, $R$, and $I$ bands to the traditional scatter quoted by the canonical TF relations in addition to known sources of scatter/outlying behavior from the TF relations. According to \cite{tc2012}, if galaxies with Hubble type Sa are included in TF calibrations, the scatter about the relation has been observed to increase (most likely due to the decreased disk contributions and increased bulge size of more early-type galaxies). For the ground-based optical and near-infrared calibrations of the TF relation, the scatter has been shown to decrease towards redder optical and near-infrared wavelengths, reaching minimal scatter at $R$ and $I$ bands \citep{pt1988,tp2000}. The scatter increases towards blue wavelengths, presumably due to increased inclination-dependent obscuration and random star formation (CF2), and also increases towards infrared ($H$ and $K$) caused by significant sky contributions in the infrared \citep{tully1982}. We do see an increase in scatter exhibited by active galaxies in the reddest infrared wavelengths as displayed in Figure~\ref{mainplot} consistent with the canonical TF relations, but the scatter does not seem to decrease from blue to red optical/near-infrared wavelengths.

While the sample is currently small, the larger observed
scatter seems to persist regardless of the type of nuclear activity. AGNs of Seyfert Type 1, 1.9, 2 all represent significant outliers from the TF relation. We find that Type 2 AGNs represent the largest percentage differences from primary distance determinations, with some approaching discrepancies of $\sim$100\%, and Type 1 AGNs and LINERs exhibiting smaller percentage differences. The Type 1 systems seem to exhibit a smaller percentage difference compared to the Type 2 systems, and we suspect the primary factor is that the nuclear contamination was removed from the integrated galaxy magnitudes for the majority of Type 1 AGNs via surface brightness decompositions (see \S\ref{section:photometry}, \citealt{misty2018}, Paper I).

Given the means of the distributions in each panel of Figure \ref{hist1}, we find that TF-based distances overestimate the primary distances for this sample by anywhere from 4\% ($V$ band) to 10\% ($B$ band), and the predicted distances overall exhibit a percentage difference of $\sim$35\%. The active galaxies in the sample with deviations from TF predictions larger than the typical quoted scatter (within the uncertainties) are: NGC 1068, NGC 1566, NGC 4051, NGC 4138, NGC 4438, NGC 4939, NGC 5194, NGC 5728, NGC 7469, and M 104. Of those, the interacting systems are NGC 4438, NGC 5194, and NGC 7469, and thus are not included in any statistics.

Though its nucleus carries a Type 2 classification, NGC 1068 was among the original sample of \cite{seyfert1943} due to its bright nucleus and high-excitation, significantly broadened emission lines. NGC 1068 was not included amongst our sample of Type 1 hosts where we modeled and removed the light contribution of the AGN, therefore the light from its nucleus remains in the corrected apparent magnitudes investigated here. We do not expect the Cepheids-based distance to NGC 1068 \citep{markham2026} to contribute to its outlying behavior, nor do we expect the light contamination from its AGN-driven outflows to significantly contribute. The ionized mass outflow in NGC 1068 has been measured to be confined to within only $\sim$200\,pc at a moderate outflow rate ($\sim8\,M_{\odot}$\,yr$^{-1}$ at a peak location of $\sim$100\,pc from the nucleus) than what has been found for similar nearby Type 2 AGN \citep{Revalski2021}. The unresolved \ion{H}{1} emission line of NGC 1068, however, exhibits broadened wings as opposed to the usual steep flanks expected from typical circular rotation \citep{ft1981,hr1989}. The high-resolution Very Large Array study by \cite{Brinks1997} revealed an atypical \ion{H}{1} rotation curve, with a slight increase in radial velocity corresponding to an \ion{H}{1} ring at $\sim$3\,kpc, and a gradual drop off in velocity at large radii as opposed to a flat outer curve. The irregularities in the rotation traced by \ion{H}{1} emission may explain the offset in absolute magnitude relative to the expected absolute magnitude at that rotation rate.

\begin{deluxetable}{lcccc}
\setlength{\tabcolsep}{3pt}
\tablecaption{Galaxy Masses and Deviations from BTF Predictions}
\tablehead{
\colhead{Target} & \colhead{$M_{\mathrm{GAS}}$} & \colhead{$M_{\star}$} & \colhead{$M_{\mathrm{BARY}}$} & \colhead{Dev.\ $M_{\mathrm{BARY}}$}
\\
& ($M/M_{\odot}$) & ($M/M_{\odot}$) & ($M/M_{\odot}$) & (dex)  \\
\colhead{(1)} & \colhead{(2)} &\colhead{(3)} & \colhead{(4)} & \colhead{(5)}
}
\startdata
NGC 1068 & 9.07 $\pm$ 0.04 & 10.81 $\pm$ 0.16 & 10.82 $\pm$ 0.16 & -0.12 $\pm$ 0.22 \\
NGC 1566 & 10.18 $\pm$ 0.09 & 10.71 $\pm$ 0.11 & 10.82 $\pm$ 0.09 & 0.62 $\pm$ 0.12 \\
NGC 3147 & 10.09 $\pm$ 0.03 & 11.34 $\pm$ 0.21 & 11.36 $\pm$ 0.20 & -0.43 $\pm$ 0.28 \\
NGC 3783 & 9.73 $\pm$ 0.24 & 10.88 $\pm$ 0.31 & 10.91 $\pm$ 0.30 & 0.18 $\pm$ 0.42 \\
NGC 3982 & 9.39 $\pm$ 0.03 & 9.26 $\pm$ 0.30 & 9.63 $\pm$ 0.16 & -0.94 $\pm$ 0.22 \\
NGC 4051 & 9.45 $\pm$ 0.02 & 10.69 $\pm$ 0.28 & 10.72 $\pm$ 0.26 & 0.72 $\pm$ 0.37 \\
NGC 4138 & 8.95 $\pm$ 0.09 & 10.17 $\pm$ 0.11 & 10.19 $\pm$ 0.11 & -0.34 $\pm$ 0.15 \\
NGC 4151 & 9.48 $\pm$ 0.02 & 10.00 $\pm$ 0.36 & 10.11 $\pm$ 0.30 & -0.13 $\pm$ 0.42 \\
NGC 4258 & 9.48 $\pm$ 0.02 & 10.62 $\pm$ 0.10 & 10.65 $\pm$ 0.10 & -0.09 $\pm$ 0.14 \\
NGC 4303 & 9.88 $\pm$ 0.01 & 10.54 $\pm$ 0.19 & 10.63 $\pm$ 0.16 & 0.42 $\pm$ 0.23 \\
NGC 4395 & 9.20 $\pm$ 0.08 & 9.09 $\pm$ 0.36 & 9.45 $\pm$ 0.20 & 0.06 $\pm$ 0.28 \\
NGC 4565 & 9.88 $\pm$ 0.02 & 10.73 $\pm$ 0.11 & 10.79 $\pm$ 0.10 & -0.16 $\pm$ 0.14 \\
NGC 4639 & 9.35 $\pm$ 0.03 & 10.24 $\pm$ 0.17 & 10.29 $\pm$ 0.15 & -0.12 $\pm$ 0.21 \\
NGC 4939 & 10.46 $\pm$ 0.02 & 10.93 $\pm$ 0.10 & 11.06 $\pm$ 0.08 & 0.12 $\pm$ 0.11 \\
NGC 4945 & 9.11 $\pm$ 0.01 & \nodata & \nodata & \nodata \\
NGC 5055 & 10.10 $\pm$ 0.03 & 10.73 $\pm$ 0.17 & 10.82 $\pm$ 0.14 & 0.07 $\pm$ 0.20 \\
NGC 5643 & 9.47 $\pm$ 0.04 & 10.48 $\pm$ 0.08 & 10.52 $\pm$ 0.07 & 0.15 $\pm$ 0.10 \\
NGC 5728 & 9.78 $\pm$ 0.03 & 11.20 $\pm$ 0.15 & 11.22 $\pm$ 0.15 & 0.41 $\pm$ 0.21 \\
NGC 6814 & 9.62 $\pm$ 0.02 & 10.94 $\pm$ 0.18 & 10.96 $\pm$ 0.17 & 0.46 $\pm$ 0.24 \\
NGC 6951 & 9.58 $\pm$ 0.04 & 10.83 $\pm$ 0.14 & 10.86 $\pm$ 0.14 & -0.25 $\pm$ 0.19 \\
M 58 & 9.08 $\pm$ 0.08 & 11.17 $\pm$ 0.13 & 11.18 $\pm$ 0.13 & 0.07 $\pm$ 0.19 \\
M 104 & 8.41 $\pm$ 0.01 & 11.30 $\pm$ 0.06 & 11.30 $\pm$ 0.06 & -0.44 $\pm$ 0.09 \\
\tableline
\multicolumn{5}{l}{\text{Interacting Galaxies}} \\
\tableline
NGC 3227* & 9.43 $\pm$ 0.09 & 10.80 $\pm$ 0.24 & 10.82 $\pm$ 0.23 & 0.08 $\pm$ 0.33 \\
NGC 3786 & 9.93 $\pm$ 0.08 & 9.70 $\pm$ 0.30 & 10.13 $\pm$ 0.15 & -0.89 $\pm$ 0.21 \\
NGC 4438 & 8.81 $\pm$ 0.06 & 10.79 $\pm$ 0.11 & 10.80 $\pm$ 0.11 & 1.04 $\pm$ 0.16 \\
NGC 5194 & 9.66 $\pm$ 0.07 & 10.43 $\pm$ 0.26 & 10.50 $\pm$ 0.23 & 1.02 $\pm$ 0.33 \\
NGC 7469* & 9.09 $\pm$ 0.04 & 10.65 $\pm$ 0.34 & 10.66 $\pm$ 0.33 & 0.58 $\pm$ 0.47 \\
NGC 7674 & 10.00 $\pm$ 0.03 & 11.07 $\pm$ 0.17 & 11.11 $\pm$ 0.16 & -0.49 $\pm$ 0.23 \\
\enddata
\tablecomments{Galaxy mass estimates and deviations from the BTF relation (derived $M_{\mathrm{BARY}}$ minus the predicted $M_{\mathrm{BARY}}$) for the sample of active galaxies. Galaxy names marked with asterisks indicate galaxies consistent with the sample of Paper I (see \S\ref{section:data}). $M_{\mathrm{BARY}}$ is calculated as $M_{\mathrm{GAS}}$ + $M_{\mathrm{\star}}$, where gas and stellar mass derivations are described in \S\ref{section:btfr}. The bottom panel denotes interacting systems, which are not included in any quoted statistics in this work.}
\label{table:btfr}
\end{deluxetable}

NGC 3227 is tidally interacting with its dwarf elliptical companion NGC 3226, and studies have revealed a richly intertwined dynamical history between the two galaxies \citep{rf1968,mundell1995,Appleton2014}. However, \cite{Schinnerer2000} and \cite{falcone2024}, who studied the kinematics of NGC 3227’s disk in cold molecular and ionized gas, respectively, did not detect unusual kinematic signatures indicative of tidal influences. Rather, \cite{falcone2024} find the ionized gas kinematics to be dominated by typical rotational motions out to $\sim$80$\arcsec$ from the SMBH, which is the approximate extent of the galactic disk. This may explain why NGC 3227, even though it is interacting, is not a significant outlier from the optical and near-infrared TF relations.

NGC 4138 carries an SA0(r) morphological classification, a much earlier type of galaxy than is used for TF calibration samples in the literature, which is the most likely reason for its outlying behavior. We do not expect distance to contribute to the large observed scatter, as the distance was determined from SBF (\citealt{tonry2001,cantiello2018}). We do not expect the disk kinematics to contribute to the scatter either, as the unresolved \ion{H}{1} 21\,cm spectrum has been shown to be symmetric about the line center \citep{hr1989,edd2005}.

For NGC 4939, we again do not treat the derived rotational properties as suspect given the 21\,cm spectra are typically symmetric \citep{dr1978,shostak1978,koribalski2004,edd2005,ct2015}. The distance method used for NGC 4939, SN type II, usually shows significantly larger scatter than Type Ia SN distances, however \cite{jaeger2017} derived an intrinsic scatter of 0.35\,mag based on 73 type II SN in a redshift range of $z=0.01-0.5$, comparable (and slightly smaller) to the observed intrinsic scatter of most TF relations. NGC 4939 lies below the TF relation outside the typical scatter at $B$ and $V$, but rises to within the typical scatter at $R$ and $I$ bands, thus NGC 4939 is fainter than expected at $B$ and $V$ wavelengths.

\begin{figure}
\centering
\includegraphics[trim={1cm 1cm 0cm 3.7cm},clip,scale=0.44]{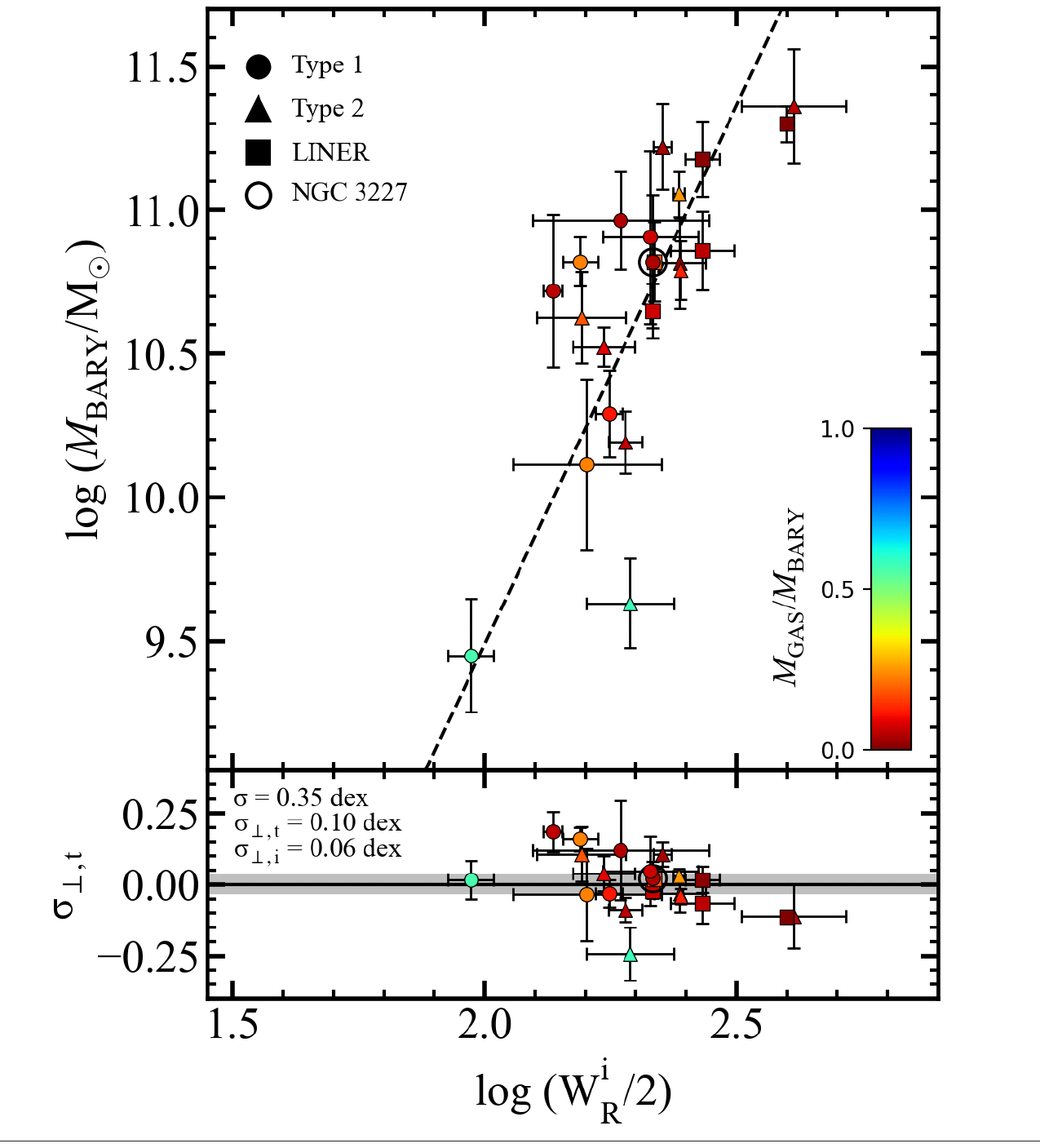}
\\
\centering
\includegraphics[trim={3cm 0cm 0cm 0cm},clip,scale=0.22]{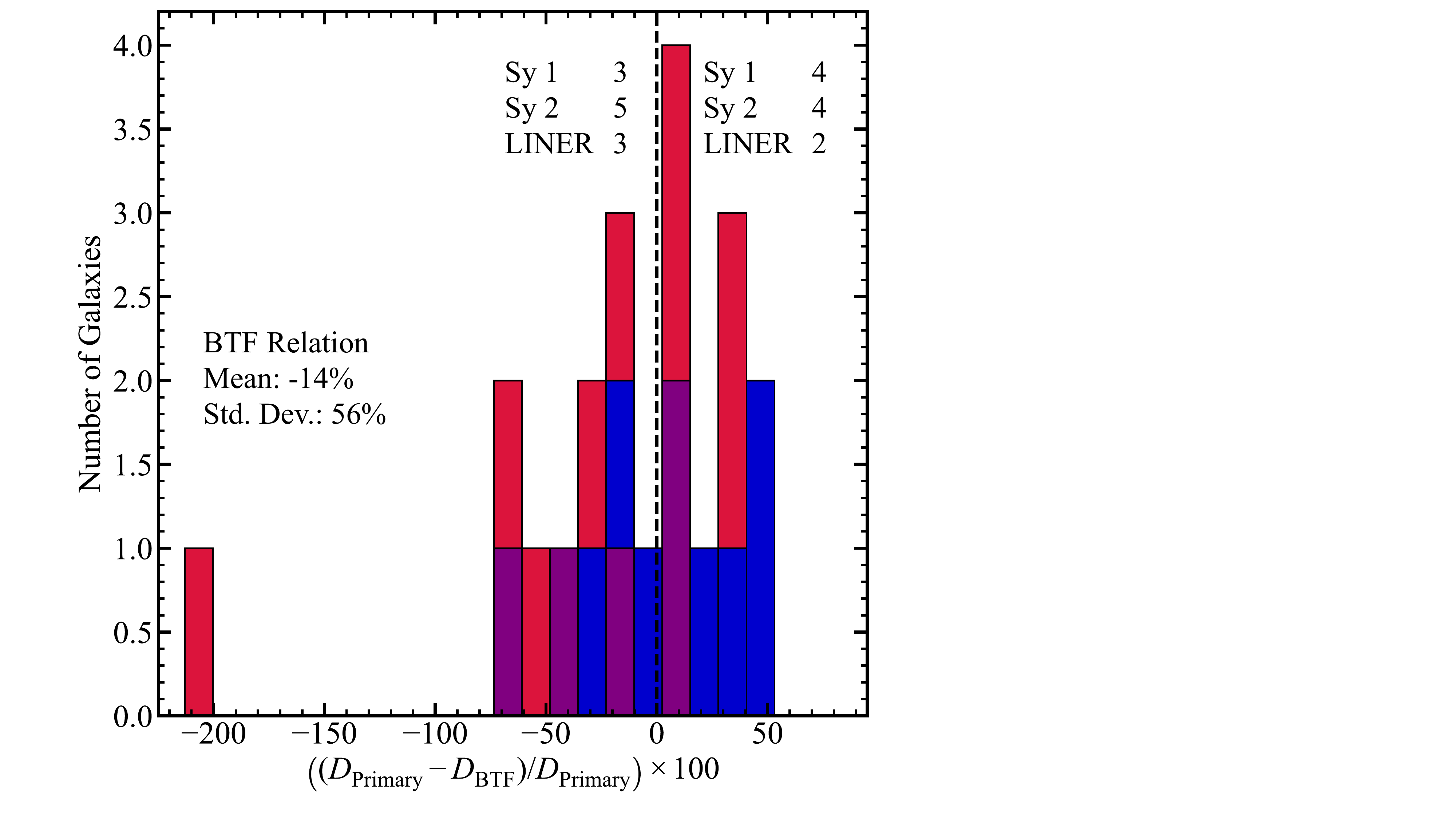}
\caption{Top panel: the BTF relation for active galaxies. $M_{\mathrm{BARY}}$ is calculated as 1.33$M_{\mathrm{HI}}+M_{\mathrm{\star}}$ (see Table~\ref{appmags_21cm}, \S\ref{section:spectroscopy}, \S\ref{section:btfr}). Each galaxy is labeled by their fraction of gas mass to total baryonic mass ($M_{\mathrm{GAS}}/M_{\mathrm{BARY}}$) indicated by the color bar. The orthogonal distance of each data point from the linear calibration ($\sigma_{\perp,\mathrm{t}}$) is plotted beneath, with several calculations of scatter indicated in the top left corner. Type 1 AGN hosts are plotted as circles, Type 1.9 and 2 hosts as triangles, LINER hosts as squares, and the interacting AGN host NGC 3227 is labeled with an open circle. Bottom panel: Histograms of the percentage deviations of distances predicted from inverting the BTF relation from each active galaxy's primary distance measurement, in the same format as Figure \ref{hist1}.} \label{fig:bary}
\end{figure}

Interestingly, the LINER host M 1042, `The Sombrero Galaxy', is included among the largest outliers, even though its distance is TRGB-based \citep{McQuinn2016} and its \ion{H}{1} spectrum has been observed to be typically symmetric \citep{Gallagher1975,Courtois2011}. M 104 is an AGN-driven LINER, as Very Long Baseline Array observations at linear scales under 0.01\,pc yielded direct evidence of a nuclear radio jet \citep{Hada2013}, thus its outlying behavior may suggest that present nuclear activity may not be a reliable indicator of the AGN's recent impact on the large-scale galaxy surface brightness, and the history of the individual AGN's feedback may need to be factored in. We do note that the adopted morphological classification for M 104 is SA(s)a, which falls under the earliest galaxy type included in TF calibrations and thus may be too early a type to reliably align with a universal TF relation.

\subsubsection{Infrared Passbands}\label{discussion:IR}
Active galaxies on the WISE TF relations in the infrared display the largest scatter with respect to all other passbands investigated in this study. As shown in Figure~\ref{mainplot}, galaxies hosting a Type 1 AGN generally lie above the relations, while Type 2 AGN generally lie closer. This is most likely because the AGN contamination for the Type 1 hosts was not modeled or removed at infrared wavelengths (see \S\ref{section:ir_photometry}). Given the means of the distributions in each panel of the top row Figure \ref{hist1}, we find that at $W_{1}$, TF-based distances are overestimated with respect to the primary distances by $\sim$40\% (the largest outlier in both filters is NGC 4395, a dwarf Type 1 AGN host), whereas at $W_{2}$ they have a small tendency to be underestimated (mean of 0.05). Both, however, exhibit significantly larger ranges of percentage differences in distance than the optical/near-infrared filters (116\% at $W_{1}$ and 70\% in $W_{2}$).

As discussed in \S\ref{section:ir_photometry}, the resolution of the WISE instrument is insufficient to separate the nuclear light from the host galaxy, thus we turned to the SEDs of \cite{rachael} to attempt to constrain the AGN light contribution through the $W_{1}$ and $W_{2}$ passbands. We represent each removal as downward facing arrows in Figure \ref{mainplot}. We note here, however, that the infrared light constrained in these filters is contributed largely by the torus \citep{lr2017}. The light emitted by a torus depends both on the AGN luminosity and physical properties of the torus itself, which are more influenced by the host galaxy rather than the central engine (e.g., \citealt{Schartmann2009}, see review by \citealt{lr2022}). Therefore, these attempts simply represent first-order estimates and should be taken with additional scrutiny.

The AGN in NGC 3227 represents the largest estimated luminosity contribution with respect to the total galaxy in the infrared ($\sim$70\% compared to $\lesssim$10\% for the remaining Type 1 hosts). We note once more that the contribution of luminosity from the AGN is estimated via extrapolation of the SED to the WISE wavelengths, and that the radiation at these wavelengths is expected to originate largely from the dusty torus. The infrared data used to construct the SEDs were largely observed at the same time with the same instrument \citep{misty2018}. NGC 3227, 4051, and 4151 were observed in April 2013, while NGC 6814 and 7469 were observed in September 2011, all with the WIYN instrument. In contrast, the operating years of the WISE telescope were 2009-2011, lending to the possibility that the AGN activity in a number of these Seyfert hosts may have been stronger during the observation period of WISE. For example, the historical light curve of NGC 4151 presented by \cite{4151_continuum1} shows that the optical nuclear continuum of NGC 4151 was approximately 0.5\,mag brighter in 2010-2011 than it was in 2013 when the infrared imaging used to construct the SED was conducted. However, even if broadband imaging happens to be timed to a pronounced period of AGN activity and luminosity output, it still may not enable sufficient removal of the effect of an AGN on integrated galaxy luminosity for use in TF distance determinations as we have shown here. 

\subsection{Active Galaxies on the BTF Relation}
As we noted in \S\ref{section:btfr}, the intrinsic scatter $\sigma_{\perp,\mathrm{i}}$ calculated by \cite{lelli2019} for the calibration for the $W_{20}$ line width definition is 0.035\,dex. The sample of (non-interacting) active galaxies exhibit an intrinsic perpendicular scatter of 0.06\,dex, a percentage difference of $\sim$55\%. The vertical scatter we find for this sample, $\sigma = 0.35$\,dex, is also slightly larger than that observed for the Spitzer Photometry and Accurate Rotation Curve (SPARC) sample of $\sigma \simeq 0.22$\,dex \citep{mcgaugh2012,lelli2016,Dutton2017}. However, the relative increase in scatter can likely be attributed to any or all of the following factors: 1) this sample is much smaller than the $W_{20}$ calibration sample of Lelli et al.\ (21 galaxies compared to 148) and only probes the high-mass end of the BTF relation at the fastest maximum rotational velocities, significantly limiting any conclusions of the universal behavior of AGN hosts on the BTF relation, 2) the line width deprojections from the $W_{20}$ calibration of \cite{lelli2019} are informed from \ion{H}{1} velocity fields as opposed to photometric axis ratios we have employed in this study, adding a source of scatter relative to the calibrating sample, and/or 3) approximately half of the axis ratios of this sample are above 0.72 (which corresponds to an inclination of 45\degree, the usual cutoff inclination for TF calibrations), and the uncertainty on $W_{\mathrm{R}}^{i}$ increases as $q_{\mathrm{d}}$ increases. 

As galaxy distance is used to calculate both $M_{\mathrm{GAS}}$ and $M_{\star}$, the BTF relation can be inverted to predict distance, and we display the percentage difference between BTF-based distances and primary distances in Figure \ref{fig:bary}. We again see Type 2 AGNs exhibit larger percentage deviations compared to Type 1 AGNs and LINERs, and find that BTF-predicted distances tend to overestimate primary distances by nearly 15\% with an overall percentage difference of nearly 60\%. However, we note once more that the method used to constrain these rotational velocities may not precisely align with the baryonic content, which is estimated independently of rotation. Therefore, these results serve as a preliminary, limited analysis of BTF-predicted distances to active galaxies.

\begin{figure}
\centering
\includegraphics[trim={1cm 1.8cm 0cm 1.8cm},clip,scale=0.545]{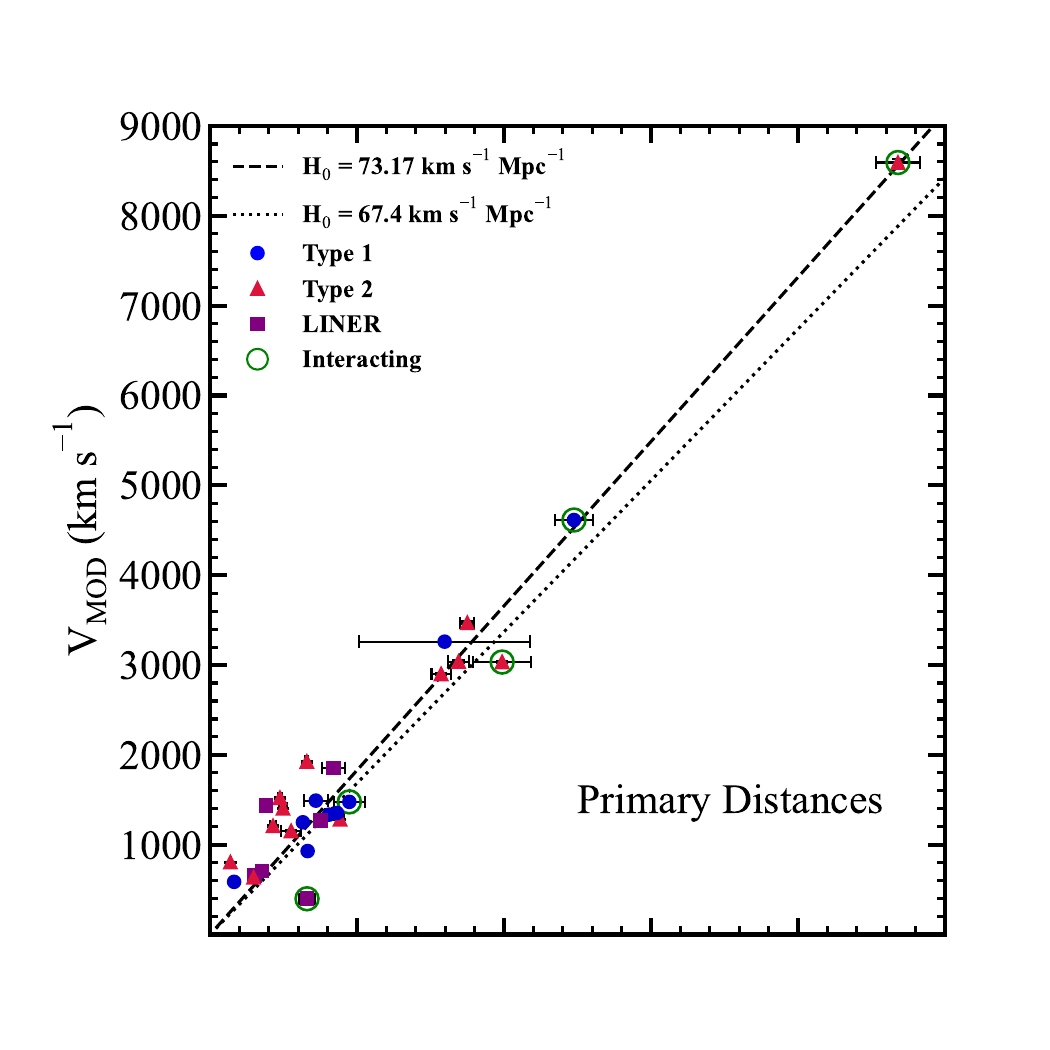}
\\
\centering
\includegraphics[trim={1cm 0cm 0cm 2.1cm},clip,scale=0.545]{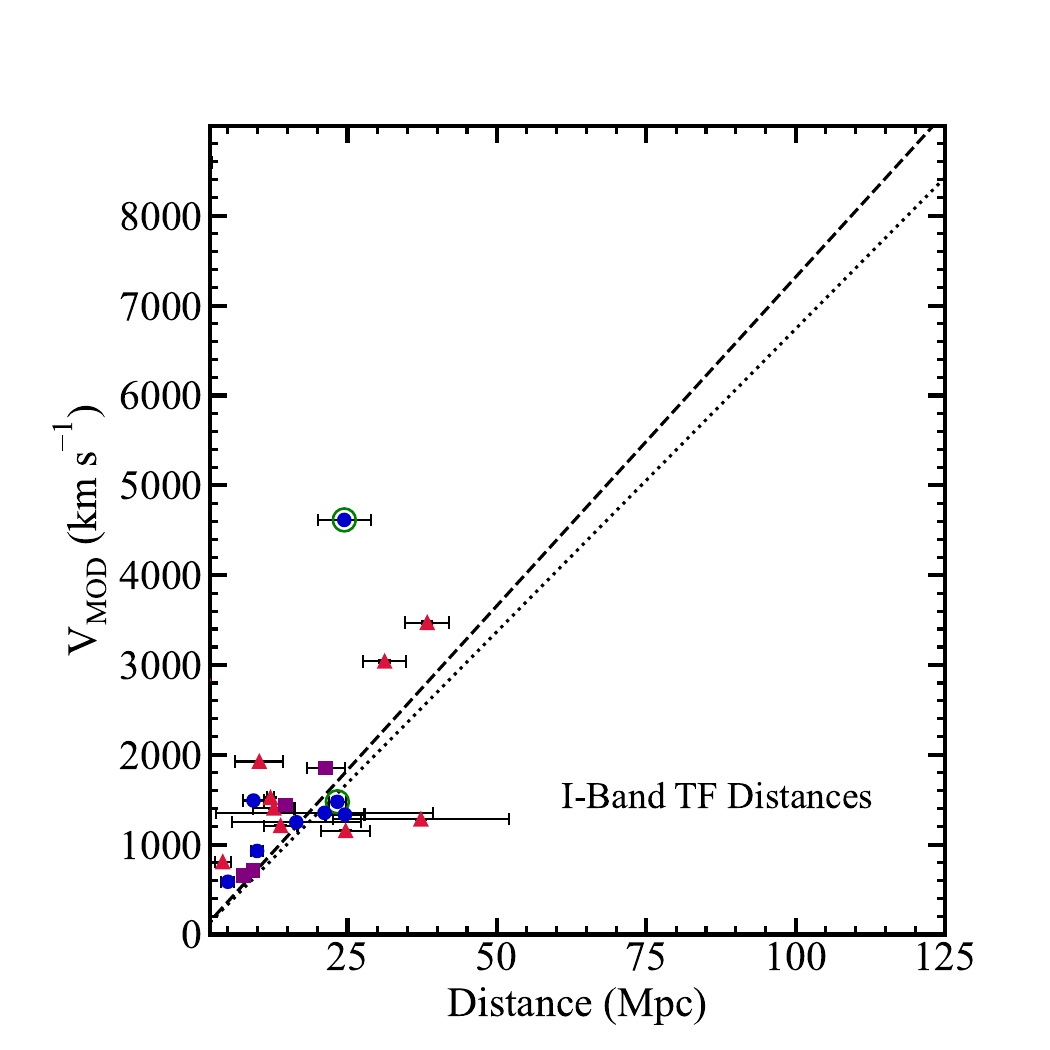}
\caption{Top panel: Hubble diagram for the sample of active galaxies with $z-$independent distances (see Table~\ref{basics}) vs.\ their cosmologically-adjusted velocity $V_{\mathrm{MOD}}$. Blue circles denote hosts of Type 1 AGNs, red triangles of Type 1.9 or 2 AGNs, purple squares of LINERs, and open green circles indicate interacting systems. We compare these to the Hubble-Lema\^itre Law, with $H_{0}$ values based on the most recent distance ladder measurements (73.17\,km\,s$^{-1}$\,Mpc$^{-1}$; \citealt{Riess2022_2,Murakami2023,Breuval2024}) and $\Lambda$CDM predictions from models to the cosmic microwave background radiation (67.4 km s$^{-1}$ Mpc$^{-1}$; \citealt{planck2018}). Bottom panel: same as the top panel, but with $I-$band TF distances employed.} 
\label{figure:hubble}
\end{figure}

There may also be an impact on galaxy-scale kinematics via AGN-driven NLR outflows, which have been observed to span up to $\sim$4\,kpc
from the central SMBH \citep{of2006,revalski2018,Revalski2018_2,Revalski2021,Revalski2022,Revalski2025}. Even nearby, low/intermediate-luminosity systems such as this sample have the capability of introducing non-gravitational gas motions at kiloparsec scales, as has been observed by \cite{Wylezalek2020}. Wylezalek et al.\ identify AGN kinematic signatures in galaxies that no longer align with photoionization from an active nucleus, termed `relic AGN' \citep{if2015}. For hosts of relic AGN, outflow and mechanical feedback signatures would be longer lived than the AGN itself. Therefore, even the hosts of nearby, low-luminosity AGNs may still carry kinematic signatures of higher-luminosity systems, and thus may be subject to galaxy-scale disturbances by AGN-driven outflows. However, resolved \ion{H}{1} 21\,cm studies such as The \ion{H}{1} Nearby Galaxy Survey \citep{walter2008} have measured flat rotation curves out to $\sim$10 $-$ 15\,kpc on average, appreciably larger than that found for NLR outflow extents and thus the bulk of the neutral gas likely remains undisturbed.

\subsection{AGN Hubble Diagram}\label{section:hubble_results}
Finally, we place active galaxies on a Hubble diagram in Figure \ref{figure:hubble}, and compare the primary distances (top panel) to distances predicted by the $I-$band TF relation (bottom panel). We correct each recessional velocity measured from the redshift of the \ion{H}{1} 21\,cm emission line (see Table~\ref{basics}) for relativistic effects assuming $\Lambda$CDM cosmology (which are small for galaxies with $z<0.1$ such as this sample). The modified, cosmologically-adjusted galaxy velocity ($V_{\textsc{MOD}}$) described by the CF program is given as
\begin{equation}
    V_{MOD} = cz[1 + 0.5(1-q_{0})z - (1/6)\left(1 - q_{0} - 3q_{0}^{2} + 1\right)z^{2}]
\end{equation}
where $z$ is the redshift with respect to the Cosmic Microwave Background rest frame, $q_{0}$ = 0.5($\Omega_{\textsc{M}}$ - 2$\Omega_{\Lambda}$), $\Omega_{\textsc{M}}$ = 0.27, and $\Omega_{\Lambda}$ = 0.73. In each panel of Figure \ref{figure:hubble}, we display the most recent three-rung distance ladder measurement of $H_{0} = 73.17 \pm 0.86$\,km\,s$^{-1}$\,Mpc$^{-1}$ \citep{Riess2022_2,Murakami2023,Breuval2024} and the predicted $H_{0}$ constrained from $\Lambda$CDM models to the cosmic microwave background radiation of $H_{0} = 67.4 \pm 0.5$\,km\,s$^{-1}$\,Mpc$^{-1}$ \citep{planck2018}. 

Both panels share a large scatter of $V_{\mathrm{MOD}}$ as a function of distance for nearby galaxies as expected, considering the contribution of $V_{\mathrm{PEC}}$ increases as redshift decreases. For primary distances, active galaxies seem well-characterized by an $H_{0}$ as measured from the cosmic distance ladder. The $I-$band TF distances for the same galaxies, however, exhibit significant deviations from the Hubble flow, even at distances closer than 50\,Mpc. The $I-$band is preferred in both the CF catalog and in observational $H_{0}$ constraints (e.g., \citealt{Sakai2000,Boubel2024,Scolnic2024}), as galaxies generally exhibit slightly less scatter at $I$ than other passbands due to the avoidance of larger internal obscuration at bluer wavelengths and a better trace of older stellar populations that contribute the majority of mass. $I-$band distances are also heavily utilized for the CF model of the local $V_{\mathrm{PEC}}$ field \citep{shaya2017,graziani2019}. Yet, the scatter from active galaxies persists even in the $I$ band, and so we suggest that AGN hosts be removed from future samples used by CF $V_{\mathrm{PEC}}$ field models.

\section{Conclusions}
We have investigated a sample of 22 active galaxies with primary distances on the TF and BTF relations. In all photometric filters explored here, we find that active galaxies exhibit $\sim$75\% larger scatter in the optical and near-infrared filters and $\sim$180\% larger scatter in the infrared compared to the (largely inactive) calibrating samples. Hosts of Type 2 AGNs represent the largest scatter at $B$, $V$, $R$, and $I$, and hosts of Type 1 AGNs contribute more to the scatter at $W_{1}$ and $W_{2}$. About the BTF relation, active galaxies once more exhibit increased scatter ($\sigma_{\perp,\mathrm{i}}=0.06$\,dex) relative to its calibration, however this may be explained by the small sample size of this study and the bias towards the highest-mass systems. This outlying behavior persists despite our efforts to constrain and remove the AGN contamination from the integrated galaxy surface brightnesses by either 1) surface brightness decompositions for the optical/near-infrared magnitudes or 2) AGN SEDs to estimate the nuclear light contribution in the WISE magnitudes, which suggests that the effect of an AGN on large-scale galaxy surface brightness may not be confined to the nucleus. 

We also calculate the percentage differences between primary and TF-based distances for this sample, and find that TF distances are biased towards overestimation of the true distances to active galaxies in every filter targeted here except for $W_{2}$. If distance to active galaxies is instead provided by inverting the BTF relation, we find a slightly larger bias towards overestimation (mean of $\sim$15\%) relative to the true distances. Even in the $I-$band, the most heavily-utilized calibration in the CF catalog, TF-predicted distances are shown to be biased further by a factor of $\sim$10\%, with an overall percentage difference of $\sim$35\%. The disk kinematics of this sample that determine the TF and BTF-predicted distances may also be suspect given the capability of AGN-driven outflows and high-velocity gas to span to kiloparsec scales (even for nearby, low-luminosity systems like this sample) introducing non-gravitational motion and even additional broadening of unresolved emission lines.

This sample is limited, however, to the closest, brightest systems of the highest stellar mass, hindering our ability to investigate the behavior of active galaxies on the TF and BTF relations for dimmer, gas-rich systems with slower rotational velocities. Systematic targeting of AGNs for $z-$independent distance measurements would provide a more statistically-significant sample to 1) enable a more robust re-analysis of this work and 2) calibrate new AGN-based rungs of the Cosmic Distance Ladder (e.g., \citealt{Watson2011}). As near-infrared and infrared TF relations are more frequently employed for large-scale $V_{\mathrm{PEC}}$ modeling of our local supercluster, we also argue for complete removal of active galaxies from future TF-based $V_{\mathrm{PEC}}$ modeling samples. 

\begin{acknowledgements}
M.C. Bentz gratefully acknowledges support from the NSF through grant AST-2407802. H.M.C. acknowledges support from the Institut Universitaire de France. This work is based on observations obtained at the MDM Observatory, operated by Dartmouth College, Columbia University, Ohio State University, Ohio University, and the
University of Michigan. This work is based on observations obtained with the Apache Point Observatory 3.5 m telescope, which is owned and operated by the Astrophysical Research Consortium, and with Apache Point Observatory’s 0.5 m Astrophysical
Research Consortium Small Aperture Telescope. This research has been supported by RECONS (www.recons.org) members Michele Silverstein, Todd Henry, and Wei-Chun Jao, who provided data as part of the long-term astrometry/photometry program at the CTIO/SMARTS 0.9 m, which is operated as part of the SMARTS Consortium. This research has made use of the NASA/IPAC Extragalactic Database (NED), which is operated by the Jet
Propulsion Laboratory, California Institute of Technology,
under contract with the National Aeronautics and Space
Administration. This publication makes use of data products from the Wide-field Infrared Survey Explorer, which is a joint project of the University of California, Los Angeles, and the Jet Propulsion Laboratory/California Institute of Technology, funded by the National Aeronautics and Space Administration.
\end{acknowledgements}

\begin{contribution}
J.\ H.\ Robinson oversaw the project and wrote the manuscript. M.\ Revalski provided extensive scientific consultation for the project on the whole, with specific attention given to resolved AGN outflow dynamics and the majority of Seyfert 2 hosts. M.\ C.\ Bentz, D.\ M.\ Crenshaw, H.\ M.\ Courtois, R.\ L.\ Merritt, and J.\ Falcone provided scientific expertise. V.\ Lahue assisted with all calculations and analyses of the optical and near-IR TF relations. I.\ Chintala assisted in the preliminary analysis of the IR TF relations. All authors contributed to the presentation of the results
\end{contribution}

\bibliographystyle{aasjournal}
\bibliography{references}

@ARTICLE{misty2009a,
   author = {{Bentz}, M.~C. and {Peterson}, B.~M. and {Netzer}, H. and {Pogge}, R.~W. and 
	{Vestergaard}, M.},
    title = "{The Radius-Luminosity Relationship for Active Galactic Nuclei: The Effect of Host-Galaxy Starlight on Luminosity Measurements. II. The Full Sample of Reverberation-Mapped AGNs}",
  journal = {\apj},
archivePrefix = "arXiv",
   eprint = {0812.2283},
     year = 2009,
    month = may,
   volume = 697,
    pages = {160-181},
      doi = {10.1088/0004-637X/697/1/160},
   adsurl = {http://adsabs.harvard.edu/abs/2009ApJ...697..160B}
}

@ARTICLE{misty2009b,
   author = {{Bentz}, M.~C. and {Walsh}, J.~L. and {Barth}, A.~J. and {Baliber}, N. and 
	{Bennert}, V.~N. and {Canalizo}, G. and {Filippenko}, A.~V. and 
	{Ganeshalingam}, M. and {Gates}, E.~L. and {Greene}, J.~E. and 
	{Hidas}, M.~G. and {Hiner}, K.~D. and {Lee}, N. and {Li}, W. and 
	{Malkan}, M.~A. and {Minezaki}, T. and {Sakata}, Y. and {Serduke}, F.~J.~D. and 
	{Silverman}, J.~M. and {Steele}, T.~N. and {Stern}, D. and {Street}, R.~A. and 
	{Thornton}, C.~E. and {Treu}, T. and {Wang}, X. and {Woo}, J.-H. and 
	{Yoshii}, Y.},
    title = "{The Lick AGN Monitoring Project: Broad-line Region Radii and Black Hole Masses from Reverberation Mapping of H{$\beta$}}",
  journal = {\apj},
archivePrefix = "arXiv",
   eprint = {0908.0003},
     year = 2009,
    month = nov,
   volume = 705,
    pages = {199-217},
      doi = {10.1088/0004-637X/705/1/199},
   adsurl = {http://adsabs.harvard.edu/abs/2009ApJ...705..199B}
}

@ARTICLE{mcgaugh2012,
   author = {{McGaugh}, S.~S.},
    title = "{The Baryonic Tully-Fisher Relation of Gas-rich Galaxies as a Test of {$\Lambda$}CDM and MOND}",
  journal = {\aj},
archivePrefix = "arXiv",
   eprint = {1107.2934},
     year = 2012,
    month = feb,
   volume = 143,
      eid = {40},
    pages = {40},
      doi = {10.1088/0004-6256/143/2/40},
   adsurl = {http://adsabs.harvard.edu/abs/2012AJ....143...40M}
}

@ARTICLE{edd2005,
   author = {{Springob}, C.~M. and {Haynes}, M.~P. and {Giovanelli}, R. and 
	{Kent}, B.~R.},
    title = "{A Digital Archive of H I 21 Centimeter Line Spectra of Optically Targeted Galaxies}",
  journal = {\apjs},
   eprint = {astro-ph/0505025},
     year = 2005,
    month = sep,
   volume = 160,
    pages = {149-162},
      doi = {10.1086/431550},
   adsurl = {http://adsabs.harvard.edu/abs/2005ApJS..160..149S}
}

@BOOK{hr1989,
   author = {{Huchtmeier}, W.~K. and {Richter}, O.-G.},
    title = "{A General Catalog of HI Observations of Galaxies. The Reference Catalog.}",
booktitle = {A General Catalog of HI Observations of Galaxies.~The Reference Catalog.~Huchtmeier, W.K., Richter, O.-G., pp.~350.~ISBN 0-387-96997-7.~Springer-Verlag Berlin Heidelberg 1989},
     year = 1989,
    pages = {350},
   adsurl = {http://adsabs.harvard.edu/abs/1989gcho.book.....H}
}

@ARTICLE{dr1978,
   author = {{Dickel}, J.~R. and {Rood}, H.~J.},
    title = "{Integrated masses of galaxies}",
  journal = {\apj},
     year = 1978,
    month = jul,
   volume = 223,
    pages = {391-409},
      doi = {10.1086/156274},
   adsurl = {http://adsabs.harvard.edu/abs/1978ApJ...223..391D}
}

@phdthesis{rachael,
  title={The Spectral Energy Distributions of Active Galactic Nuclei with Direct Black Hole Mass Measurements},
  author={Merritt, R.},
  year={2022},
  school={Georgia State Univ, \url{https://scholarworks.gsu.edu/items/c394424b-0413-4715-a703-81921e6c66bb}}
}

@ARTICLE{misty2013,
   author = {{Bentz}, M.~C. and {Denney}, K.~D. and {Grier}, C.~J. and {Barth}, A.~J. and 
	{Peterson}, B.~M. and {Vestergaard}, M. and {Bennert}, V.~N. and 
	{Canalizo}, G. and {De Rosa}, G. and {Filippenko}, A.~V. and 
	{Gates}, E.~L. and {Greene}, J.~E. and {Li}, W. and {Malkan}, M.~A. and 
	{Pogge}, R.~W. and {Stern}, D. and {Treu}, T. and {Woo}, J.-H.
	},
    title = "{The Low-luminosity End of the Radius-Luminosity Relationship for Active Galactic Nuclei}",
  journal = {\apj},
archivePrefix = "arXiv",
   eprint = {1303.1742},
     year = 2013,
    month = apr,
   volume = 767,
      eid = {149},
    pages = {149},
      doi = {10.1088/0004-637X/767/2/149},
   adsurl = {http://adsabs.harvard.edu/abs/2013ApJ...767..149B}
}

@ARTICLE{cosmicflows1,
   author = {{Tully}, R.~B. and {Shaya}, E.~J. and {Karachentsev}, I.~D. and 
	{Courtois}, H.~M. and {Kocevski}, D.~D. and {Rizzi}, L. and 
	{Peel}, A.},
    title = "{Our Peculiar Motion Away from the Local Void}",
  journal = {\apj},
archivePrefix = "arXiv",
   eprint = {0705.4139},
     year = 2008,
    month = mar,
   volume = 676,
    pages = {184-205},
      doi = {10.1086/527428},
   adsurl = {http://adsabs.harvard.edu/abs/2008ApJ...676..184T}
}

@ARTICLE{tonry2001,
   author = {{Tonry}, J.~L. and {Dressler}, A. and {Blakeslee}, J.~P. and 
	{Ajhar}, E.~A. and {Fletcher}, A.~B. and {Luppino}, G.~A. and 
	{Metzger}, M.~R. and {Moore}, C.~B.},
    title = "{The SBF Survey of Galaxy Distances. IV. SBF Magnitudes, Colors, and Distances}",
  journal = {\apj},
   eprint = {astro-ph/0011223},
     year = 2001,
    month = jan,
   volume = 546,
    pages = {681-693},
      doi = {10.1086/318301},
   adsurl = {http://adsabs.harvard.edu/abs/2001ApJ...546..681T}
}

@ARTICLE{misty2018,
   author = {{Bentz}, M.~C. and {Manne-Nicholas}, E.},
    title = "{Black Hole{\ndash}Galaxy Scaling Relationships for Active Galactic Nuclei with Reverberation Masses}",
  journal = {\apj},
archivePrefix = "arXiv",
   eprint = {1808.01329},
     year = 2018,
    month = sep,
   volume = 864,
      eid = {146},
    pages = {146},
      doi = {10.3847/1538-4357/aad808},
   adsurl = {http://adsabs.harvard.edu/abs/2018ApJ...864..146B}
}

@ARTICLE{tf1977,
   author = {{Tully}, R.~B. and {Fisher}, J.~R.},
    title = "{A new method of determining distances to galaxies}",
  journal = {\aap},
     year = 1977,
    month = feb,
   volume = 54,
    pages = {661-673},
   adsurl = {http://adsabs.harvard.edu/abs/1977A%26A....54..661T}
}

@ARTICLE{bd2001,
   author = {{Bell}, E.~F. and {de Jong}, R.~S.},
    title = "{Stellar Mass-to-Light Ratios and the Tully-Fisher Relation}",
  journal = {\apj},
   eprint = {astro-ph/0011493},
     year = 2001,
    month = mar,
   volume = 550,
    pages = {212-229},
      doi = {10.1086/319728},
   adsurl = {http://adsabs.harvard.edu/abs/2001ApJ...550..212B}
}

@ARTICLE{mcgaugh2005,
   author = {{McGaugh}, S.~S.},
    title = "{The Baryonic Tully-Fisher Relation of Galaxies with Extended Rotation Curves and the Stellar Mass of Rotating Galaxies}",
  journal = {\apj},
   eprint = {astro-ph/0506750},
     year = 2005,
    month = oct,
   volume = 632,
    pages = {859-871},
      doi = {10.1086/432968},
   adsurl = {http://adsabs.harvard.edu/abs/2005ApJ...632..859M}
}

@ARTICLE{edd,
   author = {{Tully}, R.~B. and {Rizzi}, L. and {Shaya}, E.~J. and {Courtois}, H.~M. and 
	{Makarov}, D.~I. and {Jacobs}, B.~A.},
    title = "{The Extragalactic Distance Database}",
  journal = {\aj},
     year = 2009,
    month = aug,
   volume = 138,
    pages = {323-331},
      doi = {10.1088/0004-6256/138/2/323},
   adsurl = {http://adsabs.harvard.edu/abs/2009AJ....138..323T}
}

@ARTICLE{tp2000,
   author = {{Tully}, R.~B. and {Pierce}, M.~J.},
    title = "{Distances to Galaxies from the Correlation between Luminosities and Line Widths. III. Cluster Template and Global Measurement of H$_{0}$}",
  journal = {\apj},
   eprint = {astro-ph/9911052},
     year = 2000,
    month = apr,
   volume = 533,
    pages = {744-780},
      doi = {10.1086/308700},
   adsurl = {http://adsabs.harvard.edu/abs/2000ApJ...533..744T}
}

@ARTICLE{peng2002,
   author = {{Peng}, C.~Y. and {Ho}, L.~C. and {Impey}, C.~D. and {Rix}, H.-W.
	},
    title = "{Detailed Structural Decomposition of Galaxy Images}",
  journal = {\aj},
   eprint = {astro-ph/0204182},
     year = 2002,
    month = jul,
   volume = 124,
    pages = {266-293},
      doi = {10.1086/340952},
   adsurl = {http://adsabs.harvard.edu/abs/2002AJ....124..266P}
}

@ARTICLE{peng2010,
   author = {{Peng}, C.~Y. and {Ho}, L.~C. and {Impey}, C.~D. and {Rix}, H.-W.
	},
    title = "{Detailed Decomposition of Galaxy Images. II. Beyond Axisymmetric Models}",
  journal = {\aj},
archivePrefix = "arXiv",
   eprint = {0912.0731},
     year = 2010,
    month = jun,
   volume = 139,
    pages = {2097-2129},
      doi = {10.1088/0004-6256/139/6/2097},
   adsurl = {http://adsabs.harvard.edu/abs/2010AJ....139.2097P}
}

@ARTICLE{reyes2011,
   author = {{Reyes}, R. and {Mandelbaum}, R. and {Gunn}, J.~E. and {Pizagno}, J. and 
	{Lackner}, C.~N.},
    title = "{Calibrated Tully-Fisher relations for improved estimates of disc rotation velocities}",
  journal = {\mnras},
archivePrefix = "arXiv",
   eprint = {1106.1650},
     year = 2011,
    month = nov,
   volume = 417,
    pages = {2347-2386},
      doi = {10.1111/j.1365-2966.2011.19415.x},
   adsurl = {http://adsabs.harvard.edu/abs/2011MNRAS.417.2347R}
}

@ARTICLE{sabbi2018,
   author = {{Sabbi}, E. and {Calzetti}, D. and {Ubeda}, L. and {Adamo}, A. and 
	{Cignoni}, M. and {Thilker}, D. and {Aloisi}, A. and {Elmegreen}, B.~G. and 
	{Elmegreen}, D.~M. and {Gouliermis}, D.~A. and {Grebel}, E.~K. and 
	{Messa}, M. and {Smith}, L.~J. and {Tosi}, M. and {Dolphin}, A. and 
	{Andrews}, J.~E. and {Ashworth}, G. and {Bright}, S.~N. and 
	{Brown}, T.~M. and {Chandar}, R. and {Christian}, C. and {Clayton}, G.~C. and 
	{Cook}, D.~O. and {Dale}, D.~A. and {de Mink}, S.~E. and {Dobbs}, C. and 
	{Evans}, A.~S. and {Fumagalli}, M. and {Gallagher}, III, J.~S. and 
	{Grasha}, K. and {Herrero}, A. and {Hunter}, D.~A. and {Johnson}, K.~E. and 
	{Kahre}, L. and {Kennicutt}, R.~C. and {Kim}, H. and {Krumholz}, M.~R. and 
	{Lee}, J.~C. and {Lennon}, D. and {Martin}, C. and {Nair}, P. and 
	{Nota}, A. and {{\"O}stlin}, G. and {Pellerin}, A. and {Prieto}, J. and 
	{Regan}, M.~W. and {Ryon}, J.~E. and {Sacchi}, E. and {Schaerer}, D. and 
	{Schiminovich}, D. and {Shabani}, F. and {Van Dyk}, S.~D. and 
	{Walterbos}, R. and {Whitmore}, B.~C. and {Wofford}, A.},
    title = "{The Resolved Stellar Populations in the LEGUS Galaxies1}",
  journal = {\apjs},
archivePrefix = "arXiv",
   eprint = {1801.05467},
     year = 2018,
    month = mar,
   volume = 235,
      eid = {23},
    pages = {23},
      doi = {10.3847/1538-4365/aaa8e5},
   adsurl = {http://adsabs.harvard.edu/abs/2018ApJS..235...23S}
}

@ARTICLE{dv1991,
   author = {{de Vaucouleurs}, G. and {de Vaucouleurs}, A. and {Corwin}, Jr., H.~G. and 
	{Buta}, R.~J. and {Paturel}, G. and {Fouque}, P.},
    title = "{Book-Review - Third Reference Catalogue of Bright Galaxies}",
  journal = {\skytel},
     year = 1991,
    month = dec,
   volume = 82,
    pages = {621},
   adsurl = {http://adsabs.harvard.edu/abs/1991S%26T....82Q.621D}
}

@ARTICLE{shostak1978,
       author = {{Shostak}, G.~S.},
        title = "{Integral properties of late-type galaxies derived from H I observations.}",
      journal = {\aap},
         year = 1978,
        month = Aug,
       volume = {68},
        pages = {321-341},
       adsurl = {https://ui.adsabs.harvard.edu/\#abs/1978A&A....68..321S}
}

@ARTICLE{koribalski2004,
       author = {{Koribalski}, B.~S. and {Staveley-Smith}, L. and {Kilborn}, V.~A. and
        {Ryder}, S.~D. and {Kraan-Korteweg}, R.~C. and {Ryan-Weber},
        E.~V. and {Ekers}, R.~D. and {Jerjen}, H. and {Henning}, P.~A.
        and {Putman}, M.~E. and {Zwaan}, M.~A. and {de Blok}, W.~J.~G.
        and {Calabretta}, M.~R. and {Disney}, M.~J. and {Minchin}, R.~F.
        and {Bhathal}, R. and {Boyce}, P.~J. and {Drinkwater}, M.~J. and
        {Freeman}, K.~C. and {Gibson}, B.~K. and {Green}, A.~J. and
        {Haynes}, R.~F. and {Juraszek}, S. and {Kesteven}, M.~J. and
        {Knezek}, P.~M. and {Mader}, S. and {Marquarding}, M. and
        {Meyer}, M. and {Mould}, J.~R. and {Oosterloo}, T. and
        {O'Brien}, J. and {Price}, R.~M. and {Sadler}, E.~M. and
        {Schr{\"o}der}, A. and {Stewart}, I.~M. and {Stootman}, F. and
        {Waugh}, M. and {Warren}, B.~E. and {Webster}, R.~L. and
        {Wright}, A.~E.},
        title = "{The 1000 Brightest HIPASS Galaxies: H I Properties}",
      journal = {\aj},
         year = 2004,
        month = Jul,
       volume = {128},
        pages = {16-46},
          doi = {10.1086/421744},
archivePrefix = {arXiv},
       eprint = {astro-ph/0404436},
 primaryClass = {astro-ph},
       adsurl = {https://ui.adsabs.harvard.edu/\#abs/2004AJ....128...16K}
}

@INPROCEEDINGS{dv1976,
   author = {{de Vaucouleurs}, G. and {de Vaucouleurs}, A. and {Corwin}, J.~R.
	},
    title = "{Second reference catalogue of bright galaxies}",
booktitle = {Second reference catalogue of bright galaxies, Vol. 1976, p. Austin: University of Texas Press.},
     year = 1976,
   volume = 1976,
   adsurl = {http://adsabs.harvard.edu/abs/1976RC2...C......0D}
}

@ARTICLE{roberts1969,
   author = {{Roberts}, M.~S.},
    title = "{Integral Properties of Spiral and Irregular Galaxies}",
  journal = {\aj},
     year = 1969,
    month = sep,
   volume = 74,
    pages = {859-876},
      doi = {10.1086/110874},
   adsurl = {http://adsabs.harvard.edu/abs/1969AJ.....74..859R}
}

@ARTICLE{epstein1964,
   author = {{Epstein}, E.~E.},
    title = "{Atomic hydrogen in galaxies.}",
  journal = {\aj},
     year = 1964,
    month = sep,
   volume = 69,
    pages = {490-520},
      doi = {10.1086/109305},
   adsurl = {http://adsabs.harvard.edu/abs/1964AJ.....69..490E}
}

@ARTICLE{seyfert1943,
   author = {{Seyfert}, C.~K.},
    title = "{Nuclear Emission in Spiral Nebulae.}",
  journal = {\apj},
     year = 1943,
    month = jan,
   volume = 97,
    pages = {28},
      doi = {10.1086/144488},
   adsurl = {http://adsabs.harvard.edu/abs/1943ApJ....97...28S}
}

@ARTICLE{mcmahon2013,
   author = {{McMahon}, R.~G. and {Banerji}, M. and {Gonzalez}, E. and {Koposov}, S.~E. and 
	{Bejar}, V.~J. and {Lodieu}, N. and {Rebolo}, R. and {VHS Collaboration}
	},
    title = "{First Scientific Results from the VISTA Hemisphere Survey (VHS)}",
  journal = {The Messenger},
     year = 2013,
    month = dec,
   volume = 154,
    pages = {35-37},
   adsurl = {http://adsabs.harvard.edu/abs/2013Msngr.154...35M}
}

@ARTICLE{mcgaugh2000,
   author = {{McGaugh}, S.~S. and {Schombert}, J.~M. and {Bothun}, G.~D. and 
	{de Blok}, W.~J.~G.},
    title = "{The Baryonic Tully-Fisher Relation}",
  journal = {\apjl},
   eprint = {astro-ph/0003001},
     year = 2000,
    month = apr,
   volume = 533,
    pages = {L99-L102},
      doi = {10.1086/312628},
   adsurl = {http://adsabs.harvard.edu/abs/2000ApJ...533L..99M}
}

@ARTICLE{thim2004,
   author = {{Thim}, F. and {Hoessel}, J.~G. and {Saha}, A. and {Claver}, J. and 
	{Dolphin}, A. and {Tammann}, G.~A.},
    title = "{Cepheids and Long-Period Variables in NGC 4395}",
  journal = {\aj},
   eprint = {astro-ph/0401558},
     year = 2004,
    month = apr,
   volume = 127,
    pages = {2322-2343},
      doi = {10.1086/382244},
   adsurl = {http://adsabs.harvard.edu/abs/2004AJ....127.2322T}
}

@article{mundell1995,
    author = {Mundell, C. and Pedlar, A. and G. and Axon, D. J. and Meaburn, J. and Unger, S. W.},
    title = "{Neutral hydrogen studies of the Seyfert galaxy NGC 3227}",
    journal = {Monthly Notices of the Royal Astronomical Society},
    volume = {277},
    number = {2},
    pages = {641-654},
    year = {1995},
    month = {11},
    issn = {0035-8711},
    eprint = {http://oup.prod.sis.lan/mnras/article-pdf/277/2/641/18200196/mnras277-0641.pdf},
}

@ARTICLE{cosmicflows3,
       author = {{Tully}, R. Brent and {Courtois}, H{\'e}l{\`e}ne M. and
         {Sorce}, Jenny G.},
        title = "{Cosmicflows-3}",
      journal = {\aj},
         year = "2016",
        month = "Aug",
       volume = {152},
       number = {2},
          eid = {50},
        pages = {50},
          doi = {10.3847/0004-6256/152/2/50},
archivePrefix = {arXiv},
       eprint = {1605.01765},
 primaryClass = {astro-ph.CO},
       adsurl = {https://ui.adsabs.harvard.edu/abs/2016AJ....152...50T}
}

@ARTICLE{shaya2017,
       author = {{Shaya}, Edward J. and {Tully}, R. Brent and {Hoffman}, Yehuda and
         {Pomar{\`e}de}, Daniel},
        title = "{Action Dynamics of the Local Supercluster}",
      journal = {\apj},
         year = "2017",
        month = "Dec",
       volume = {850},
       number = {2},
          eid = {207},
        pages = {207},
          doi = {10.3847/1538-4357/aa9525},
archivePrefix = {arXiv},
       eprint = {1710.08935},
 primaryClass = {astro-ph.CO},
       adsurl = {https://ui.adsabs.harvard.edu/abs/2017ApJ...850..207S}
}

@ARTICLE{misty2019,
       author = {{Bentz}, Misty C. and {Ferrarese}, Laura and {Onken}, Christopher A. and
         {Peterson}, Bradley M. and {Valluri}, Monica},
        title = "{A Cepheid-based Distance to the Seyfert Galaxy NGC 6814}",
      journal = {\apj},
         year = "2019",
        month = "Nov",
       volume = {885},
       number = {2},
          eid = {161},
        pages = {161},
          doi = {10.3847/1538-4357/ab48fb},
       adsurl = {https://ui.adsabs.harvard.edu/abs/2019ApJ...885..161B}
}

@ARTICLE{mundell1999,
       author = {{Mundell}, C.~G. and {Pedlar}, A. and {Shone}, D.~L. and {Robinson}, A.},
        title = "{Gas dynamics in the barred Seyfert galaxy NGC 4151 - II. High-resolution HI study}",
      journal = {\mnras},
         year = "1999",
        month = "Apr",
       volume = {304},
       number = {3},
        pages = {481-494},
          doi = {10.1046/j.1365-8711.1999.02331.x},
archivePrefix = {arXiv},
       eprint = {astro-ph/9812183},
 primaryClass = {astro-ph},
       adsurl = {https://ui.adsabs.harvard.edu/abs/1999MNRAS.304..481M}
}

@ARTICLE{schlafly2011,
       author = {{Schlafly}, Edward F. and {Finkbeiner}, Douglas P.},
        title = "{Measuring Reddening with Sloan Digital Sky Survey Stellar Spectra and Recalibrating SFD}",
      journal = {\apj},
         year = "2011",
        month = "Aug",
       volume = {737},
       number = {2},
          eid = {103},
        pages = {103},
          doi = {10.1088/0004-637X/737/2/103},
archivePrefix = {arXiv},
       eprint = {1012.4804},
 primaryClass = {astro-ph.GA},
       adsurl = {https://ui.adsabs.harvard.edu/abs/2011ApJ...737..103S}
}

@ARTICLE{schlegel1998,
       author = {{Schlegel}, David J. and {Finkbeiner}, Douglas P. and {Davis}, Marc},
        title = "{Maps of Dust Infrared Emission for Use in Estimation of Reddening and Cosmic Microwave Background Radiation Foregrounds}",
      journal = {\apj},
         year = "1998",
        month = "Jun",
       volume = {500},
       number = {2},
        pages = {525-553},
          doi = {10.1086/305772},
archivePrefix = {arXiv},
       eprint = {astro-ph/9710327},
 primaryClass = {astro-ph},
       adsurl = {https://ui.adsabs.harvard.edu/abs/1998ApJ...500..525S}
}

@ARTICLE{tf1985,
       author = {{Tully}, R.~B. and {Fouque}, P.},
        title = "{The extragalactic distance scale. I - Corrections to fundamental observables.}",
      journal = {\apjs},
         year = "1985",
        month = "May",
       volume = {58},
        pages = {67-80},
          doi = {10.1086/191029},
       adsurl = {https://ui.adsabs.harvard.edu/abs/1985ApJS...58...67T}
}

@ARTICLE{courtois2009,
       author = {{Courtois}, H{\'e}l{\`e}ne M. and {Tully}, R. Brent and
         {Fisher}, J. Richard and {Bonhomme}, Nicolas and {Zavodny}, Maximilian and
         {Barnes}, Austin},
        title = "{The Extragalactic Distance Database: All Digital H I Profile Catalog}",
      journal = {\aj},
         year = "2009",
        month = "Dec",
       volume = {138},
       number = {6},
        pages = {1938-1956},
          doi = {10.1088/0004-6256/138/6/1938},
archivePrefix = {arXiv},
       eprint = {0902.3670},
 primaryClass = {astro-ph.CO},
       adsurl = {https://ui.adsabs.harvard.edu/abs/2009AJ....138.1938C}
}

@ARTICLE{cosmicflows2,
       author = {{Tully}, R. Brent and {Courtois}, H{\'e}l{\`e}ne M. and
         {Dolphin}, Andrew E. and {Fisher}, J. Richard and
         {H{\'e}raudeau}, Philippe and {Jacobs}, Bradley A. and
         {Karachentsev}, Igor D. and {Makarov}, Dmitry and {Makarova}, Lidia and
         {Mitronova}, Sofia and {Rizzi}, Luca and {Shaya}, Edward J. and
         {Sorce}, Jenny G. and {Wu}, Po-Feng},
        title = "{Cosmicflows-2: The Data}",
      journal = {\aj},
         year = "2013",
        month = "Oct",
       volume = {146},
       number = {4},
          eid = {86},
        pages = {86},
          doi = {10.1088/0004-6256/146/4/86},
archivePrefix = {arXiv},
       eprint = {1307.7213},
 primaryClass = {astro-ph.CO},
       adsurl = {https://ui.adsabs.harvard.edu/abs/2013AJ....146...86T}
}

@ARTICLE{tully1998,
       author = {{Tully}, R. Brent and {Pierce}, Michael J. and {Huang}, Jia-Sheng and
         {Saunders}, Will and {Verheijen}, Marc A.~W. and {Witchalls}, Peter L.},
        title = "{Global Extinction in Spiral Galaxies}",
      journal = {\aj},
         year = 1998,
        month = jun,
       volume = {115},
       number = {6},
        pages = {2264-2272},
          doi = {10.1086/300379},
archivePrefix = {arXiv},
       eprint = {astro-ph/9802247},
 primaryClass = {astro-ph},
       adsurl = {https://ui.adsabs.harvard.edu/abs/1998AJ....115.2264T}
}

@ARTICLE{Kourkchi2020,
       author = {{Kourkchi}, Ehsan and {Tully}, R. Brent and {Anand}, Gagandeep S. and
         {Courtois}, H{\'e}l{\`e}ne M. and {Dupuy}, Alexandra and
         {Neill}, James D. and {Rizzi}, Luca and {Seibert}, Mark},
        title = "{Cosmicflows-4: The Calibration of Optical and Infrared Tully-Fisher Relations}",
      journal = {\apj},
         year = 2020,
        month = jun,
       volume = {896},
       number = {1},
          eid = {3},
        pages = {3},
          doi = {10.3847/1538-4357/ab901c},
archivePrefix = {arXiv},
       eprint = {2004.14499},
 primaryClass = {astro-ph.GA},
       adsurl = {https://ui.adsabs.harvard.edu/abs/2020ApJ...896....3K}
}

@ARTICLE{graziani2019,
       author = {{Graziani}, R. and {Courtois}, H.~M. and {Lavaux}, G. and {Hoffman}, Y. and
         {Tully}, R.~B. and {Copin}, Y. and {Pomar{\`e}de}, D.},
        title = "{The peculiar velocity field up to z {\ensuremath{\sim}} 0.05 by forward-modelling Cosmicflows-3 data}",
      journal = {\mnras},
         year = 2019,
        month = oct,
       volume = {488},
       number = {4},
        pages = {5438-5451},
          doi = {10.1093/mnras/stz078},
archivePrefix = {arXiv},
       eprint = {1901.01818},
 primaryClass = {astro-ph.CO},
       adsurl = {https://ui.adsabs.harvard.edu/abs/2019MNRAS.488.5438G}
}

@ARTICLE{koshida2017,
       author = {{Koshida}, Shintaro and {Yoshii}, Yuzuru and {Kobayashi}, Yukiyasu and
         {Minezaki}, Takeo and {Enya}, Keigo and {Suganuma}, Masahiro and
         {Tomita}, Hiroyuki and {Aoki}, Tsutomu and {Peterson}, Bruce A.},
        title = "{Calibration of AGN Reverberation Distance Measurements}",
      journal = {\apjl},
         year = 2017,
        month = jun,
       volume = {842},
       number = {2},
          eid = {L13},
        pages = {L13},
          doi = {10.3847/2041-8213/aa7553},
archivePrefix = {arXiv},
       eprint = {1705.09757},
 primaryClass = {astro-ph.GA},
       adsurl = {https://ui.adsabs.harvard.edu/abs/2017ApJ...842L..13K}
}

@ARTICLE{Ganeshalingam2013,
       author = {{Ganeshalingam}, Mohan and {Li}, Weidong and {Filippenko}, Alexei V.},
        title = "{Constraints on dark energy with the LOSS SN Ia sample}",
      journal = {\mnras},
         year = 2013,
        month = aug,
       volume = {433},
       number = {3},
        pages = {2240-2258},
          doi = {10.1093/mnras/stt893},
archivePrefix = {arXiv},
       eprint = {1307.0824},
 primaryClass = {astro-ph.CO},
       adsurl = {https://ui.adsabs.harvard.edu/abs/2013MNRAS.433.2240G}
}

@ARTICLE{lp1912,
       author = {{Leavitt}, Henrietta S. and {Pickering}, Edward C.},
        title = "{Periods of 25 Variable Stars in the Small Magellanic Cloud.}",
      journal = {Harvard College Observatory Circular},
         year = 1912,
        month = mar,
       volume = {173},
        pages = {1-3},
       adsurl = {https://ui.adsabs.harvard.edu/abs/1912HarCi.173....1L}
}

@ARTICLE{iben1983,
       author = {{Iben}, I., Jr. and {Renzini}, A.},
        title = "{Asymptotic giant branch evolution and beyond.}",
      journal = {\araa},
         year = 1983,
        month = jan,
       volume = {21},
        pages = {271-342},
          doi = {10.1146/annurev.aa.21.090183.001415},
       adsurl = {https://ui.adsabs.harvard.edu/abs/1983ARA&A..21..271I}
}

@ARTICLE{salaris1997,
       author = {{Salaris}, Maurizio and {Cassisi}, Santi},
        title = "{The `tip' of the red giant branch as a distance indicator: results from evolutionary models}",
      journal = {\mnras},
         year = 1997,
        month = aug,
       volume = {289},
       number = {2},
        pages = {406-414},
          doi = {10.1093/mnras/289.2.406},
archivePrefix = {arXiv},
       eprint = {astro-ph/9703186},
 primaryClass = {astro-ph},
       adsurl = {https://ui.adsabs.harvard.edu/abs/1997MNRAS.289..406S}
}

@ARTICLE{riess2016,
       author = {{Riess}, Adam G. and {Macri}, Lucas M. and {Hoffmann}, Samantha L. and
         {Scolnic}, Dan and {Casertano}, Stefano and {Filippenko}, Alexei V. and
         {Tucker}, Brad E. and {Reid}, Mark J. and {Jones}, David O. and
         {Silverman}, Jeffrey M. and {Chornock}, Ryan and {Challis}, Peter and
         {Yuan}, Wenlong and {Brown}, Peter J. and {Foley}, Ryan J.},
        title = "{A 2.4\% Determination of the Local Value of the Hubble Constant}",
      journal = {\apj},
         year = 2016,
        month = jul,
       volume = {826},
       number = {1},
          eid = {56},
        pages = {56},
          doi = {10.3847/0004-637X/826/1/56},
archivePrefix = {arXiv},
       eprint = {1604.01424},
 primaryClass = {astro-ph.CO},
       adsurl = {https://ui.adsabs.harvard.edu/abs/2016ApJ...826...56R}
}

@article{Blakeslee2010,
	doi = {10.1088/0004-637x/724/1/657},
	url = {https://doi.org/10.1088%2F0004-637x%2F724%2F1%2F657},
	year = 2010,
	month = {nov},
	publisher = {{IOP} Publishing},
	volume = {724},
	number = {1},
	pages = {657--668},
	author = {John P. Blakeslee and Michele Cantiello and Simona Mei and Patrick C{\^{o}}t{\'{e}} and Regina Barber DeGraaff and Laura Ferrarese and Andr{\'{e}}s Jord{\'{a}}n and Eric W. Peng and John L. Tonry and Guy Worthey},
	title = {{SURFACE} {BRIGHTNESS} {FLUCTUATIONS} {IN} {THEHUBBLE} {SPACE} {TELESCOPEACS}/{WFC} F814W {BANDPASS} {AND} {AN} {UPDATE} {ON} {GALAXY} {DISTANCES}},
	journal = {The Astrophysical Journal}
}

@ARTICLE{walter2008,
       author = {{Walter}, Fabian and {Brinks}, Elias and {de Blok}, W.~J.~G. and
         {Bigiel}, Frank and {Kennicutt}, Robert C., Jr. and
         {Thornley}, Michele D. and {Leroy}, Adam},
        title = "{THINGS: The H I Nearby Galaxy Survey}",
      journal = {\aj},
         year = 2008,
        month = dec,
       volume = {136},
       number = {6},
        pages = {2563-2647},
          doi = {10.1088/0004-6256/136/6/2563},
archivePrefix = {arXiv},
       eprint = {0810.2125},
 primaryClass = {astro-ph},
       adsurl = {https://ui.adsabs.harvard.edu/abs/2008AJ....136.2563W}
}

@ARTICLE{tc2012,
       author = {{Tully}, R. Brent and {Courtois}, H{\'e}l{\`e}ne M.},
        title = "{Cosmicflows-2: I-band Luminosity-H I Linewidth Calibration}",
      journal = {\apj},
         year = 2012,
        month = apr,
       volume = {749},
       number = {1},
          eid = {78},
        pages = {78},
          doi = {10.1088/0004-637X/749/1/78},
archivePrefix = {arXiv},
       eprint = {1202.3191},
 primaryClass = {astro-ph.CO},
       adsurl = {https://ui.adsabs.harvard.edu/abs/2012ApJ...749...78T}
}

@article{Jha2007,
	doi = {10.1086/512054},
	url = {https://doi.org/10.1086%2F512054},
	year = 2007,
	month = {apr},
	publisher = {{IOP} Publishing},
	volume = {659},
	number = {1},
	pages = {122--148},
	author = {Saurabh Jha and Adam G. Riess and Robert P. Kirshner},
	title = {Improved Distances to Type Ia Supernovae with Multicolor Light-Curve Shapes: {MLCS}2k2},
	journal = {The Astrophysical Journal}
}

@ARTICLE{Lelli2016,
       author = {{Lelli}, Federico and {McGaugh}, Stacy S. and {Schombert}, James M.},
        title = "{The Small Scatter of the Baryonic Tully-Fisher Relation}",
      journal = {\apjl},
         year = 2016,
        month = jan,
       volume = {816},
       number = {1},
          eid = {L14},
        pages = {L14},
          doi = {10.3847/2041-8205/816/1/L14},
archivePrefix = {arXiv},
       eprint = {1512.04543},
 primaryClass = {astro-ph.GA},
       adsurl = {https://ui.adsabs.harvard.edu/abs/2016ApJ...816L..14L}
}

@ARTICLE{iorio2017,
       author = {{Iorio}, G. and {Fraternali}, F. and {Nipoti}, C. and {Di Teodoro}, E. and
         {Read}, J.~I. and {Battaglia}, G.},
        title = "{LITTLE THINGS in 3D: robust determination of the circular velocity of dwarf irregular galaxies}",
      journal = {\mnras},
         year = 2017,
        month = apr,
       volume = {466},
       number = {4},
        pages = {4159-4192},
          doi = {10.1093/mnras/stw3285},
archivePrefix = {arXiv},
       eprint = {1611.03865},
 primaryClass = {astro-ph.GA},
       adsurl = {https://ui.adsabs.harvard.edu/abs/2017MNRAS.466.4159I}
}

@article{lelli2019,
    author = {Lelli, Federico and McGaugh, Stacy S and Schombert, James M and Desmond, Harry and Katz, Harley},
    title = "{The baryonic Tully–Fisher relation for different velocity definitions and implications for galaxy angular momentum}",
    journal = {Monthly Notices of the Royal Astronomical Society},
    volume = {484},
    number = {3},
    pages = {3267-3278},
    year = {2019},
    month = {01},
    issn = {0035-8711},
    doi = {10.1093/mnras/stz205},
    url = {https://doi.org/10.1093/mnras/stz205},
    eprint = {https://academic.oup.com/mnras/article-pdf/484/3/3267/27711809/stz205.pdf},
}

@article{Sakai2000,
	doi = {10.1086/308305},
	url = {https://doi.org/10.1086%2F308305},
	year = 2000,
	month = {feb},
	publisher = {{IOP} Publishing},
	volume = {529},
	number = {2},
	pages = {698--722},
	author = {Shoko Sakai and Jeremy R. Mould and Shaun M. G. Hughes and John P. Huchra and Lucas M. Macri and Robert C. Kennicutt, Jr. and Brad K. Gibson and Laura Ferrarese and Wendy L. Freedman and Mingsheng Han and Holland C. Ford and John A. Graham and Garth D. Illingworth and Daniel D. Kelson and Barry F. Madore and Kim Sebo and N. A. Silbermann and Peter B. Stetson},
	title = {{TheHubble} Space {TelescopeKey} Project on the Extragalactic Distance Scale. {XXIV}. The Calibration of Tully-Fisher Relations and the Value of the Hubble Constant},
	journal = {The Astrophysical Journal}
}

@ARTICLE{planck2018,
       author = {{Planck Collaboration} and {Aghanim}, N. and {Akrami}, Y. and
         {Ashdown}, M. and {Aumont}, J. and {Baccigalupi}, C. and
         {Ballardini}, M. and {Banday}, A.~J. and {Barreiro}, R.~B. and
         {Bartolo}, N. and {Basak}, S. and {Battye}, R. and {Benabed}, K. and
         {Bernard}, J. -P. and {Bersanelli}, M. and {Bielewicz}, P. and
         {Bock}, J.~J. and {Bond}, J.~R. and {Borrill}, J. and {Bouchet}, F.~R. and
         {Boulanger}, F. and {Bucher}, M. and {Burigana}, C. and
         {Butler}, R.~C. and {Calabrese}, E. and {Cardoso}, J. -F. and
         {Carron}, J. and {Challinor}, A. and {Chiang}, H.~C. and {Chluba}, J. and
         {Colombo}, L.~P.~L. and {Combet}, C. and {Contreras}, D. and
         {Crill}, B.~P. and {Cuttaia}, F. and {de Bernardis}, P. and
         {de Zotti}, G. and {Delabrouille}, J. and {Delouis}, J. -M. and
         {Di Valentino}, E. and {Diego}, J.~M. and {Dor{\'e}}, O. and
         {Douspis}, M. and {Ducout}, A. and {Dupac}, X. and {Dusini}, S. and
         {Efstathiou}, G. and {Elsner}, F. and {En{\ss}lin}, T.~A. and
         {Eriksen}, H.~K. and {Fantaye}, Y. and {Farhang}, M. and
         {Fergusson}, J. and {Fernandez-Cobos}, R. and {Finelli}, F. and
         {Forastieri}, F. and {Frailis}, M. and {Fraisse}, A.~A. and
         {Franceschi}, E. and {Frolov}, A. and {Galeotta}, S. and {Galli}, S. and
         {Ganga}, K. and {G{\'e}nova-Santos}, R.~T. and {Gerbino}, M. and
         {Ghosh}, T. and {Gonz{\'a}lez-Nuevo}, J. and {G{\'o}rski}, K.~M. and
         {Gratton}, S. and {Gruppuso}, A. and {Gudmundsson}, J.~E. and
         {Hamann}, J. and {Handley}, W. and {Hansen}, F.~K. and {Herranz}, D. and
         {Hildebrandt}, S.~R. and {Hivon}, E. and {Huang}, Z. and
         {Jaffe}, A.~H. and {Jones}, W.~C. and {Karakci}, A. and
         {Keih{\"a}nen}, E. and {Keskitalo}, R. and {Kiiveri}, K. and {Kim}, J. and
         {Kisner}, T.~S. and {Knox}, L. and {Krachmalnicoff}, N. and {Kunz}, M. and
         {Kurki-Suonio}, H. and {Lagache}, G. and {Lamarre}, J. -M. and
         {Lasenby}, A. and {Lattanzi}, M. and {Lawrence}, C.~R. and
         {Le Jeune}, M. and {Lemos}, P. and {Lesgourgues}, J. and {Levrier}, F. and
         {Lewis}, A. and {Liguori}, M. and {Lilje}, P.~B. and {Lilley}, M. and
         {Lindholm}, V. and {L{\'o}pez-Caniego}, M. and {Lubin}, P.~M. and
         {Ma}, Y. -Z. and {Mac{\'\i}as-P{\'e}rez}, J.~F. and {Maggio}, G. and
         {Maino}, D. and {Mandolesi}, N. and {Mangilli}, A. and
         {Marcos-Caballero}, A. and {Maris}, M. and {Martin}, P.~G. and
         {Martinelli}, M. and {Mart{\'\i}nez-Gonz{\'a}lez}, E. and
         {Matarrese}, S. and {Mauri}, N. and {McEwen}, J.~D. and
         {Meinhold}, P.~R. and {Melchiorri}, A. and {Mennella}, A. and
         {Migliaccio}, M. and {Millea}, M. and {Mitra}, S. and
         {Miville-Desch{\^e}nes}, M. -A. and {Molinari}, D. and {Montier}, L. and
         {Morgante}, G. and {Moss}, A. and {Natoli}, P. and
         {N{\o}rgaard-Nielsen}, H.~U. and {Pagano}, L. and {Paoletti}, D. and
         {Partridge}, B. and {Patanchon}, G. and {Peiris}, H.~V. and
         {Perrotta}, F. and {Pettorino}, V. and {Piacentini}, F. and
         {Polastri}, L. and {Polenta}, G. and {Puget}, J. -L. and
         {Rachen}, J.~P. and {Reinecke}, M. and {Remazeilles}, M. and
         {Renzi}, A. and {Rocha}, G. and {Rosset}, C. and {Roudier}, G. and
         {Rubi{\~n}o-Mart{\'\i}n}, J.~A. and {Ruiz-Granados}, B. and
         {Salvati}, L. and {Sandri}, M. and {Savelainen}, M. and {Scott}, D. and
         {Shellard}, E.~P.~S. and {Sirignano}, C. and {Sirri}, G. and
         {Spencer}, L.~D. and {Sunyaev}, R. and {Suur-Uski}, A. -S. and
         {Tauber}, J.~A. and {Tavagnacco}, D. and {Tenti}, M. and
         {Toffolatti}, L. and {Tomasi}, M. and {Trombetti}, T. and
         {Valenziano}, L. and {Valiviita}, J. and {Van Tent}, B. and
         {Vibert}, L. and {Vielva}, P. and {Villa}, F. and {Vittorio}, N. and {Wand
        elt}, B.~D. and {Wehus}, I.~K. and {White}, M. and {White}, S.~D.~M. and
         {Zacchei}, A. and {Zonca}, A.},
        title = "{Planck 2018 results. VI. Cosmological parameters}",
      journal = {\aap},
         year = 2020,
        month = sep,
       volume = {641},
          eid = {A6},
        pages = {A6},
          doi = {10.1051/0004-6361/201833910},
archivePrefix = {arXiv},
       eprint = {1807.06209},
 primaryClass = {astro-ph.CO},
       adsurl = {https://ui.adsabs.harvard.edu/abs/2020A&A...641A...6P}
}

@article{Chilingarian2010,
    author = {Chilingarian, Igor V. and Melchior, Anne-Laure and Zolotukhin, Ivan Yu.},
    title = "{Analytical approximations of K-corrections in optical and near-infrared bands}",
    journal = {Monthly Notices of the Royal Astronomical Society},
    volume = {405},
    number = {3},
    pages = {1409-1420},
    year = {2010},
    month = {06},
    issn = {0035-8711},
    doi = {10.1111/j.1365-2966.2010.16506.x},
    url = {https://doi.org/10.1111/j.1365-2966.2010.16506.x},
    eprint = {https://academic.oup.com/mnras/article-pdf/405/3/1409/2868030/mnras0405-1409.pdf},
}

@ARTICLE{pt1988,
       author = {{Pierce}, Michael J. and {Tully}, R. Brent},
        title = "{Distances to the Virgo and Ursa Major Clusters and a Determination of H 0}",
      journal = {\apj},
         year = 1988,
        month = jul,
       volume = {330},
        pages = {579},
          doi = {10.1086/166495},
       adsurl = {https://ui.adsabs.harvard.edu/abs/1988ApJ...330..579P}
}

@ARTICLE{yuan2020,
       author = {{Yuan}, W. and {Fausnaugh}, M.~M. and {Hoffmann}, S.~L. and
         {Macri}, L.~M. and {Peterson}, B.~M. and {Riess}, A.~G. and
         {Bentz}, M.~C. and {Brown}, J.~S. and {Bont{\`a}}, E. Dalla and
         {Davies}, R.~I. and {Rosa}, G. De and {Ferrarese}, L. and
         {Grier}, C.~J. and {Hicks}, E.~K.~S. and {Onken}, C.~A. and
         {Pogge}, R.~W. and {Storchi-Bergmann}, T. and {Vestergaard}, M.},
        title = "{The Cepheid Distance to the Seyfert 1 Galaxy NGC 4151}",
      journal = {\apj},
         year = 2020,
        month = oct,
       volume = {902},
       number = {1},
          eid = {26},
        pages = {26},
          doi = {10.3847/1538-4357/abb377},
archivePrefix = {arXiv},
       eprint = {2007.07888},
 primaryClass = {astro-ph.GA},
       adsurl = {https://ui.adsabs.harvard.edu/abs/2020ApJ...902...26Y}
}

@article{Barton2001,
	doi = {10.1086/318759},
	url = {https://doi.org/10.1086/318759},
	year = 2001,
	month = {feb},
	publisher = {American Astronomical Society},
	volume = {121},
	number = {2},
	pages = {625--648},
	author = {Elizabeth J. Barton and Margaret J. Geller and Benjamin C. Bromley and Liese van Zee and Scott J. Kenyon},
	title = {The Tully-Fisher Relation as a Measure of Luminosity Evolution: A Low-Redshift Baseline for Evolving Galaxies},
	journal = {The Astronomical Journal}
}

@ARTICLE{yuan4051,
       author = {{Yuan}, Wenlong and {Macri}, Lucas M. and {Peterson}, Bradley M. and {Riess}, Adam G. and {Fausnaugh}, Michael M. and {Hoffmann}, Samantha L. and {Anand}, Gagandeep S. and {Bentz}, Misty C. and {Dalla Bont{\`a}}, Elena and {Davies}, Richard I. and {de Rosa}, Gisella and {Ferrarese}, Laura and {Grier}, Catherine J. and {Hicks}, Erin K.~S. and {Onken}, Christopher A. and {Pogge}, Richard W. and {Storchi-Bergmann}, Thaisa and {Vestergaard}, Marianne},
        title = "{The Cepheid Distance to the Narrow-Line Seyfert 1 Galaxy NGC 4051}",
      journal = {arXiv e-prints},
         year = 2020,
        month = dec,
          eid = {arXiv:2012.05931},
        pages = {arXiv:2012.05931},
archivePrefix = {arXiv},
       eprint = {2012.05931},
 primaryClass = {astro-ph.GA},
       adsurl = {https://ui.adsabs.harvard.edu/abs/2020arXiv201205931Y}
}

@ARTICLE{misty2021,
       author = {{Bentz}, Misty C. and {Williams}, Peter R. and {Street}, Rachel and {Onken}, Christopher A. and {Valluri}, Monica and {Treu}, Tommaso},
        title = "{A Detailed View of the Broad-line Region in NGC 3783 from Velocity-resolved Reverberation Mapping}",
      journal = {\apj},
         year = 2021,
        month = oct,
       volume = {920},
       number = {2},
          eid = {112},
        pages = {112},
          doi = {10.3847/1538-4357/ac19af},
archivePrefix = {arXiv},
       eprint = {2108.00482},
 primaryClass = {astro-ph.GA},
       adsurl = {https://ui.adsabs.harvard.edu/abs/2021ApJ...920..112B}
}

@ARTICLE{me2,
       author = {{Robinson}, Justin H. and {Bentz}, Misty C. and {Courtois}, H{\'e}l{\`e}ne M. and {Johnson}, Megan C. and {Crenshaw}, D.~M. and {Meena}, Beena and {Polack}, Garrett E. and {Silverstein}, Michele L. and {Chen}, Dading},
        title = "{Tully-Fisher Distances and Dynamical Mass Constraints for 24 Host Galaxies of Reverberation-mapped AGNs}",
      journal = {\apj},
         year = 2021,
        month = may,
       volume = {912},
       number = {2},
          eid = {160},
        pages = {160},
          doi = {10.3847/1538-4357/abedaa},
archivePrefix = {arXiv},
       eprint = {2103.07000},
 primaryClass = {astro-ph.GA},
       adsurl = {https://ui.adsabs.harvard.edu/abs/2021ApJ...912..160R}
}

@ARTICLE{antonucci1993,
       author = {{Antonucci}, Robert},
        title = "{Unified models for active galactic nuclei and quasars.}",
      journal = {\araa},
         year = 1993,
        month = jan,
       volume = {31},
        pages = {473-521},
          doi = {10.1146/annurev.aa.31.090193.002353},
       adsurl = {https://ui.adsabs.harvard.edu/abs/1993ARA&A..31..473A}
}

@ARTICLE{up1995,
       author = {{Urry}, C. Megan and {Padovani}, Paolo},
        title = "{Unified Schemes for Radio-Loud Active Galactic Nuclei}",
      journal = {\pasp},
         year = 1995,
        month = sep,
       volume = {107},
        pages = {803},
          doi = {10.1086/133630},
archivePrefix = {arXiv},
       eprint = {astro-ph/9506063},
 primaryClass = {astro-ph},
       adsurl = {https://ui.adsabs.harvard.edu/abs/1995PASP..107..803U}
}

@ARTICLE{netzer2015,
       author = {{Netzer}, Hagai},
        title = "{Revisiting the Unified Model of Active Galactic Nuclei}",
      journal = {\araa},
         year = 2015,
        month = aug,
       volume = {53},
        pages = {365-408},
          doi = {10.1146/annurev-astro-082214-122302},
archivePrefix = {arXiv},
       eprint = {1505.00811},
 primaryClass = {astro-ph.GA},
       adsurl = {https://ui.adsabs.harvard.edu/abs/2015ARA&A..53..365N}
}

@ARTICLE{padovani2017,
       author = {{Padovani}, P. and {Alexander}, D.~M. and {Assef}, R.~J. and {De Marco}, B. and {Giommi}, P. and {Hickox}, R.~C. and {Richards}, G.~T. and {Smol{\v{c}}i{\'c}}, V. and {Hatziminaoglou}, E. and {Mainieri}, V. and {Salvato}, M.},
        title = "{Active galactic nuclei: what's in a name?}",
      journal = {\aapr},
         year = 2017,
        month = aug,
       volume = {25},
       number = {1},
          eid = {2},
        pages = {2},
          doi = {10.1007/s00159-017-0102-9},
archivePrefix = {arXiv},
       eprint = {1707.07134},
 primaryClass = {astro-ph.GA},
       adsurl = {https://ui.adsabs.harvard.edu/abs/2017A&ARv..25....2P}
}

@ARTICLE{kw1974,
       author = {{Khachikian}, E.~Y. and {Weedman}, D.~W.},
        title = "{An atlas of Seyfert galaxies.}",
      journal = {\apj},
         year = 1974,
        month = sep,
       volume = {192},
        pages = {581-589},
          doi = {10.1086/153093},
       adsurl = {https://ui.adsabs.harvard.edu/abs/1974ApJ...192..581K}
}

@ARTICLE{roberts1962,
       author = {{Roberts}, M.~S.},
        title = "{The neutral hydrogen content of late-type spiral galaxies.}",
      journal = {\aj},
         year = 1962,
        month = jan,
       volume = {67},
        pages = {437-446},
          doi = {10.1086/108752},
       adsurl = {https://ui.adsabs.harvard.edu/abs/1962AJ.....67..437R}
}

@ARTICLE{Baumgartner2013,
       author = {{Baumgartner}, W.~H. and {Tueller}, J. and {Markwardt}, C.~B. and {Skinner}, G.~K. and {Barthelmy}, S. and {Mushotzky}, R.~F. and {Evans}, P.~A. and {Gehrels}, N.},
        title = "{The 70 Month Swift-BAT All-sky Hard X-Ray Survey}",
      journal = {\apjs},
         year = 2013,
        month = aug,
       volume = {207},
       number = {2},
          eid = {19},
        pages = {19},
          doi = {10.1088/0067-0049/207/2/19},
archivePrefix = {arXiv},
       eprint = {1212.3336},
 primaryClass = {astro-ph.HE},
       adsurl = {https://ui.adsabs.harvard.edu/abs/2013ApJS..207...19B}
}

@ARTICLE{4151_continuum1,
       author = {{Oknyanskij}, V.~L. and {Metlova}, N.~V. and {Artamonov}, B.~P. and {Lyuty}, A.~V. and {Lyuty}, V.~M.},
        title = "{Variability of NGC4151 During 2008-2013}",
      journal = {Odessa Astronomical Publications},
         year = 2013,
        month = jan,
       volume = {26},
        pages = {212},
       adsurl = {https://ui.adsabs.harvard.edu/abs/2013OAP....26..212O}
}

@ARTICLE{koss2022,
       author = {{Koss}, Michael J. and {Ricci}, Claudio and {Trakhtenbrot}, Benny and {Oh}, Kyuseok and {den Brok}, Jakob S. and {Mej{\'\i}a-Restrepo}, Julian E. and {Stern}, Daniel and {Privon}, George C. and {Treister}, Ezequiel and {Powell}, Meredith C. and {Mushotzky}, Richard and {Bauer}, Franz E. and {Ananna}, Tonima T. and {Balokovi{\'c}}, Mislav and {B{\"a}r}, Rudolf E. and {Becker}, George and {Bessiere}, Patricia and {Burtscher}, Leonard and {Caglar}, Turgay and {Congiu}, Enrico and {Evans}, Phil and {Harrison}, Fiona and {Heida}, Marianne and {Ichikawa}, Kohei and {Kamraj}, Nikita and {Lamperti}, Isabella and {Pacucci}, Fabio and {Ricci}, Federica and {Riffel}, Rog{\'e}rio and {Rojas}, Alejandra F. and {Schawinski}, Kevin and {Temple}, Matthew J. and {Urry}, C. Megan and {Veilleux}, Sylvain and {Williams}, Jonathan},
        title = "{BASS. XXII. The BASS DR2 AGN Catalog and Data}",
      journal = {\apjs},
         year = 2022,
        month = jul,
       volume = {261},
       number = {1},
          eid = {2},
        pages = {2},
          doi = {10.3847/1538-4365/ac6c05},
archivePrefix = {arXiv},
       eprint = {2207.12432},
 primaryClass = {astro-ph.GA},
       adsurl = {https://ui.adsabs.harvard.edu/abs/2022ApJS..261....2K}
}

@ARTICLE{panessa2002,
       author = {{Panessa}, F. and {Bassani}, L.},
        title = "{Unabsorbed Seyfert 2 galaxies}",
      journal = {\aap},
         year = 2002,
        month = nov,
       volume = {394},
        pages = {435-442},
          doi = {10.1051/0004-6361:20021161},
archivePrefix = {arXiv},
       eprint = {astro-ph/0208496},
 primaryClass = {astro-ph},
       adsurl = {https://ui.adsabs.harvard.edu/abs/2002A&A...394..435P}
}

@article{Courtois2012,
doi = {10.1088/0004-637X/749/2/174},
url = {https://dx.doi.org/10.1088/0004-637X/749/2/174},
year = {2012},
month = {apr},
publisher = {The American Astronomical Society},
volume = {749},
number = {2},
pages = {174},
author = {Hélène M. Courtois and R. Brent Tully},
title = {COSMICFLOWS-2: TYPE Ia SUPERNOVA CALIBRATION AND H0},
journal = {The Astrophysical Journal}
}

@ARTICLE{riess2022,
       author = {{Riess}, Adam G. and {Yuan}, Wenlong and {Macri}, Lucas M. and {Scolnic}, Dan and {Brout}, Dillon and {Casertano}, Stefano and {Jones}, David O. and {Murakami}, Yukei and {Anand}, Gagandeep S. and {Breuval}, Louise and {Brink}, Thomas G. and {Filippenko}, Alexei V. and {Hoffmann}, Samantha and {Jha}, Saurabh W. and {D'arcy Kenworthy}, W. and {Mackenty}, John and {Stahl}, Benjamin E. and {Zheng}, WeiKang},
        title = "{A Comprehensive Measurement of the Local Value of the Hubble Constant with 1 km s$^{-1}$ Mpc$^{-1}$ Uncertainty from the Hubble Space Telescope and the SH0ES Team}",
      journal = {\apjl},
         year = 2022,
        month = jul,
       volume = {934},
       number = {1},
          eid = {L7},
        pages = {L7},
          doi = {10.3847/2041-8213/ac5c5b},
archivePrefix = {arXiv},
       eprint = {2112.04510},
 primaryClass = {astro-ph.CO},
       adsurl = {https://ui.adsabs.harvard.edu/abs/2022ApJ...934L...7R}
}

@ARTICLE{falcone2024,
       author = {{Falcone}, Julia and {Crenshaw}, D. Michael and {Fischer}, Travis C. and {Meena}, Beena and {Revalski}, Mitchell and {Shea}, Maura Kathleen and {Riffel}, Rogemar A. and {Chapman}, Zo and {Ferree}, Nicolas and {Tutterow}, Jacob and {Davis}, Madeline},
        title = "{An Analysis of AGN-Driven Outflows in the Seyfert 1 Galaxy NGC 3227}",
      journal = {arXiv e-prints},
         year = 2024,
        month = may,
          eid = {arXiv:2405.20162},
        pages = {arXiv:2405.20162},
          doi = {10.48550/arXiv.2405.20162},
archivePrefix = {arXiv},
       eprint = {2405.20162},
 primaryClass = {astro-ph.GA},
       adsurl = {https://ui.adsabs.harvard.edu/abs/2024arXiv240520162F}
}

@ARTICLE{gravity2021,
       author = {{GRAVITY Collaboration} and {Amorim}, A. and {Baub{\"o}ck}, M. and {Bentz}, M.~C. and {Brandner}, W. and {Bolzer}, M. and {Cl{\'e}net}, Y. and {Davies}, R. and {de Zeeuw}, P.~T. and {Dexter}, J. and {Drescher}, A. and {Eckart}, A. and {Eisenhauer}, F. and {F{\"o}rster Schreiber}, N.~M. and {Garcia}, P.~J.~V. and {Genzel}, R. and {Gillessen}, S. and {Gratadour}, D. and {H{\"o}nig}, S. and {Kaltenbrunner}, D. and {Kishimoto}, M. and {Lacour}, S. and {Lutz}, D. and {Millour}, F. and {Netzer}, H. and {Onken}, C.~A. and {Ott}, T. and {Paumard}, T. and {Perraut}, K. and {Perrin}, G. and {Petrucci}, P.~O. and {Pfuhl}, O. and {Prieto}, M.~A. and {Rouan}, D. and {Shangguan}, J. and {Shimizu}, T. and {Stadler}, J. and {Sternberg}, A. and {Straub}, O. and {Straubmeier}, C. and {Street}, R. and {Sturm}, E. and {Tacconi}, L.~J. and {Tristram}, K.~R.~W. and {Vermot}, P. and {von Fellenberg}, S. and {Widmann}, F. and {Woillez}, J.},
        title = "{A geometric distance to the supermassive black Hole of NGC 3783}",
      journal = {\aap},
         year = 2021,
        month = oct,
       volume = {654},
          eid = {A85},
        pages = {A85},
          doi = {10.1051/0004-6361/202141426},
archivePrefix = {arXiv},
       eprint = {2107.14262},
 primaryClass = {astro-ph.GA},
       adsurl = {https://ui.adsabs.harvard.edu/abs/2021A&A...654A..85G}
}

@ARTICLE{ft1981,
       author = {{Fisher}, J.~R. and {Tully}, R.~B.},
        title = "{Neutral hydrogen observations of a large sample of galaxies.}",
      journal = {\apjs},
         year = 1981,
        month = dec,
       volume = {47},
        pages = {139-200},
          doi = {10.1086/190755},
       adsurl = {https://ui.adsabs.harvard.edu/abs/1981ApJS...47..139F}
}

@ARTICLE{cantiello2018,
       author = {{Cantiello}, Michele and {Blakeslee}, John P. and {Ferrarese}, Laura and {C{\^o}t{\'e}}, Patrick and {Roediger}, Joel C. and {Raimondo}, Gabriella and {Peng}, Eric W. and {Gwyn}, Stephen and {Durrell}, Patrick R. and {Cuillandre}, Jean-Charles},
        title = "{The Next Generation Virgo Cluster Survey (NGVS). XVIII. Measurement and Calibration of Surface Brightness Fluctuation Distances for Bright Galaxies in Virgo (and Beyond)}",
      journal = {\apj},
         year = 2018,
        month = apr,
       volume = {856},
       number = {2},
          eid = {126},
        pages = {126},
          doi = {10.3847/1538-4357/aab043},
archivePrefix = {arXiv},
       eprint = {1802.05526},
 primaryClass = {astro-ph.GA},
       adsurl = {https://ui.adsabs.harvard.edu/abs/2018ApJ...856..126C}
}

@ARTICLE{Radburn-Smith2011,
       author = {{Radburn-Smith}, D.~J. and {de Jong}, R.~S. and {Seth}, A.~C. and {Bailin}, J. and {Bell}, E.~F. and {Brown}, T.~M. and {Bullock}, J.~S. and {Courteau}, S. and {Dalcanton}, J.~J. and {Ferguson}, H.~C. and {Goudfrooij}, P. and {Holfeltz}, S. and {Holwerda}, B.~W. and {Purcell}, C. and {Sick}, J. and {Streich}, D. and {Vlajic}, M. and {Zucker}, D.~B.},
        title = "{The GHOSTS Survey. I. Hubble Space Telescope Advanced Camera for Surveys Data}",
      journal = {\apjs},
         year = 2011,
        month = aug,
       volume = {195},
       number = {2},
          eid = {18},
        pages = {18},
          doi = {10.1088/0067-0049/195/2/18},
       adsurl = {https://ui.adsabs.harvard.edu/abs/2011ApJS..195...18R}
}

@ARTICLE{ho1997,
       author = {{Ho}, Luis C. and {Filippenko}, Alexei V. and {Sargent}, Wallace L.~W.},
        title = "{A Search for ``Dwarf'' Seyfert Nuclei. III. Spectroscopic Parameters and Properties of the Host Galaxies}",
      journal = {\apjs},
         year = 1997,
        month = oct,
       volume = {112},
       number = {2},
        pages = {315-390},
          doi = {10.1086/313041},
archivePrefix = {arXiv},
       eprint = {astro-ph/9704107},
 primaryClass = {astro-ph},
       adsurl = {https://ui.adsabs.harvard.edu/abs/1997ApJS..112..315H}
}

@ARTICLE{jaeger2017,
       author = {{de Jaeger}, T. and {Gonz{\'a}lez-Gait{\'a}n}, S. and {Hamuy}, M. and {Galbany}, L. and {Anderson}, J.~P. and {Phillips}, M.~M. and {Stritzinger}, M.~D. and {Carlberg}, R.~G. and {Sullivan}, M. and {Guti{\'e}rrez}, C.~P. and {Hook}, I.~M. and {Howell}, D. Andrew and {Hsiao}, E.~Y. and {Kuncarayakti}, H. and {Ruhlmann-Kleider}, V. and {Folatelli}, G. and {Pritchet}, C. and {Basa}, S.},
        title = "{A Type II Supernova Hubble Diagram from the CSP-I, SDSS-II, and SNLS Surveys}",
      journal = {\apj},
         year = 2017,
        month = feb,
       volume = {835},
       number = {2},
          eid = {166},
        pages = {166},
          doi = {10.3847/1538-4357/835/2/166},
archivePrefix = {arXiv},
       eprint = {1612.05636},
 primaryClass = {astro-ph.CO},
       adsurl = {https://ui.adsabs.harvard.edu/abs/2017ApJ...835..166D}
}

@ARTICLE{anand2021,
       author = {{Anand}, Gagandeep S. and {Rizzi}, Luca and {Tully}, R. Brent and {Shaya}, Edward J. and {Karachentsev}, Igor D. and {Makarov}, Dmitry I. and {Makarova}, Lidia and {Wu}, Po-Feng and {Dolphin}, Andrew E. and {Kourkchi}, Ehsan},
        title = "{The Extragalactic Distance Database: The Color-Magnitude Diagrams/Tip of the Red Giant Branch Distance Catalog}",
      journal = {\aj},
         year = 2021,
        month = aug,
       volume = {162},
       number = {2},
          eid = {80},
        pages = {80},
          doi = {10.3847/1538-3881/ac0440},
archivePrefix = {arXiv},
       eprint = {2104.02649},
 primaryClass = {astro-ph.GA},
       adsurl = {https://ui.adsabs.harvard.edu/abs/2021AJ....162...80A}
}

@ARTICLE{McQuinn2017,
       author = {{McQuinn}, Kristen. B.~W. and {Skillman}, Evan D. and {Dolphin}, Andrew E. and {Berg}, Danielle and {Kennicutt}, Robert},
        title = "{Accurate Distances to Important Spiral Galaxies: M63, M74, NGC 1291, NGC 4559, NGC 4625, and NGC 5398}",
      journal = {\aj},
         year = 2017,
        month = aug,
       volume = {154},
       number = {2},
          eid = {51},
        pages = {51},
          doi = {10.3847/1538-3881/aa7aad},
archivePrefix = {arXiv},
       eprint = {1706.06586},
 primaryClass = {astro-ph.GA},
       adsurl = {https://ui.adsabs.harvard.edu/abs/2017AJ....154...51M}
}

@ARTICLE{ford1985,
       author = {{Ford}, H.~C. and {Crane}, P.~C. and {Jacoby}, G.~H. and {Lawrie}, D.~G. and {van der Hulst}, J.~M.},
        title = "{Bubbles and jets in the center of M 51.}",
      journal = {\apj},
         year = 1985,
        month = jun,
       volume = {293},
        pages = {132-147},
          doi = {10.1086/163220},
       adsurl = {https://ui.adsabs.harvard.edu/abs/1985ApJ...293..132F}
}

@ARTICLE{Burns2018,
       author = {{Burns}, Christopher R. and {Parent}, Emilie and {Phillips}, M.~M. and {Stritzinger}, Maximilian and {Krisciunas}, Kevin and {Suntzeff}, Nicholas B. and {Hsiao}, Eric Y. and {Contreras}, Carlos and {Anais}, Jorge and {Boldt}, Luis and {Busta}, Luis and {Campillay}, Abdo and {Castell{\'o}n}, Sergio and {Folatelli}, Gast{\'o}n and {Freedman}, Wendy L. and {Gonz{\'a}lez}, Consuelo and {Hamuy}, Mario and {Heoflich}, Peter and {Krzeminski}, Wojtek and {Madore}, Barry F. and {Morrell}, Nidia and {Persson}, S.~E. and {Roth}, Miguel and {Salgado}, Francisco and {Ser{\'o}n}, Jacqueline and {Torres}, Sim{\'o}n},
        title = "{The Carnegie Supernova Project: Absolute Calibration and the Hubble Constant}",
      journal = {\apj},
         year = 2018,
        month = dec,
       volume = {869},
       number = {1},
          eid = {56},
        pages = {56},
          doi = {10.3847/1538-4357/aae51c},
archivePrefix = {arXiv},
       eprint = {1809.06381},
 primaryClass = {astro-ph.CO},
       adsurl = {https://ui.adsabs.harvard.edu/abs/2018ApJ...869...56B}
}

@ARTICLE{perez2000,
       author = {{P{\'e}rez}, E. and {M{\'a}rquez}, I. and {Marrero}, I. and {Durret}, F. and {Gonz{\'a}lez Delgado}, R.~M. and {Masegosa}, J. and {Maza}, J. and {Moles}, M.},
        title = "{Circumnuclear structure and kinematics in the active galaxy NGC 6951}",
      journal = {\aap},
         year = 2000,
        month = jan,
       volume = {353},
        pages = {893-909},
          doi = {10.48550/arXiv.astro-ph/9909495},
archivePrefix = {arXiv},
       eprint = {astro-ph/9909495},
 primaryClass = {astro-ph},
       adsurl = {https://ui.adsabs.harvard.edu/abs/2000A&A...353..893P}
}

@ARTICLE{Weyant2014,
       author = {{Weyant}, Anja and {Wood-Vasey}, W. Michael and {Allen}, Lori and {Garnavich}, Peter M. and {Jha}, Saurabh W. and {Joyce}, Richard and {Matheson}, Thomas},
        title = "{SweetSpot: Near-infrared Observations of 13 Type Ia Supernovae from a New NOAO Survey Probing the Nearby Smooth Hubble Flow}",
      journal = {\apj},
         year = 2014,
        month = apr,
       volume = {784},
       number = {2},
          eid = {105},
        pages = {105},
          doi = {10.1088/0004-637X/784/2/105},
archivePrefix = {arXiv},
       eprint = {1310.2259},
 primaryClass = {astro-ph.CO},
       adsurl = {https://ui.adsabs.harvard.edu/abs/2014ApJ...784..105W}
}

@ARTICLE{Osterbrock1993,
       author = {{Osterbrock}, Donald E. and {Martel}, Andre},
        title = "{Spectroscopic Study of the CfA Sample of Seyfert Galaxies}",
      journal = {\apj},
         year = 1993,
        month = sep,
       volume = {414},
        pages = {552},
          doi = {10.1086/173102},
       adsurl = {https://ui.adsabs.harvard.edu/abs/1993ApJ...414..552O}
}

@ARTICLE{Ruiz-Lapuente1996,
       author = {{Ruiz-Lapuente}, Pilar},
        title = "{The Hubble Constant from 56Co-powered Nebular Candles}",
      journal = {\apjl},
         year = 1996,
        month = jul,
       volume = {465},
        pages = {L83},
          doi = {10.1086/310155},
archivePrefix = {arXiv},
       eprint = {astro-ph/9604044},
 primaryClass = {astro-ph},
       adsurl = {https://ui.adsabs.harvard.edu/abs/1996ApJ...465L..83R}
}

@ARTICLE{cosmicflows4,
       author = {{Tully}, R. Brent and {Kourkchi}, Ehsan and {Courtois}, H{\'e}l{\`e}ne M. and {Anand}, Gagandeep S. and {Blakeslee}, John P. and {Brout}, Dillon and {Jaeger}, Thomas de and {Dupuy}, Alexandra and {Guinet}, Daniel and {Howlett}, Cullan and {Jensen}, Joseph B. and {Pomar{\`e}de}, Daniel and {Rizzi}, Luca and {Rubin}, David and {Said}, Khaled and {Scolnic}, Daniel and {Stahl}, Benjamin E.},
        title = "{Cosmicflows-4}",
      journal = {\apj},
         year = 2023,
        month = feb,
       volume = {944},
       number = {1},
          eid = {94},
        pages = {94},
          doi = {10.3847/1538-4357/ac94d8},
archivePrefix = {arXiv},
       eprint = {2209.11238},
 primaryClass = {astro-ph.CO},
       adsurl = {https://ui.adsabs.harvard.edu/abs/2023ApJ...944...94T}
}

@ARTICLE{McQuinn2016,
       author = {{McQuinn}, Kristen. B.~W. and {Skillman}, Evan D. and {Dolphin}, Andrew E. and {Berg}, Danielle and {Kennicutt}, Robert},
        title = "{The Distance to M104}",
      journal = {\aj},
         year = 2016,
        month = nov,
       volume = {152},
       number = {5},
          eid = {144},
        pages = {144},
          doi = {10.3847/0004-6256/152/5/144},
       adsurl = {https://ui.adsabs.harvard.edu/abs/2016AJ....152..144M}
}

@ARTICLE{wise,
       author = {{Wright}, Edward L. and {Eisenhardt}, Peter R.~M. and {Mainzer}, Amy K. and {Ressler}, Michael E. and {Cutri}, Roc M. and {Jarrett}, Thomas and {Kirkpatrick}, J. Davy and {Padgett}, Deborah and {McMillan}, Robert S. and {Skrutskie}, Michael and {Stanford}, S.~A. and {Cohen}, Martin and {Walker}, Russell G. and {Mather}, John C. and {Leisawitz}, David and {Gautier}, Thomas N., III and {McLean}, Ian and {Benford}, Dominic and {Lonsdale}, Carol J. and {Blain}, Andrew and {Mendez}, Bryan and {Irace}, William R. and {Duval}, Valerie and {Liu}, Fengchuan and {Royer}, Don and {Heinrichsen}, Ingolf and {Howard}, Joan and {Shannon}, Mark and {Kendall}, Martha and {Walsh}, Amy L. and {Larsen}, Mark and {Cardon}, Joel G. and {Schick}, Scott and {Schwalm}, Mark and {Abid}, Mohamed and {Fabinsky}, Beth and {Naes}, Larry and {Tsai}, Chao-Wei},
        title = "{The Wide-field Infrared Survey Explorer (WISE): Mission Description and Initial On-orbit Performance}",
      journal = {\aj},
         year = 2010,
        month = dec,
       volume = {140},
       number = {6},
        pages = {1868-1881},
          doi = {10.1088/0004-6256/140/6/1868},
archivePrefix = {arXiv},
       eprint = {1008.0031},
 primaryClass = {astro-ph.IM},
       adsurl = {https://ui.adsabs.harvard.edu/abs/2010AJ....140.1868W}
}

@ARTICLE{cecil2000,
       author = {{Cecil}, G. and {Greenhill}, L.~J. and {DePree}, C.~G. and {Nagar}, N. and {Wilson}, A.~S. and {Dopita}, M.~A. and {P{\'e}rez-Fournon}, I. and {Argon}, A.~L. and {Moran}, J.~M.},
        title = "{The Active Jet in NGC 4258 and Its Associated Shocks}",
      journal = {\apj},
         year = 2000,
        month = jun,
       volume = {536},
       number = {2},
        pages = {675-696},
          doi = {10.1086/308959},
       adsurl = {https://ui.adsabs.harvard.edu/abs/2000ApJ...536..675C}
}

@article{
page2001,
author = {M. J. Page  and J. A. Stevens  and J. P. D. Mittaz  and F. J. Carrera },
title = {Submillimeter Evidence for the Coeval Growth of Massive Black Holes and Galaxy Bulges},
journal = {Science},
volume = {294},
number = {5551},
pages = {2516-2518},
year = {2001},
doi = {10.1126/science.1065880},
URL = {https://www.science.org/doi/abs/10.1126/science.1065880},
eprint = {https://www.science.org/doi/pdf/10.1126/science.1065880}}

@ARTICLE{alexander2005,
       author = {{Alexander}, D.~M. and {Smail}, I. and {Bauer}, F.~E. and {Chapman}, S.~C. and {Blain}, A.~W. and {Brandt}, W.~N. and {Ivison}, R.~J.},
        title = "{Rapid growth of black holes in massive star-forming galaxies}",
      journal = {\nat},
         year = 2005,
        month = mar,
       volume = {434},
       number = {7034},
        pages = {738-740},
          doi = {10.1038/nature03473},
archivePrefix = {arXiv},
       eprint = {astro-ph/0503453},
 primaryClass = {astro-ph},
       adsurl = {https://ui.adsabs.harvard.edu/abs/2005Natur.434..738A}
}

@ARTICLE{Netzer2007,
       author = {{Netzer}, Hagai and {Lutz}, Dieter and {Schweitzer}, Mario and {Contursi}, Alessandra and {Sturm}, Eckhard and {Tacconi}, Linda J. and {Veilleux}, Sylvain and {Kim}, D. -C. and {Rupke}, David and {Baker}, Andrew J. and {Dasyra}, Kalliopi and {Mazzarella}, Joseph and {Lord}, Steven},
        title = "{Spitzer Quasar and ULIRG Evolution Study (QUEST). II. The Spectral Energy Distributions of Palomar-Green Quasars}",
      journal = {\apj},
         year = 2007,
        month = sep,
       volume = {666},
       number = {2},
        pages = {806-816},
          doi = {10.1086/520716},
archivePrefix = {arXiv},
       eprint = {0706.0818},
 primaryClass = {astro-ph},
       adsurl = {https://ui.adsabs.harvard.edu/abs/2007ApJ...666..806N}
}

@ARTICLE{Silverman2009,
       author = {{Silverman}, J.~D. and {Lamareille}, F. and {Maier}, C. and {Lilly}, S.~J. and {Mainieri}, V. and {Brusa}, M. and {Cappelluti}, N. and {Hasinger}, G. and {Zamorani}, G. and {Scodeggio}, M. and {Bolzonella}, M. and {Contini}, T. and {Carollo}, C.~M. and {Jahnke}, K. and {Kneib}, J. -P. and {Le F{\`e}vre}, O. and {Merloni}, A. and {Bardelli}, S. and {Bongiorno}, A. and {Brunner}, H. and {Caputi}, K. and {Civano}, F. and {Comastri}, A. and {Coppa}, G. and {Cucciati}, O. and {de la Torre}, S. and {de Ravel}, L. and {Elvis}, M. and {Finoguenov}, A. and {Fiore}, F. and {Franzetti}, P. and {Garilli}, B. and {Gilli}, R. and {Iovino}, A. and {Kampczyk}, P. and {Knobel}, C. and {Kova{\v{c}}}, K. and {Le Borgne}, J. -F. and {Le Brun}, V. and {Mignoli}, M. and {Pello}, R. and {Peng}, Y. and {Perez Montero}, E. and {Ricciardelli}, E. and {Tanaka}, M. and {Tasca}, L. and {Tresse}, L. and {Vergani}, D. and {Vignali}, C. and {Zucca}, E. and {Bottini}, D. and {Cappi}, A. and {Cassata}, P. and {Fumana}, M. and {Griffiths}, R. and {Kartaltepe}, J. and {Koekemoer}, A. and {Marinoni}, C. and {McCracken}, H.~J. and {Memeo}, P. and {Meneux}, B. and {Oesch}, P. and {Porciani}, C. and {Salvato}, M.},
        title = "{Ongoing and Co-Evolving Star Formation in zCOSMOS Galaxies Hosting Active Galactic Nuclei}",
      journal = {\apj},
         year = 2009,
        month = may,
       volume = {696},
       number = {1},
        pages = {396-410},
          doi = {10.1088/0004-637X/696/1/396},
archivePrefix = {arXiv},
       eprint = {0810.3653},
 primaryClass = {astro-ph},
       adsurl = {https://ui.adsabs.harvard.edu/abs/2009ApJ...696..396S}
}

@ARTICLE{Santini2012,
       author = {{Santini}, P. and {Rosario}, D.~J. and {Shao}, L. and {Lutz}, D. and {Maiolino}, R. and {Alexander}, D.~M. and {Altieri}, B. and {Andreani}, P. and {Aussel}, H. and {Bauer}, F.~E. and {Berta}, S. and {Bongiovanni}, A. and {Brandt}, W.~N. and {Brusa}, M. and {Cepa}, J. and {Cimatti}, A. and {Daddi}, E. and {Elbaz}, D. and {Fontana}, A. and {F{\"o}rster Schreiber}, N.~M. and {Genzel}, R. and {Grazian}, A. and {Le Floc'h}, E. and {Magnelli}, B. and {Mainieri}, V. and {Nordon}, R. and {P{\'e}rez Garcia}, A.~M. and {Poglitsch}, A. and {Popesso}, P. and {Pozzi}, F. and {Riguccini}, L. and {Rodighiero}, G. and {Salvato}, M. and {Sanchez-Portal}, M. and {Sturm}, E. and {Tacconi}, L.~J. and {Valtchanov}, I. and {Wuyts}, S.},
        title = "{Enhanced star formation rates in AGN hosts with respect to inactive galaxies from PEP-Herschel observations}",
      journal = {\aap},
         year = 2012,
        month = apr,
       volume = {540},
          eid = {A109},
        pages = {A109},
          doi = {10.1051/0004-6361/201118266},
archivePrefix = {arXiv},
       eprint = {1201.4394},
 primaryClass = {astro-ph.CO},
       adsurl = {https://ui.adsabs.harvard.edu/abs/2012A&A...540A.109S}
}

@ARTICLE{Azadi2015,
       author = {{Azadi}, Mojegan and {Aird}, James and {Coil}, Alison L. and {Moustakas}, John and {Mendez}, Alexander J. and {Blanton}, Michael R. and {Cool}, Richard J. and {Eisenstein}, Daniel J. and {Wong}, Kenneth C. and {Zhu}, Guangtun},
        title = "{PRIMUS: The Relationship between Star Formation and AGN Accretion}",
      journal = {\apj},
         year = 2015,
        month = jun,
       volume = {806},
       number = {2},
          eid = {187},
        pages = {187},
          doi = {10.1088/0004-637X/806/2/187},
archivePrefix = {arXiv},
       eprint = {1407.1975},
 primaryClass = {astro-ph.GA},
       adsurl = {https://ui.adsabs.harvard.edu/abs/2015ApJ...806..187A}
}

@ARTICLE{Kannappan2002,
       author = {{Kannappan}, Sheila J. and {Fabricant}, Daniel G. and {Franx}, Marijn},
        title = "{Physical Sources of Scatter in the Tully-Fisher Relation}",
      journal = {\aj},
         year = 2002,
        month = may,
       volume = {123},
       number = {5},
        pages = {2358-2386},
          doi = {10.1086/339972},
archivePrefix = {arXiv},
       eprint = {astro-ph/0202111},
 primaryClass = {astro-ph},
       adsurl = {https://ui.adsabs.harvard.edu/abs/2002AJ....123.2358K}
}

@ARTICLE{Ristea2024,
       author = {{Ristea}, A. and {Cortese}, L. and {Fraser-McKelvie}, A. and {Catinella}, B. and {van de Sande}, J. and {Croom}, S.~M. and {Swinbank}, A.~M.},
        title = "{The Tully-Fisher relation from SDSS-MaNGA: physical causes of scatter and variation at different radii}",
      journal = {\mnras},
         year = 2024,
        month = jan,
       volume = {527},
       number = {3},
        pages = {7438-7458},
          doi = {10.1093/mnras/stad3638},
archivePrefix = {arXiv},
       eprint = {2311.13251},
 primaryClass = {astro-ph.GA},
       adsurl = {https://ui.adsabs.harvard.edu/abs/2024MNRAS.527.7438R}
}

@ARTICLE{Buchalter2001,
       author = {{Buchalter}, Ari and {Jimenez}, Raul and {Kamionkowski}, Marc},
        title = "{Galactosynthesis: halo histories, star formation and discs}",
      journal = {\mnras},
         year = 2001,
        month = mar,
       volume = {322},
       number = {1},
        pages = {43-66},
          doi = {10.1046/j.1365-8711.2001.04031.x},
archivePrefix = {arXiv},
       eprint = {astro-ph/0006032},
 primaryClass = {astro-ph},
       adsurl = {https://ui.adsabs.harvard.edu/abs/2001MNRAS.322...43B}
}

@ARTICLE{Pizagno2007,
       author = {{Pizagno}, James and {Prada}, Francisco and {Weinberg}, David H. and {Rix}, Hans-Walter and {Pogge}, Richard W. and {Grebel}, Eva K. and {Harbeck}, Daniel and {Blanton}, Michael and {Brinkmann}, J. and {Gunn}, James E.},
        title = "{The Tully-Fisher Relation and its Residuals for a Broadly Selected Sample of Galaxies}",
      journal = {\aj},
         year = 2007,
        month = sep,
       volume = {134},
       number = {3},
        pages = {945-972},
          doi = {10.1086/519522},
archivePrefix = {arXiv},
       eprint = {astro-ph/0608472},
 primaryClass = {astro-ph},
       adsurl = {https://ui.adsabs.harvard.edu/abs/2007AJ....134..945P}
}

@ARTICLE{Ouellette2017,
       author = {{Ouellette}, Nathalie N. -Q. and {Courteau}, St{\'e}phane and {Holtzman}, Jon A. and {Dutton}, Aaron A. and {Cappellari}, Michele and {Dalcanton}, Julianne J. and {McDonald}, Michael and {Roediger}, Joel C. and {Taylor}, James E. and {Tully}, R. Brent and {C{\^o}t{\'e}}, Patrick and {Ferrarese}, Laura and {Peng}, Eric W.},
        title = "{The Spectroscopy and H-band Imaging of Virgo Cluster Galaxies (SHIVir) Survey: Scaling Relations and the Stellar-to-total Mass Relation}",
      journal = {\apj},
         year = 2017,
        month = jul,
       volume = {843},
       number = {1},
          eid = {74},
        pages = {74},
          doi = {10.3847/1538-4357/aa74b1},
archivePrefix = {arXiv},
       eprint = {1705.10794},
 primaryClass = {astro-ph.GA},
       adsurl = {https://ui.adsabs.harvard.edu/abs/2017ApJ...843...74O}
}

@ARTICLE{Mountrichas2023_2,
       author = {{Mountrichas}, George},
        title = "{The co-evolution of supermassive black holes and galaxies in luminous AGN over a wide range of redshift}",
      journal = {\aap},
         year = 2023,
        month = apr,
       volume = {672},
          eid = {A98},
        pages = {A98},
          doi = {10.1051/0004-6361/202345924},
archivePrefix = {arXiv},
       eprint = {2302.10937},
 primaryClass = {astro-ph.GA},
       adsurl = {https://ui.adsabs.harvard.edu/abs/2023A&A...672A..98M}
}

@ARTICLE{Prieto2006,
       author = {{Prieto}, Jos{\'e} Luis and {Rest}, Armin and {Suntzeff}, Nicholas B.},
        title = "{A New Method to Calibrate the Magnitudes of Type Ia Supernovae at Maximum Light}",
      journal = {\apj},
         year = 2006,
        month = aug,
       volume = {647},
       number = {1},
        pages = {501-512},
          doi = {10.1086/504307},
archivePrefix = {arXiv},
       eprint = {astro-ph/0603407},
 primaryClass = {astro-ph},
       adsurl = {https://ui.adsabs.harvard.edu/abs/2006ApJ...647..501P}
}

@ARTICLE{Hicken2009,
       author = {{Hicken}, Malcolm and {Wood-Vasey}, W. Michael and {Blondin}, Stephane and {Challis}, Peter and {Jha}, Saurabh and {Kelly}, Patrick L. and {Rest}, Armin and {Kirshner}, Robert P.},
        title = "{Improved Dark Energy Constraints from \raisebox{-0.5ex}\textasciitilde100 New CfA Supernova Type Ia Light Curves}",
      journal = {\apj},
         year = 2009,
        month = aug,
       volume = {700},
       number = {2},
        pages = {1097-1140},
          doi = {10.1088/0004-637X/700/2/1097},
archivePrefix = {arXiv},
       eprint = {0901.4804},
 primaryClass = {astro-ph.CO},
       adsurl = {https://ui.adsabs.harvard.edu/abs/2009ApJ...700.1097H}
}

@ARTICLE{Folatelli2010,
       author = {{Folatelli}, Gast{\'o}n and {Phillips}, M.~M. and {Burns}, Christopher R. and {Contreras}, Carlos and {Hamuy}, Mario and {Freedman}, W.~L. and {Persson}, S.~E. and {Stritzinger}, Maximilian and {Suntzeff}, Nicholas B. and {Krisciunas}, Kevin and {Boldt}, Luis and {Gonz{\'a}lez}, Sergio and {Krzeminski}, Wojtek and {Morrell}, Nidia and {Roth}, Miguel and {Salgado}, Francisco and {Madore}, Barry F. and {Murphy}, David and {Wyatt}, Pamela and {Li}, Weidong and {Filippenko}, Alexei V. and {Miller}, Nicole},
        title = "{The Carnegie Supernova Project: Analysis of the First Sample of Low-Redshift Type-Ia Supernovae}",
      journal = {\aj},
         year = 2010,
        month = jan,
       volume = {139},
       number = {1},
        pages = {120-144},
          doi = {10.1088/0004-6256/139/1/120},
archivePrefix = {arXiv},
       eprint = {0910.3317},
 primaryClass = {astro-ph.CO},
       adsurl = {https://ui.adsabs.harvard.edu/abs/2010AJ....139..120F}
}

@ARTICLE{vv2006,
       author = {{V{\'e}ron-Cetty}, M. -P. and {V{\'e}ron}, P.},
        title = "{A catalogue of quasars and active nuclei: 12th edition}",
      journal = {\aap},
         year = 2006,
        month = aug,
       volume = {455},
       number = {2},
        pages = {773-777},
          doi = {10.1051/0004-6361:20065177},
       adsurl = {https://ui.adsabs.harvard.edu/abs/2006A&A...455..773V}
}

@ARTICLE{Humphreys2013,
       author = {{Humphreys}, E.~M.~L. and {Reid}, M.~J. and {Moran}, J.~M. and {Greenhill}, L.~J. and {Argon}, A.~L.},
        title = "{Toward a New Geometric Distance to the Active Galaxy NGC 4258. III. Final Results and the Hubble Constant}",
      journal = {\apj},
         year = 2013,
        month = sep,
       volume = {775},
       number = {1},
          eid = {13},
        pages = {13},
          doi = {10.1088/0004-637X/775/1/13},
archivePrefix = {arXiv},
       eprint = {1307.6031},
 primaryClass = {astro-ph.CO},
       adsurl = {https://ui.adsabs.harvard.edu/abs/2013ApJ...775...13H}
}

@ARTICLE{Courtois2011,
       author = {{Courtois}, H{\'e}l{\`e}ne M. and {Tully}, R. Brent and {Makarov}, D.~I. and {Mitronova}, S. and {Koribalski}, B. and {Karachentsev}, I.~D. and {Fisher}, J. Richard},
        title = "{Cosmic Flows: Green Bank Telescope and Parkes H I observations}",
      journal = {\mnras},
         year = 2011,
        month = jul,
       volume = {414},
       number = {3},
        pages = {2005-2016},
          doi = {10.1111/j.1365-2966.2011.18515.x},
archivePrefix = {arXiv},
       eprint = {1101.3802},
 primaryClass = {astro-ph.CO},
       adsurl = {https://ui.adsabs.harvard.edu/abs/2011MNRAS.414.2005C}
}

@ARTICLE{jaeger2017_2,
       author = {{de Jaeger}, T. and {Galbany}, L. and {Filippenko}, A.~V. and {Gonz{\'a}lez-Gait{\'a}n}, S. and {Yasuda}, N. and {Maeda}, K. and {Tanaka}, M. and {Morokuma}, T. and {Moriya}, T.~J. and {Tominaga}, N. and {Nomoto}, K. and {Komiyama}, Y. and {Anderson}, J.~P. and {Brink}, T.~G. and {Carlberg}, R.~G. and {Folatelli}, G. and {Hamuy}, M. and {Pignata}, G. and {Zheng}, W.},
        title = "{SN 2016jhj at redshift 0.34: extending the Type II supernova Hubble diagram using the standard candle method}",
      journal = {\mnras},
         year = 2017,
        month = dec,
       volume = {472},
       number = {4},
        pages = {4233-4243},
          doi = {10.1093/mnras/stx2300},
archivePrefix = {arXiv},
       eprint = {1709.01513},
 primaryClass = {astro-ph.HE},
       adsurl = {https://ui.adsabs.harvard.edu/abs/2017MNRAS.472.4233D}
}

@ARTICLE{ct2015,
       author = {{Courtois}, H{\'e}l{\`e}ne M. and {Tully}, R. Brent},
        title = "{Update on H I data collection from Green Bank, Parkes and Arecibo telescopes for the Cosmic Flows project}",
      journal = {\mnras},
         year = 2015,
        month = feb,
       volume = {447},
       number = {2},
        pages = {1531-1534},
          doi = {10.1093/mnras/stu2405},
archivePrefix = {arXiv},
       eprint = {1411.5719},
 primaryClass = {astro-ph.GA},
       adsurl = {https://ui.adsabs.harvard.edu/abs/2015MNRAS.447.1531C}
}

@ARTICLE{Gallagher1975,
       author = {{Gallagher}, J.~S. and {Faber}, S.~M. and {Balick}, B.},
        title = "{H I in early-type galaxies. I. Observations.}",
      journal = {\apj},
         year = 1975,
        month = nov,
       volume = {202},
        pages = {7-21},
          doi = {10.1086/153948},
       adsurl = {https://ui.adsabs.harvard.edu/abs/1975ApJ...202....7G}
}

@ARTICLE{phillips1993,
       author = {{Phillips}, M.~M.},
        title = "{The Absolute Magnitudes of Type IA Supernovae}",
      journal = {\apjl},
         year = 1993,
        month = aug,
       volume = {413},
        pages = {L105},
          doi = {10.1086/186970},
       adsurl = {https://ui.adsabs.harvard.edu/abs/1993ApJ...413L.105P}
}

@ARTICLE{Hamuy1996,
       author = {{Hamuy}, Mario and {Phillips}, M.~M. and {Suntzeff}, Nicholas B. and {Schommer}, Robert A. and {Maza}, Jose and {Aviles}, R.},
        title = "{The Absolute Luminosities of the Calan/Tololo Type IA Supernovae}",
      journal = {\aj},
         year = 1996,
        month = dec,
       volume = {112},
        pages = {2391},
          doi = {10.1086/118190},
archivePrefix = {arXiv},
       eprint = {astro-ph/9609059},
 primaryClass = {astro-ph},
       adsurl = {https://ui.adsabs.harvard.edu/abs/1996AJ....112.2391H}
}

@ARTICLE{Riess1996,
       author = {{Riess}, Adam G. and {Press}, William H. and {Kirshner}, Robert P.},
        title = "{A Precise Distance Indicator: Type IA Supernova Multicolor Light-Curve Shapes}",
      journal = {\apj},
         year = 1996,
        month = dec,
       volume = {473},
        pages = {88},
          doi = {10.1086/178129},
archivePrefix = {arXiv},
       eprint = {astro-ph/9604143},
 primaryClass = {astro-ph},
       adsurl = {https://ui.adsabs.harvard.edu/abs/1996ApJ...473...88R}
}

@ARTICLE{Perlmutter1997,
       author = {{Perlmutter}, S. and {Gabi}, S. and {Goldhaber}, G. and {Goobar}, A. and {Groom}, D.~E. and {Hook}, I.~M. and {Kim}, A.~G. and {Kim}, M.~Y. and {Lee}, J.~C. and {Pain}, R. and {Pennypacker}, C.~R. and {Small}, I.~A. and {Ellis}, R.~S. and {McMahon}, R.~G. and {Boyle}, B.~J. and {Bunclark}, P.~S. and {Carter}, D. and {Irwin}, M.~J. and {Glazebrook}, K. and {Newberg}, H.~J.~M. and {Filippenko}, A.~V. and {Matheson}, T. and {Dopita}, M. and {Couch}, W.~J.},
        title = "{Measurements of the Cosmological Parameters {\ensuremath{\Omega}} and {\ensuremath{\Lambda}} from the First Seven Supernovae at z >= 0.35}",
      journal = {\apj},
         year = 1997,
        month = jul,
       volume = {483},
       number = {2},
        pages = {565-581},
          doi = {10.1086/304265},
archivePrefix = {arXiv},
       eprint = {astro-ph/9608192},
 primaryClass = {astro-ph},
       adsurl = {https://ui.adsabs.harvard.edu/abs/1997ApJ...483..565P}
}

@ARTICLE{woosley1987,
       author = {{Woosley}, S.~E. and {Pinto}, P.~A. and {Martin}, P.~G. and {Weaver}, Thomas A.},
        title = "{Supernova 1987A in the Large Magellanic Cloud: The Explosion of a approximately 20 M$_{sun}$ Star Which Has Experienced Mass Loss?}",
      journal = {\apj},
         year = 1987,
        month = jul,
       volume = {318},
        pages = {664},
          doi = {10.1086/165402},
       adsurl = {https://ui.adsabs.harvard.edu/abs/1987ApJ...318..664W}
}

@ARTICLE{Schlegel1990,
       author = {{Schlegel}, Eric M.},
        title = "{A new subclass of type II supernovae ?}",
      journal = {\mnras},
         year = 1990,
        month = may,
       volume = {244},
        pages = {269-271},
       adsurl = {https://ui.adsabs.harvard.edu/abs/1990MNRAS.244..269S}
}

@ARTICLE{Barbon1979,
       author = {{Barbon}, R. and {Ciatti}, F. and {Rosino}, L.},
        title = "{Photometric properties of type II supernovae.}",
      journal = {\aap},
         year = 1979,
        month = feb,
       volume = {72},
        pages = {287-292},
       adsurl = {https://ui.adsabs.harvard.edu/abs/1979A&A....72..287B}
}

@ARTICLE{Doggett1985,
       author = {{Doggett}, J.~B. and {Branch}, D.},
        title = "{A comparative study of supernova light curves.}",
      journal = {\aj},
         year = 1985,
        month = nov,
       volume = {90},
        pages = {2303-2311},
          doi = {10.1086/113934},
       adsurl = {https://ui.adsabs.harvard.edu/abs/1985AJ.....90.2303D}
}

@ARTICLE{kourkchi2019,
       author = {{Kourkchi}, Ehsan and {Tully}, R. Brent and {Neill}, J. Don and {Seibert}, Mark and {Courtois}, H{\'e}l{\`e}ne M. and {Dupuy}, Alexandra},
        title = "{Global Attenuation in Spiral Galaxies in Optical and Infrared Bands}",
      journal = {\apj},
         year = 2019,
        month = oct,
       volume = {884},
       number = {1},
          eid = {82},
        pages = {82},
          doi = {10.3847/1538-4357/ab4192},
archivePrefix = {arXiv},
       eprint = {1909.01572},
 primaryClass = {astro-ph.GA},
       adsurl = {https://ui.adsabs.harvard.edu/abs/2019ApJ...884...82K}
}

@ARTICLE{Kourkchi2020_2,
       author = {{Kourkchi}, Ehsan and {Tully}, R. Brent and {Eftekharzadeh}, Sarah and {Llop}, Jordan and {Courtois}, H{\'e}l{\`e}ne M. and {Guinet}, Daniel and {Dupuy}, Alexandra and {Neill}, James D. and {Seibert}, Mark and {Andrews}, Michael and {Chuang}, Juana and {Danesh}, Arash and {Gonzalez}, Randy and {Holthaus}, Alexandria and {Mokelke}, Amber and {Schoen}, Devin and {Urasaki}, Chase},
        title = "{Cosmicflows-4: The Catalog of {\ensuremath{\sim}}10,000 Tully-Fisher Distances}",
      journal = {\apj},
         year = 2020,
        month = oct,
       volume = {902},
       number = {2},
          eid = {145},
        pages = {145},
          doi = {10.3847/1538-4357/abb66b},
archivePrefix = {arXiv},
       eprint = {2009.00733},
 primaryClass = {astro-ph.GA},
       adsurl = {https://ui.adsabs.harvard.edu/abs/2020ApJ...902..145K}
}

@ARTICLE{os1968,
       author = {{Oke}, J.~B. and {Sandage}, Allan},
        title = "{Energy Distributions, K Corrections, and the Stebbins-Whitford Effect for Giant Elliptical Galaxies}",
      journal = {\apj},
         year = 1968,
        month = oct,
       volume = {154},
        pages = {21},
          doi = {10.1086/149737},
       adsurl = {https://ui.adsabs.harvard.edu/abs/1968ApJ...154...21O}
}

@ARTICLE{Huang2007,
       author = {{Huang}, J. -S. and {Ashby}, M.~L.~N. and {Barmby}, P. and {Brodwin}, M. and {Brown}, M.~J.~I. and {Caldwell}, N. and {Cool}, R.~J. and {Eisenhardt}, P. and {Eisenstein}, D. and {Fazio}, G.~G. and {Le Floc'h}, E. and {Green}, P. and {Kochanek}, C.~S. and {Lu}, Nanyao and {Pahre}, M.~A. and {Rigopoulou}, D. and {Rosenberg}, J.~L. and {Smith}, H.~A. and {Wang}, Z. and {Willmer}, C.~N.~A. and {Willner}, S.~P.},
        title = "{The Local Galaxy 8 {\ensuremath{\mu}}m Luminosity Function}",
      journal = {\apj},
         year = 2007,
        month = aug,
       volume = {664},
       number = {2},
        pages = {840-849},
          doi = {10.1086/519241},
archivePrefix = {arXiv},
       eprint = {0704.3609},
 primaryClass = {astro-ph},
       adsurl = {https://ui.adsabs.harvard.edu/abs/2007ApJ...664..840H}
}

@ARTICLE{Riess2022_2,
       author = {{Riess}, Adam G. and {Yuan}, Wenlong and {Macri}, Lucas M. and {Scolnic}, Dan and {Brout}, Dillon and {Casertano}, Stefano and {Jones}, David O. and {Murakami}, Yukei and {Anand}, Gagandeep S. and {Breuval}, Louise and {Brink}, Thomas G. and {Filippenko}, Alexei V. and {Hoffmann}, Samantha and {Jha}, Saurabh W. and {D'arcy Kenworthy}, W. and {Mackenty}, John and {Stahl}, Benjamin E. and {Zheng}, WeiKang},
        title = "{A Comprehensive Measurement of the Local Value of the Hubble Constant with 1 km s$^{-1}$ Mpc$^{-1}$ Uncertainty from the Hubble Space Telescope and the SH0ES Team}",
      journal = {\apjl},
         year = 2022,
        month = jul,
       volume = {934},
       number = {1},
          eid = {L7},
        pages = {L7},
          doi = {10.3847/2041-8213/ac5c5b},
archivePrefix = {arXiv},
       eprint = {2112.04510},
 primaryClass = {astro-ph.CO},
       adsurl = {https://ui.adsabs.harvard.edu/abs/2022ApJ...934L...7R}
}

@ARTICLE{Murakami2023,
       author = {{Murakami}, Yukei S. and {Riess}, Adam G. and {Stahl}, Benjamin E. and {D'Arcy Kenworthy}, W. and {Pluck}, Dahne-More A. and {Macoretta}, Antonella and {Brout}, Dillon and {Jones}, David O. and {Scolnic}, Dan M. and {Filippenko}, Alexei V.},
        title = "{Leveraging SN Ia spectroscopic similarity to improve the measurement of H $_{0}$}",
      journal = {\jcap},
         year = 2023,
        month = nov,
       volume = {2023},
       number = {11},
          eid = {046},
        pages = {046},
          doi = {10.1088/1475-7516/2023/11/046},
archivePrefix = {arXiv},
       eprint = {2306.00070},
 primaryClass = {astro-ph.CO},
       adsurl = {https://ui.adsabs.harvard.edu/abs/2023JCAP...11..046M}
}

@ARTICLE{Breuval2024,
       author = {{Breuval}, Louise and {Riess}, Adam G. and {Casertano}, Stefano and {Yuan}, Wenlong and {Macri}, Lucas M. and {Romaniello}, Martino and {Murakami}, Yukei S. and {Scolnic}, Daniel and {Anand}, Gagandeep S. and {Soszy{\'n}ski}, Igor},
        title = "{Small Magellanic Cloud Cepheids Observed with the Hubble Space Telescope Provide a New Anchor for the SH0ES Distance Ladder}",
      journal = {\apj},
         year = 2024,
        month = sep,
       volume = {973},
       number = {1},
          eid = {30},
        pages = {30},
          doi = {10.3847/1538-4357/ad630e},
archivePrefix = {arXiv},
       eprint = {2404.08038},
 primaryClass = {astro-ph.CO},
       adsurl = {https://ui.adsabs.harvard.edu/abs/2024ApJ...973...30B}
}

@ARTICLE{revalski2018,
       author = {{Revalski}, M. and {Crenshaw}, D.~M. and {Kraemer}, S.~B. and {Fischer}, T.~C. and {Schmitt}, H.~R. and {Machuca}, C.},
        title = "{Quantifying Feedback from Narrow Line Region Outflows in Nearby Active Galaxies. I. Spatially Resolved Mass Outflow Rates for the Seyfert 2 Galaxy Markarian 573}",
      journal = {\apj},
         year = 2018,
        month = mar,
       volume = {856},
       number = {1},
          eid = {46},
        pages = {46},
          doi = {10.3847/1538-4357/aab107},
archivePrefix = {arXiv},
       eprint = {1802.07734},
 primaryClass = {astro-ph.GA},
       adsurl = {https://ui.adsabs.harvard.edu/abs/2018ApJ...856...46R}
}

@ARTICLE{Revalski2018_2,
       author = {{Revalski}, M. and {Dashtamirova}, D. and {Crenshaw}, D.~M. and {Kraemer}, S.~B. and {Fischer}, T.~C. and {Schmitt}, H.~R. and {Gnilka}, C.~L. and {Schmidt}, J. and {Elvis}, M. and {Fabbiano}, G. and {Storchi-Bergmann}, T. and {Maksym}, W.~P. and {Gandhi}, P.},
        title = "{Quantifying Feedback from Narrow Line Region Outflows in Nearby Active Galaxies. II. Spatially Resolved Mass Outflow Rates for the QSO2 Markarian 34}",
      journal = {\apj},
         year = 2018,
        month = nov,
       volume = {867},
       number = {2},
          eid = {88},
        pages = {88},
          doi = {10.3847/1538-4357/aae3e6},
archivePrefix = {arXiv},
       eprint = {1809.09105},
 primaryClass = {astro-ph.GA},
       adsurl = {https://ui.adsabs.harvard.edu/abs/2018ApJ...867...88R}
}

@ARTICLE{Revalski2021,
       author = {{Revalski}, Mitchell and {Meena}, Beena and {Martinez}, Francisco and {Polack}, Garrett E. and {Crenshaw}, D. Michael and {Kraemer}, Steven B. and {Collins}, Nicholas R. and {Fischer}, Travis C. and {Schmitt}, Henrique R. and {Schmidt}, Judy and {Maksym}, W. Peter and {Rafelski}, Marc},
        title = "{Quantifying Feedback from Narrow Line Region Outflows in Nearby Active Galaxies. III. Results for the Seyfert 2 Galaxies Markarian 3, Markarian 78, and NGC 1068}",
      journal = {\apj},
         year = 2021,
        month = apr,
       volume = {910},
       number = {2},
          eid = {139},
        pages = {139},
          doi = {10.3847/1538-4357/abdcad},
archivePrefix = {arXiv},
       eprint = {2101.06270},
 primaryClass = {astro-ph.GA},
       adsurl = {https://ui.adsabs.harvard.edu/abs/2021ApJ...910..139R}
}

@ARTICLE{Revalski2022,
       author = {{Revalski}, Mitchell and {Crenshaw}, D. Michael and {Rafelski}, Marc and {Kraemer}, Steven B. and {Polack}, Garrett E. and {Falc{\~a}o}, Anna Trindade and {Fischer}, Travis C. and {Meena}, Beena and {Martinez}, Francisco and {Schmitt}, Henrique R. and {Collins}, Nicholas R. and {Falcone}, Julia},
        title = "{Quantifying Feedback from Narrow Line Region Outflows in Nearby Active Galaxies. IV. The Effects of Different Density Estimates on the Ionized Gas Masses and Outflow Rates}",
      journal = {\apj},
         year = 2022,
        month = may,
       volume = {930},
       number = {1},
          eid = {14},
        pages = {14},
          doi = {10.3847/1538-4357/ac5f3d},
archivePrefix = {arXiv},
       eprint = {2203.07387},
 primaryClass = {astro-ph.GA},
       adsurl = {https://ui.adsabs.harvard.edu/abs/2022ApJ...930...14R}
}

@ARTICLE{Revalski2025,
       author = {{Revalski}, Mitchell and {Crenshaw}, D. Michael and {Polack}, Garrett E. and {Rafelski}, Marc and {Kraemer}, Steven B. and {Fischer}, Travis C. and {Meena}, Beena and {Schmitt}, Henrique R. and {Trindade Falc{\~a}o}, Anna and {Falcone}, Julia and {Shea}, Maura Kathleen},
        title = "{Quantifying Feedback from Narrow Line Region Outflows in Nearby Active Galaxies. V. The Expanded Sample}",
      journal = {\apj},
         year = 2025,
        month = may,
       volume = {984},
       number = {1},
          eid = {32},
        pages = {32},
          doi = {10.3847/1538-4357/adc131},
archivePrefix = {arXiv},
       eprint = {2503.17444},
 primaryClass = {astro-ph.GA},
       adsurl = {https://ui.adsabs.harvard.edu/abs/2025ApJ...984...32R}
}

@book{of2006,
title={Astrophysics of Gaseous Nebulae and Active Galactic Nuclei},
author={Osterbrock, Donald E and Ferland, Gary J},
year={2006},
publisher={University Science Books},
address={Sausalito, CA}
}

@ARTICLE{tully1982,
       author = {{Tully}, R.~B. and {Mould}, J.~R. and {Aaronson}, M.},
        title = "{A color-magnitude relation for spiral galaxies}",
      journal = {\apj},
         year = 1982,
        month = jun,
       volume = {257},
        pages = {527-537},
          doi = {10.1086/160009},
       adsurl = {https://ui.adsabs.harvard.edu/abs/1982ApJ...257..527T}
}

@ARTICLE{rf1968,
       author = {{Rubin}, Vera C. and {Ford}, Jr., W. Kent},
        title = "{Spectrographic Study of the Seyfert Galaxy NGC 3227}",
      journal = {\apj},
         year = 1968,
        month = nov,
       volume = {154},
        pages = {431},
          doi = {10.1086/149773},
       adsurl = {https://ui.adsabs.harvard.edu/abs/1968ApJ...154..431R}
}

@ARTICLE{Appleton2014,
       author = {{Appleton}, P.~N. and {Mundell}, C. and {Bitsakis}, T. and {Lacy}, M. and {Alatalo}, K. and {Armus}, L. and {Charmandaris}, V. and {Duc}, P. -A. and {Lisenfeld}, U. and {Ogle}, P.},
        title = "{Accretion-Inhibited Star Formation in the Warm Molecular Disk of the Green-valley Elliptical Galaxy NGC 3226?}",
      journal = {\apj},
         year = 2014,
        month = dec,
       volume = {797},
       number = {2},
          eid = {117},
        pages = {117},
          doi = {10.1088/0004-637X/797/2/117},
archivePrefix = {arXiv},
       eprint = {1410.7347},
 primaryClass = {astro-ph.GA},
       adsurl = {https://ui.adsabs.harvard.edu/abs/2014ApJ...797..117A}
}

@ARTICLE{Schinnerer2000,
       author = {{Schinnerer}, E. and {Eckart}, A. and {Tacconi}, L.~J.},
        title = "{Distribution and Kinematics of the Circumnuclear Molecular Gas in the Seyfert 1 Galaxy NGC 3227}",
      journal = {\apj},
         year = 2000,
        month = apr,
       volume = {533},
       number = {2},
        pages = {826-849},
          doi = {10.1086/308703},
archivePrefix = {arXiv},
       eprint = {astro-ph/9911487},
 primaryClass = {astro-ph},
       adsurl = {https://ui.adsabs.harvard.edu/abs/2000ApJ...533..826S}
}

@ARTICLE{Watson2011,
       author = {{Watson}, D. and {Denney}, K.~D. and {Vestergaard}, M. and {Davis}, T.~M.},
        title = "{A New Cosmological Distance Measure Using Active Galactic Nuclei}",
      journal = {\apjl},
         year = 2011,
        month = oct,
       volume = {740},
       number = {2},
          eid = {L49},
        pages = {L49},
          doi = {10.1088/2041-8205/740/2/L49},
archivePrefix = {arXiv},
       eprint = {1109.4632},
 primaryClass = {astro-ph.CO},
       adsurl = {https://ui.adsabs.harvard.edu/abs/2011ApJ...740L..49W}
}

@ARTICLE{sorce2013,
       author = {{Sorce}, Jenny G. and {Courtois}, H{\'e}l{\`e}ne M. and {Tully}, R. Brent and {Seibert}, Mark and {Scowcroft}, Victoria and {Freedman}, Wendy L. and {Madore}, Barry F. and {Persson}, S. Eric and {Monson}, Andy and {Rigby}, Jane},
        title = "{Calibration of the Mid-infrared Tully-Fisher Relation}",
      journal = {\apj},
         year = 2013,
        month = mar,
       volume = {765},
       number = {2},
          eid = {94},
        pages = {94},
          doi = {10.1088/0004-637X/765/2/94},
archivePrefix = {arXiv},
       eprint = {1301.4833},
 primaryClass = {astro-ph.CO},
       adsurl = {https://ui.adsabs.harvard.edu/abs/2013ApJ...765...94S}
}

@ARTICLE{neill2014,
       author = {{Neill}, J.~D. and {Seibert}, Mark and {Tully}, R. Brent and {Courtois}, H{\'e}l{\`e}ne and {Sorce}, Jenny G. and {Jarrett}, T.~H. and {Scowcroft}, Victoria and {Masci}, Frank J.},
        title = "{The Calibration of the WISE W1 and W2 Tully-Fisher Relation}",
      journal = {\apj},
         year = 2014,
        month = sep,
       volume = {792},
       number = {2},
          eid = {129},
        pages = {129},
          doi = {10.1088/0004-637X/792/2/129},
archivePrefix = {arXiv},
       eprint = {1407.7528},
 primaryClass = {astro-ph.CO},
       adsurl = {https://ui.adsabs.harvard.edu/abs/2014ApJ...792..129N}
}

@ARTICLE{Torres-Flores2013,
       author = {{Torres-Flores}, S. and {Mendes de Oliveira}, C. and {Plana}, H. and {Amram}, P. and {Epinat}, B.},
        title = "{The Tully-Fisher relations for Hickson compact group galaxies}",
      journal = {\mnras},
         year = 2013,
        month = jul,
       volume = {432},
       number = {4},
        pages = {3085-3096},
          doi = {10.1093/mnras/stt663},
archivePrefix = {arXiv},
       eprint = {1304.4493},
 primaryClass = {astro-ph.CO},
       adsurl = {https://ui.adsabs.harvard.edu/abs/2013MNRAS.432.3085T}
}

@ARTICLE{Frosst2022,
       author = {{Frosst}, Matthew and {Courteau}, St{\'e}phane and {Arora}, Nikhil and {Stone}, Connor and {Macci{\`o}}, Andrea V. and {Blank}, Marvin},
        title = "{The diversity of spiral galaxies explained}",
      journal = {\mnras},
         year = 2022,
        month = aug,
       volume = {514},
       number = {3},
        pages = {3510-3531},
          doi = {10.1093/mnras/stac1497},
archivePrefix = {arXiv},
       eprint = {2204.02412},
 primaryClass = {astro-ph.GA},
       adsurl = {https://ui.adsabs.harvard.edu/abs/2022MNRAS.514.3510F}
}

@ARTICLE{Manzano-King2019,
       author = {{Manzano-King}, Christina M. and {Canalizo}, Gabriela and {Sales}, Laura V.},
        title = "{AGN-Driven Outflows in Dwarf Galaxies}",
      journal = {\apj},
         year = 2019,
        month = oct,
       volume = {884},
       number = {1},
          eid = {54},
        pages = {54},
          doi = {10.3847/1538-4357/ab4197},
archivePrefix = {arXiv},
       eprint = {1905.09287},
 primaryClass = {astro-ph.GA},
       adsurl = {https://ui.adsabs.harvard.edu/abs/2019ApJ...884...54M}
}

@article{Manzano-King2020,
    author = {Manzano-King, Christina M and Canalizo, Gabriela},
    title = {Active galactic nucleus and dwarf galaxy gas kinematics},
    journal = {Monthly Notices of the Royal Astronomical Society},
    volume = {498},
    number = {3},
    pages = {4562-4576},
    year = {2020},
    month = {09},
    issn = {0035-8711},
    doi = {10.1093/mnras/staa2654},
    url = {https://doi.org/10.1093/mnras/staa2654},
    eprint = {https://academic.oup.com/mnras/article-pdf/498/3/4562/33798783/staa2654.pdf},
}

@ARTICLE{Boubel2024,
       author = {{Boubel}, Paula and {Colless}, Matthew and {Said}, Khaled and {Staveley-Smith}, Lister},
        title = "{An improved Tully-Fisher estimate of H$_{0}$}",
      journal = {\mnras},
         year = 2024,
        month = sep,
       volume = {533},
       number = {2},
        pages = {1550-1559},
          doi = {10.1093/mnras/stae1925},
archivePrefix = {arXiv},
       eprint = {2408.03660},
 primaryClass = {astro-ph.CO},
       adsurl = {https://ui.adsabs.harvard.edu/abs/2024MNRAS.533.1550B}
}

@ARTICLE{Freedman2001,
       author = {{Freedman}, Wendy L. and {Madore}, Barry F. and {Gibson}, Brad K. and {Ferrarese}, Laura and {Kelson}, Daniel D. and {Sakai}, Shoko and {Mould}, Jeremy R. and {Kennicutt}, Jr., Robert C. and {Ford}, Holland C. and {Graham}, John A. and {Huchra}, John P. and {Hughes}, Shaun M.~G. and {Illingworth}, Garth D. and {Macri}, Lucas M. and {Stetson}, Peter B.},
        title = "{Final Results from the Hubble Space Telescope Key Project to Measure the Hubble Constant}",
      journal = {\apj},
         year = 2001,
        month = may,
       volume = {553},
       number = {1},
        pages = {47-72},
          doi = {10.1086/320638},
archivePrefix = {arXiv},
       eprint = {astro-ph/0012376},
 primaryClass = {astro-ph},
       adsurl = {https://ui.adsabs.harvard.edu/abs/2001ApJ...553...47F}
}

@ARTICLE{Masters2006,
       author = {{Masters}, Karen L. and {Springob}, Christopher M. and {Haynes}, Martha P. and {Giovanelli}, Riccardo},
        title = "{SFI++ I: A New I-Band Tully-Fisher Template, the Cluster Peculiar Velocity Dispersion, and H$_{0}$}",
      journal = {\apj},
         year = 2006,
        month = dec,
       volume = {653},
       number = {2},
        pages = {861-880},
          doi = {10.1086/508924},
archivePrefix = {arXiv},
       eprint = {astro-ph/0609249},
 primaryClass = {astro-ph},
       adsurl = {https://ui.adsabs.harvard.edu/abs/2006ApJ...653..861M}
}

@ARTICLE{Scolnic2024,
       author = {{Scolnic}, Daniel and {Boubel}, Paula and {Byrne}, Jakob and {Riess}, Adam G. and {Anand}, Gagandeep S.},
        title = "{Calibrating the Tully-Fisher Relation to Measure the Hubble Constant}",
      journal = {arXiv e-prints},
         year = 2024,
        month = dec,
          eid = {arXiv:2412.08449},
        pages = {arXiv:2412.08449},
          doi = {10.48550/arXiv.2412.08449},
archivePrefix = {arXiv},
       eprint = {2412.08449},
 primaryClass = {astro-ph.CO},
       adsurl = {https://ui.adsabs.harvard.edu/abs/2024arXiv241208449S}
}

@ARTICLE{Woo2002,
       author = {{Woo}, Jong-Hak and {Urry}, C. Megan},
        title = "{Active Galactic Nucleus Black Hole Masses and Bolometric Luminosities}",
      journal = {\apj},
         year = 2002,
        month = nov,
       volume = {579},
       number = {2},
        pages = {530-544},
          doi = {10.1086/342878},
archivePrefix = {arXiv},
       eprint = {astro-ph/0207249},
 primaryClass = {astro-ph},
       adsurl = {https://ui.adsabs.harvard.edu/abs/2002ApJ...579..530W}
}

@ARTICLE{Amanullah2010,
       author = {{Amanullah}, R. and {Lidman}, C. and {Rubin}, D. and {Aldering}, G. and {Astier}, P. and {Barbary}, K. and {Burns}, M.~S. and {Conley}, A. and {Dawson}, K.~S. and {Deustua}, S.~E. and {Doi}, M. and {Fabbro}, S. and {Faccioli}, L. and {Fakhouri}, H.~K. and {Folatelli}, G. and {Fruchter}, A.~S. and {Furusawa}, H. and {Garavini}, G. and {Goldhaber}, G. and {Goobar}, A. and {Groom}, D.~E. and {Hook}, I. and {Howell}, D.~A. and {Kashikawa}, N. and {Kim}, A.~G. and {Knop}, R.~A. and {Kowalski}, M. and {Linder}, E. and {Meyers}, J. and {Morokuma}, T. and {Nobili}, S. and {Nordin}, J. and {Nugent}, P.~E. and {{\"O}stman}, L. and {Pain}, R. and {Panagia}, N. and {Perlmutter}, S. and {Raux}, J. and {Ruiz-Lapuente}, P. and {Spadafora}, A.~L. and {Strovink}, M. and {Suzuki}, N. and {Wang}, L. and {Wood-Vasey}, W.~M. and {Yasuda}, N. and {Supernova Cosmology Project}, The},
        title = "{Spectra and Hubble Space Telescope Light Curves of Six Type Ia Supernovae at 0.511 < z < 1.12 and the Union2 Compilation}",
      journal = {\apj},
         year = 2010,
        month = jun,
       volume = {716},
       number = {1},
        pages = {712-738},
          doi = {10.1088/0004-637X/716/1/712},
archivePrefix = {arXiv},
       eprint = {1004.1711},
 primaryClass = {astro-ph.CO},
       adsurl = {https://ui.adsabs.harvard.edu/abs/2010ApJ...716..712A}
}

@ARTICLE{Tiley2016,
       author = {{Tiley}, Alfred L. and {Bureau}, Martin and {Saintonge}, Am{\'e}lie and {Topal}, Selcuk and {Davis}, Timothy A. and {Torii}, Kazufumi},
        title = "{The Tully-Fisher relation of COLD GASS Galaxies}",
      journal = {\mnras},
         year = 2016,
        month = oct,
       volume = {461},
       number = {4},
        pages = {3494-3515},
          doi = {10.1093/mnras/stw1545},
archivePrefix = {arXiv},
       eprint = {1607.01393},
 primaryClass = {astro-ph.GA},
       adsurl = {https://ui.adsabs.harvard.edu/abs/2016MNRAS.461.3494T}
}

@ARTICLE{Baldwin1981,
       author = {{Baldwin}, J.~A. and {Phillips}, M.~M. and {Terlevich}, R.},
        title = "{Classification parameters for the emission-line spectra of extragalactic objects.}",
      journal = {\pasp},
         year = 1981,
        month = feb,
       volume = {93},
        pages = {5-19},
          doi = {10.1086/130766},
       adsurl = {https://ui.adsabs.harvard.edu/abs/1981PASP...93....5B}
}

@ARTICLE{Keel1980,
       author = {{Keel}, W.~C.},
        title = "{Inclination effects on the recognition of Seyfert galaxies.}",
      journal = {\aj},
         year = 1980,
        month = mar,
       volume = {85},
        pages = {198-203},
          doi = {10.1086/112662},
       adsurl = {https://ui.adsabs.harvard.edu/abs/1980AJ.....85..198K}
}
\end{document}